\documentclass{article}
\usepackage[top=32mm, bottom=30mm, left=25mm, right=25mm]{geometry}
\usepackage{amsmath}
\usepackage{amsthm}
\usepackage{amssymb}
\usepackage{graphicx}
\usepackage{epstopdf}
\usepackage{cleveref}
\usepackage[utf8]{inputenc}
\usepackage{amsfonts}
\usepackage{float}
\usepackage[final]{changes}

\crefname{equation}{Eq.\!\!}{Eqs.\!}
\crefname{figure}{Fig.\!\!}{Figs.\!}
\crefname{appendix}{Appendix}{Appendices}
\begin{document}

\title{Gaussian regularization for resonant states:\\ open and dispersive optical systems}
\author{B. Stout$^{1}$, R. Colom$^{2,4}$, N. Bonod$^{1}$ and R.C. McPhedran$^{3}$,\\$^{1}$Aix-Marseille Univ, CNRS, Centrale Marseille, Institut Fresnel, 13397 Marseille, France. \\$^{2}$Avignon Université, UMR 1114 EMMAH, Avignon Cedex 84018, France,\\ $^{3}$IPOS, School of Physics, University of Sydney, 2006, Australia \\ $^{4}$Zuse Institute Berlin, Takustra{\ss}e 7, 14195 Berlin, Germany}
\maketitle

\begin{abstract}
Resonant States (RS), also known as Quasi-Normal Modes (QNMs), are eigenstates that arise in spectral expansions of linear response functions of open systems. Manipulation of these spatially `divergent' oscillating functions requires a departure from the usual definitions of inner product, normalization and orthogonality typically encountered in the studies of closed systems. We show that once RS fields are expanded on a multipole basis, Gaussian regularization methods provide \emph{analytical} results for crucial RS inner product integrals \added{in the problematic region exterior to the scattering system}. Our demonstrations are carried out in the context of light scattering by spatially bounded objects composed of both electrically and magnetically dispersive media, with demonstrative analytic calculations being shown to \emph{completely} retrieve the results of exact Mie theory.
\end{abstract}

\section{Introduction}

The eigenstates of open systems, known as Resonant States (RSs), can be viewed as a means of simultaneously describing both the spatial and temporal behavior of the oscillatory `eigenmodes' of an open system. As such, RSs have been a fruitful, albeit problematic, concept in many branches of physics for over a century,~\cite{lamb00} where they have been given a wide variety of names like: \emph{quasi-normal modes}, \emph{transient modes}, and the famous \emph{leaky modes} of wave-guides. The RS concept was widely developed in quantum mechanics~\cite{Resonancebook} after its pioneering success by Gamow in describing alpha decay~\cite{Gamow28}, and later by Siegert for characterizing nuclear reactions~\cite{Siegert39}.  Throughout the 20$^{\rm th}$ century, quantum RS descriptions continued to draw the attention of prominent physicists like Wigner~\cite{wig55},  Peierls~\cite{CalderonPeierls76}, and Zel'dovich~\cite{zeld61,perelzeld}, the latter of whom described the mathematical difficulties associated with RSs as follows:
``\emph{An exponentially decaying state that describes, for example, the phenomenon of a decay, is characterized by a complex value of the energy, the imaginary part of the energy giving the decay probability. The wave function of this state increases exponentially in absolute value at large distances, and therefore the usual methods of normalization, of perturbation theory, and of expansion in terms of eigenfunctions do not apply to this state}''.
For us, the pivotal words in Zel'dovich's remark are `usual methods'. Indeed, this work shows that `non-traditional' normalization methods allow the RSs to quantitatively express open system spectral response functions.

The utility of RSs is not limited to the quantum realm, and in the 1990s Leung {\it et al.}\cite{leung94a,leung94b} studied various mathematical aspects of the RS states in `classical field' applications, renaming them  ``Quasi-Normal-Modes" (QNMs), a terminology often adopted in photonics and colliding black holes.\cite{QNM_Black_Berti_2009} It thus appears that RSs are a nearly universal tool for describing open systems in arbitrary spatial dimensions, and we restrict attention here to 3D electromagnetic systems for the purpose of discussion. 

Combining the assumption that energy outside a system is `radiated' into a \emph{background media} governed by a Helmholtz type equation satisfying \emph{outgoing} boundary conditions, leads to frequency domain eigenstates  characterized by complex eigenfrequencies, $\omega_{\alpha}= \omega_{\alpha}^{\prime} + i\omega_{\alpha}^{\prime\prime}$, with causality constraining $\omega_{\alpha}^{\prime\prime}$ to be negative (given that we adopted an $\exp(-i\omega t)$ inverse temporal Fourier transform/time harmonic convention). 
In terms of wavenumber, $k_{\alpha}  = \omega_{\alpha}/c$, the associated far-field radial dependence is proportional to 
$\exp\left[ik_{\alpha}(r-ct)\right]/k_{\alpha}r$, giving RS excitations a physically desirable exponential temporal decay,  but necessarily accompanied by exponentially increasing spatial oscillations at large distances, 
(sometimes referred to as the `exponential catastrophe'~\cite{Calderon2010}).
An example of the generality of the RS approach is found in electromagnetic scattering problems~\cite{baum71,haus86,hu09}, where one encounters both vector and scalar wave RS problems. Mathematical descriptions of \emph{vector} RSs are more complex than those for scalar fields, and this was a major motivation for this work (as well as refs.~\cite{McPhStout20,StMcPh17,summie}, where the mathematical groundwork and physical motivation are developed).

For a long time, the mathematical difficulties posed by the spatial divergence of RSs led to a widely held belief they were principally of phenomenological interest, but it has become increasingly appreciated in recent years that RSs can be used as the basis for \emph{quantitative} calculations. Notably, causality considerations regularize the exponentially diverging behavior when one transforms from frequency domain descriptions back into the time domain.\cite{lamb00,Beck1960,Nussenzveig1961,Nuss72,Colom18} Determining the correct normalization of the RSs in the frequency domain long remained a matter of debate with a number of recent papers in photonics advocating various RS normalization schemes, many of which involve a regularization of the RS inner product integrals (see refs.~\cite{Sauvan2013,Krist2015,Muljar2016a,Muljar2016b,yan2018rigorous,Mulj_comment2017} and articles cited in two recent reviews \cite{LalanneRev18,Kristensen:20}). A notable feature of the RS paradigm is that normalization adjusts the eigenstate \emph{phases} in addition to their amplitudes.

For the propose of discussion, normalization schemes can be classified into three principal categories. A first technique relies on the use of perfect matched layers (PML) to regularize the normalization integral \cite{Sauvan2013,yan2018rigorous,LalanneRev18}, which can be interpreted as a deformation of spatial integration paths into the complex plane. This technique is strongly linked to the complex scaling approach developed in the context of  quantum scattering theory \cite{Resonancebook,moiseyev2011non}.
Two other methods make use of a surface integral to complement the volume integral ~\cite{Krist2015,Muljar2016a,Muljar2016b,Kristensen:20}. The definitions for the surface integral in these two approaches differ however and they therefore lead to two different definitions for the norm. The first definition \cite{Krist2015,Kristensen:20} has been inspired by an earlier work of Lei {\it et al.} \cite{lai1990time} which makes use of the Silver-Muller radiation condition that describes the asymptotic behavior of outgoing waves to evaluate the surface term. The volume integral then needs to be calculated over a volume that is sufficiently large for the surface integral to be located in the far-field region. Another definition of the norm integral makes use of analytic continuation of the RS field along with some properties of vector analysis to express the surface term.\cite{Muljar2016a,Muljar2016b}

Here we illustrate yet another regularization scheme, along the lines first proposed by Zel'dovich~\cite{zeld61}, to regularize the inner product by introducing a Gaussian `killing' function, $\exp(-\eta r^{2})$ into RS inner product integrals and then taking the limit $\eta \rightarrow 0$ after integration (a technique which nowadays is considered as belonging to the general distribution techniques of regularization by `good functions' \cite{Jones82}, and which agrees with the old technique of fractional calculus dating back to Euler and Leibnitz, and with that of Mellin transforms \cite{Zag06}).  One advantage of this technique is that it rigorously leads to analytic expressions for RS type integrals in the exterior regions where the field divergences occur, (shown in a 1992 paper by one of the present authors and two colleagues~cite{MDS}). This technique was recently extended to 3D electromagnetic RSs where it was referred to somewhat picturesquely (or melodically) as ``Killing Mie Softly'' with the derivations involving distribution theory mathematics by two of the present authors~\cite{McPhStout20}.  Besides its value in replacing numerical integrals with analytic formulas, we will see that this approach also sheds new light on the other regularization schemes, and helps justify their physical justification and equivalence.

This article is organized as follows: \cref{sect:Resstates} reviews our RS formulation of  electromagnetic scattering problems followed by a presentation of RS multipolar representations. In \cref{ssect:RSorth}, we derive orthogonality relations for the RSs of scatterers possessing  electric \emph{and/or} magnetic dispersion. We then show that the analytical formulas for Gaussian regularized multipole integrals given in Ref.\cite{McPhStout20} and \cref{KMS}, verify the general orthogonality formulas in the particular case of spherical scatterers. Next, we derive resonant state normalization formulas for fully dispersive scattering materials, followed by a Gaussian multipole application in the case of  electrically and magnetically dispersive spheres. These analytic formulas open the door to RS studies of meta-materials and illustrate electromagnetic duality in a manner that was not achieved previously. The physical significance of these results and their relationship with other formulations and results in the literature are also discussed in this section, with additional comparisons with the literature being detailed in \cref{app:comp}.

The application of RS spectral expansions to Mie theory, which we have developed and exploited for a number of years now\cite{StMcPh17,summie,Colom18,Grig2013,St15,Lass18}, is reviewed in \cref{sect:ResMieTh}. Advantages of this approach is that it enables RS expansions to remain accurate even far from the resonant frequencies by including the crucial  \emph{non-resonant} terms in the RS expansion. \Cref{Examples} is dedicated to high precision numerical results and applications for dispersive scatterers. These are intended to serve as benchmarks for those using our methods or elaborating  purely numerical models for nanophotonic and metamaterial problems. The conclusion discusses some of the avenues for future investigations and applications. The appendices \ref{app:comp}-\ref{DrudeMod}, elaborate on the mathematical definitions and derivations presented in the main text.

\section{Resonant states}
\label{sect:Resstates}

When formulating Resonant States (RSs) in electromagnetism, it is convenient to
express the electromagnetic field as a six component `ket' state, $\left\vert \Psi\right\rangle$, and appropriately dimensioned source currents, $\left\vert J\right\rangle$, as multi-component fields,
\begin{equation} \left\vert \Psi\right\rangle \equiv\left\vert \begin{matrix} \boldsymbol{E} \\ \boldsymbol{H} \end{matrix} \right\rangle \quad , \quad  \left\vert J\right\rangle \equiv\left\vert \begin{matrix} \boldsymbol{j} \\ \boldsymbol{0} \end{matrix} \right\rangle \;,\label{Psivect} \end{equation}
where $\boldsymbol{E}$ and $\boldsymbol{H}$ are respectively the vector electric and magnetic fields. We found it convenient to adopt, `field theoretic' units of m$^{-3/2}$, but Gaussian units or impedance adjusted SI units are acceptable alternatives since these also assign the same units to both electric and magnetic fields (although we found that m$^{-3/2}$ field dimensions are  particularly well adapted to  Green's functions, energy density, LDOS~\cite{LDOS} applications). Six component electromagnetic field formulations are nothing new, and notably $\left\langle \boldsymbol{r} \vert \Psi\right\rangle$ in our notation is quite similar (but not quite identical) to the 6-component field, $\overrightarrow{\mathbb{F}}(\boldsymbol{r})$, that appeared in an RS study during the writing of an initial version of this manuscript.\cite{Muljarov:18}

The scattering particles are taken to be characterized by spatially local materials with sharp boundaries and temporally dispersive constitutive parameters, $\varepsilon(\boldsymbol{r},\omega)$, and/or $\mu(\boldsymbol{r},\omega)$. The scatterers are assumed to be immersed in an isotropic non-dispersive background medium, described by real-valued constitutive parameters $\varepsilon_{b}$ or $\mu_{b}$, with the background media wave velocity given by $c_{b}=c_{\rm v}\sqrt{\varepsilon_{b}\mu_{b}}$, where $c_{\rm v}$ is the vacuum speed of light. Henceforth, the constitutive parameters of the scatterers will be consistently \emph{normalized} with respect to the background medium properties, \begin{align}
\varepsilon(\boldsymbol{r},\omega)
\equiv\varepsilon_{s}(\boldsymbol{r},\omega)/\varepsilon_{b} \qquad,\qquad \mu(\boldsymbol{r},\omega)  \equiv\mu_{s}(\boldsymbol{r},\omega)/\mu_{b}  \;,
\end{align}
which notably simplifies various derivations, and renders almost all further formulas independent of unit conventions.

Using the above definitions and normalizations, the frequency domain Maxwell equations can be written as a single convenient equation,
\begin{subequations} \begin{equation} \omega\Gamma(\omega)\left\vert \Psi\right\rangle =\mathbb{L}\left\vert \Psi\right\rangle +\frac{1}{i}\left\vert J\right\rangle\;, \end{equation}
where the medium metric, $\Gamma(\omega)$, and the linear 
differential operator, $\mathbb{L}$, are symmetric $6\times 6$ matrix operators whose spatial representations are,
\begin{equation}
\Gamma(\omega)\equiv
\begin{bmatrix}
\varepsilon(\boldsymbol{r},\omega)  & 0\\
0 & -\mu(\boldsymbol{r},\omega)
\end{bmatrix}
  \qquad {\rm and} \qquad 
\mathbb{L}\equiv i c_{b}
\begin{bmatrix}
0 & \nabla\times\\
\nabla\times & 0
\end{bmatrix}  \;.
\end{equation}\label{Maxeq}\end{subequations}
Resonant states are defined as source-free eigensolutions of \cref{Maxeq},
\begin{equation}
\omega_{\alpha}\Gamma(\omega_{\alpha}) \left\vert \Psi_{\alpha}\right\rangle
=\mathbb{L}\left\vert \Psi_{\alpha}\right\rangle \;, \label{RSeq}
\end{equation}
obeying outgoing boundary conditions with \emph{complex} eigenfrequencies,
\begin{equation} \omega_{\alpha}\equiv \omega_{\alpha}^{\prime}+ i \omega_{\alpha}^{\prime\prime}\;, \qquad \omega_{\alpha}^{\prime\prime} < 0 \quad \;, \label{Eqres} \end{equation}
with $\omega_{\alpha}^{\prime\prime} < 0$ a necessary condition for exponential temporal decay.

The frequency dependence of the medium metric, $\Gamma(\omega)$, is required to satisfy Kramers-Kronig relations and  $\Gamma^{\ast}(\omega) = \Gamma(-\omega^{\ast})$, so that fields in the time domain are both causal and real valued. Consequently, even though negative frequency RSs must be included in our analysis, they are not independent of 
the positive frequency solutions. One remarks that
\cref{RSeq} is a self-consistent eigenvalue equation in terms of the complex wavenumber (frequency), and consequently the determination of RS eigenvalues must almost always be carried out numerically.

Taking the complex conjugate of \cref{Eqres} for a given eigenstate $\omega_{\alpha}$, and remarking that $\mathbb{L}^{\ast} = -\mathbb{L}$, one obtains,
\begin{align} -\omega_{\alpha}^{\ast}\Gamma(-\omega^{\ast}_{\alpha}) \left\vert \Psi_{\alpha}^{\ast}\right\rangle =\mathbb{L}\left\vert \Psi_{\alpha}^{\ast}\right\rangle \;, \label{negfreq} \end{align}
which is of the same form as \cref{RSeq}, so that for any RS eigenvalue, $\omega_{\alpha}$, there is an 
associated RS with eigenvalue, $-\omega_{\alpha}^{\ast}$,
which also satisfies outgoing boundary conditions.

Since the RS frequencies are discrete, their index, $\alpha$, can be assigned integer values, but when symmetries permit additional quantum numbers, following the discussion in \cref{ssect:RSdefs} below, we will designate RS indices by, $\alpha(q,...,\ell)$ where $q,...$, denotes symmetry based quantum numbers, while the $\ell$ number adopts both positive and negative integer indices, $\ell=\pm \left((0),1,2,\ldots\right)$ ; the presence of an  $\ell=0$ index depends on the mode symmetry properties. From the discussion of the previous paragraph, we assign the $\ell$ indices such that, $\omega_{\alpha(q,...;-\ell)} = -\omega^{\ast}_{\alpha(q,...,\ell)}$, which requires all $\omega_{\alpha(q,...;0)}$ to have  purely imaginary values.
In the cases considered here, there will generally be at most one RS eigenvalue on the negative imaginary axis for a given set of quantum numbers, 
so we can denote such states $\omega_{\alpha(q,...,\ell = 0)}$ when they occur.

A Green function response operator, $\mathbb{G}$, by definition, produces the electromagnetic field equations of \cref{Maxeq} when acting on arbitrary source currents, $\left\vert J\right\rangle$. Adopting a non-conventional `bra' and `ket' notation discussed below in \cref{ssect:RSorth}, \added{and using a first order Taylor expansion of $\Gamma(\omega)$ about $\Gamma(\omega_{\alpha})$,} the above RS formalism tells us that the \emph{spectral expansion} of the Green function operator in the frequency domain takes the form,
\begin{align} \mathbb{G}(\omega) \rightarrow i \sum_{\alpha} \frac{1}{\langle \Psi_{\alpha}^{\rm (arb.)} |\left[ \omega \Gamma\right]_{\alpha}^{\prime}|\Psi_{\alpha}^{\rm (arb.)}\rangle } \frac{| \Psi_{\alpha}^{\rm (arb.)}\rangle\langle \Gamma_{\alpha} \Psi_{\alpha}^{\rm (arb.)}| }{\omega-\omega_{\alpha}} + \mathbb{G}_{\rm n.r.}(\omega) \;, \label{Gspect}
\end{align}
which employs the shorthand notations, $\Gamma_{\alpha}\equiv \Gamma(\omega_{\alpha})$ and $\left[ \omega \Gamma\right]_{\alpha}^{\prime}\equiv \frac{d}{d\omega} \left[ \omega \Gamma(\omega)\right]_{\omega=\omega_{\alpha}}$. The operator, $\mathbb{G}_{\rm n.r.}(\omega)$, in \cref{Gspect} \added{is taken to include contributions like static poles \cite{Muljstat2020} and other  \emph{non-resonant state} contributions}  associated with the non-uniqueness of the Green operator and boundary conditions.

The state, $|\Psi_{\alpha}^{\rm (arb.)}\rangle$, in \cref{Gspect} stands for a state which is `arbitrarily' normalized (in the sense that neither \cref{RSeq} nor \cref{Gspect} determines how RSs should be normalized). The Green's function can however be made to `determine' the RS normalization if we assign a dimensionless predetermined number to $\langle \Psi_{\alpha}^{\rm (arb.)} |\left[ \omega \Gamma\right]_{\alpha}^{\prime}|\Psi_{\alpha}^{\rm (arb.)}\rangle$ (a value of `$1$' appearing to be the `natural' choice since this leads to a particularly simple expression for $\mathbb{G}$ in \cref{Gspect}). We will adopt a different normalization choice, in \cref{NormSect}, which facilitates the expressions of linear response theory which possibly give the RS a more physically intuitive interpretation. 

In the following sections, we first show how RSs can be developed on a multipole basis, and how RS product integrals can be evaluated analytically via Gaussian regularization. This methodology is then used to validate a number of RS properties in this section, finally arriving at analytic RS orthogonality and normalization formulas in the case of spherical scatterers.
 
\subsection{Multipolar expansions of resonant states}
\label{ssect:RSdefs}

\begin{figure}[htb]\begin{center} \includegraphics[width=7cm]{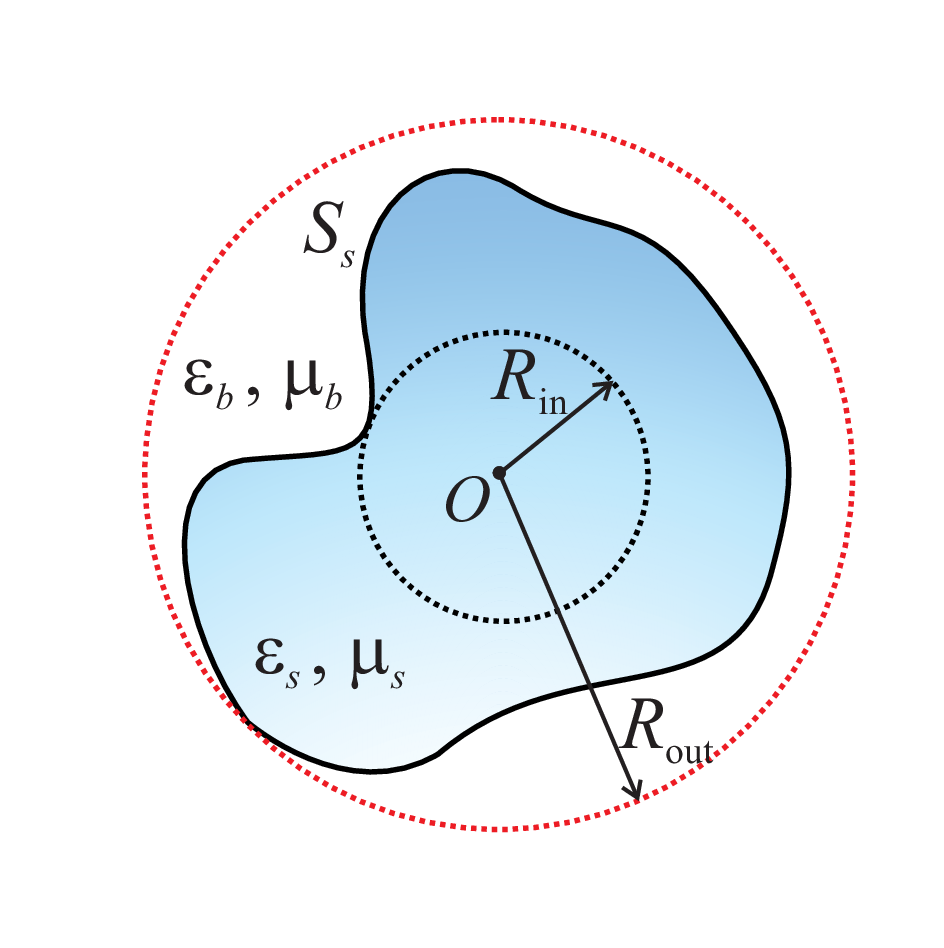} \caption{\label{fig:RinRout}Scattering system (a particle with exterior surface $S_{s}$ in this example) immersed in a background medium of  $\varepsilon_{b}$ and permeability $\mu_{b}$. Only the background material is present outside a sphere of radius, $R_{\rm out}$, surrounding the scattering particle, described by a permittivity $\varepsilon_{s}$ and permeability $\mu_{s}$. A sphere of radius, $R_{\rm in}$, contains only the scattering medium.}
\end{center}
\end{figure}

In any homogeneous region of the scattering system, any three dimensional RS can be expanded in terms of multipole wave functions evaluated at the RS frequency, $\omega_{\alpha}$. Specifically, let us consider the generalizable example of a single scattering particle as shown in \cref{fig:RinRout}. In the homogeneous  \emph{background} media lying outside a sphere of radius $R_{\rm out}$, a RS field, $\Psi_{\alpha}(\boldsymbol{r})$, can be expanded in terms of multipolar basis functions, $\Phi_{q,n,m}^{(+)}(\omega,\boldsymbol{r})$:
\begin{align}
 \langle \boldsymbol{r} \vert \Psi_{\alpha} \rangle \equiv \Psi_{\alpha}(\boldsymbol{r} ) = \sum_{q=0}^{1} \sum_{n=1}^{\infty} \sum_{m=0}^{n} c_{q,n,m}^{(\alpha)} \Phi_{q,n,m}^{(+)}(\omega_{\alpha},\boldsymbol{r}) \qquad \qquad r>R_{\rm out}\;, \label{RSdev}
\end{align}
where $c_{q,n,m}^{(\alpha)}$ are the RS expansion coefficients.

The six component functions, $\Phi_{q,n,m}^{(+)}(\omega,\boldsymbol{r})$, can be characterized according to being either of magnetic ($h$) field types (denoted by a $q=0$ index), or of electric ($e$) field types (denoted by the index $q=1$). The other indices are the  multipole (angular momentum) numbers, $n\ge 1$, and the azimuthal (projection) numbers $m<n$. Using multipole analysis, the $\Phi_{q,n,m}^{(+)}$ functions can be defined as:
\begin{align}\begin{split}
  \Phi_{h,n,m}^{(+)}(\omega,\boldsymbol{r})  &\equiv \Phi_{0,n,m}^{(+)}(\omega,\boldsymbol{r}) \equiv k^{3/2}i^{n} \begin{pmatrix} \boldsymbol{M}_{h,n,m}^{(+)}(k\boldsymbol{r}) \\ -i \boldsymbol{N}_{h,n,m}^{(+)}(k\boldsymbol{r}) \end{pmatrix} \quad \\  \Phi_{e,n,m}^{(+)}(\omega,\boldsymbol{r}) &\equiv \Phi_{1,n,m}^{(+)}(\omega,\boldsymbol{r}) \equiv k^{3/2} i^{n-1} \begin{pmatrix} \boldsymbol{N}_{e,n,m}^{(+)}(k\boldsymbol{r}) \\ -i \boldsymbol{M}_{e,n,m}^{(+)}(k\boldsymbol{r}) \end{pmatrix} \;, \label{Psiout} \end{split}\end{align}
where $k\equiv\omega/c_{b}$ is the exterior medium wavenumber, while  $\boldsymbol{M}^{(+)}_{q,n,m}(k\boldsymbol{r})$ and $\boldsymbol{N}^{(+)}_{q,n,m}(k\boldsymbol{r})$, are the dimensionless outgoing Vector Partial Waves (VPWs), (defined in \cref{VPWs} with additional details and references).
 
If one chooses the origin of the coordinate system to lie inside a scattering particle of interest, one can determine a homogeneous spherical region of radius $R_{\rm in}$ containing only the scattering medium. Inside this region, the RS fields can be expressed,
\begin{align} \Psi_{\alpha}(\boldsymbol{r})  = \sum_{q=0}^{1} \sum_{n=1}^{\infty}\sum_{m=0}^{n} d_{q,n,m}^{(\alpha)} \Phi_{q,n,m}^{(1)}(\omega_{\alpha},\boldsymbol{r})  \qquad \qquad r<R_{\rm in}\;, \label{RSin} \end{align}
where the real-valued RS expansion coefficients are designated $d_{q,n,m}^{(\alpha)}$ and the $\Phi_{q,n,m}^{(1)}$ are formulated in terms of the \emph{`regular'} ({\rm i.e.} singularity free) VPWs, and the normalized (possibly dispersive) constitutive parameters of the scattering media:
\begin{align} \varepsilon(\omega) \equiv \frac{\varepsilon_{s}(\omega)}{\varepsilon_{b}} \quad, \quad \mu(\omega) \equiv \frac{\mu_{s}(\omega)}{\mu_{b}} \quad,\quad \rho(\omega)\equiv\sqrt{\frac{\varepsilon_{s}(\omega) \mu_{s}(\omega)}{\varepsilon_{b}\mu_{b}}}\;, \label{param} \end{align}
so that the regular, medium dependent media wave functions, $\Phi_{q,n,m}^{(1)}$ can be defined as:
\begin{align}\begin{split} \Phi_{h,n,m}^{(1)}\left(\omega,\boldsymbol{r} \right) &\equiv k^{3/2}i^{n}
\begin{pmatrix}
\boldsymbol{M}_{h,n,m}^{(1)}(k\boldsymbol{r}\rho(\omega) ) \\ -i \sqrt{\frac{\varepsilon(\omega)}{\mu(\omega)}}\boldsymbol{N}_{n,m}^{(1)}( k\boldsymbol{r}\rho(\omega))
\end{pmatrix}  \\\Phi^{(1)}_{e,n,m}(\omega,\boldsymbol{r}) 
&\equiv k^{3/2}i^{n-1}
\begin{pmatrix}
\boldsymbol{N}_{n,m}^{(1)}( k\boldsymbol{r}\rho(\omega)) \\
 -i \sqrt{\frac{\varepsilon(\omega)}{\mu(\omega)}}\boldsymbol{M}_{n,m}^{(1)} (k\boldsymbol{r}\rho(\omega) )
\end{pmatrix}  \;, \label{Psint}
\end{split}\end{align}
where $\boldsymbol{M}_{n,m}^{(1)}$ and $\boldsymbol{N}_{n,m}^{(1)}$ are regular ({\it i.e.} singularity free) VPWs.

One way to solve the RS eigenstate values, $\omega_{\alpha}$, would be to numerically solve the propagation of each $\Phi^{(1)}_{q,n,m}(\omega,\boldsymbol{r})$ function through the inhomogeneous region, $R_{\rm in}<r<R_{\rm out}$, and then determine the values of  $\omega_{\alpha}$ which allow sets of coefficients $c_{q,n,m}^{(\alpha)}$ and $d_{q,n,m}^{(\alpha)}$ to be determined which satisfy both electric and magnetic boundary values at $R_{\rm in}$ and $R_{\rm out}$. An example of this procedure for spherical particles is given in \cref{ssect:RSsphere} below, but for particles of arbitrary shape, alternative schemes, employing other types of basis functions are generally preferred (cf. articles cited in \cite{LalanneRev18}). 

The generality of the multipole technique stems from the fact that regardless of the means by which RS eigensolutions have been obtained, their solutions in homogeneous regions can be reexpressed in terms of the spherical wave basis described in this section (using for example the techniques described in ref. \cite{Demesy:18}). As mentioned earlier, solving an eigenvalue problems like \cref{RSeq}, only determines the coefficients $c_{q,n,m}^{(\alpha)}$ and $d_{q,n,m}^{(\alpha)}$ up to an overall `normalization' factor, henceforth designated $\mathcal{N}_{\alpha}$, which must be determined by additional criteria which will be specified in \cref{NormSect}.

\subsection{Resonant states of spherical particles} \label{ssect:RSsphere}

For spherically symmetric particles, there is no mixing of the multipole indices and for each distinct set of multipole numbers  ($q$, $n$, $m$), there exists an infinite discrete set of resonant states (degenerate with respect to the $m$ number). As discussed after \cref{negfreq} above, an additional `quantum' index, $\ell$, enumerates the different RS eigenfrequencies, $\omega_{\alpha}$ within a given multipole set so that $\alpha(q, n, m, \ell)$ designates one and only one RS. Due to the symmetries, the RS multipole developments of \cref{RSdev,RSin} simplify here to:
\begin{equation} \Psi_{\alpha}(\boldsymbol{r}) =\left\{ \begin{array}[c]{c} d_{\alpha}\Phi_{q,n,m}^{(1)}( \omega_{\alpha},\boldsymbol{r}) \; \ ;\qquad r<R\\ \; c_{\alpha}\Phi_{q,n,m}^{(+)}( \omega_{\alpha},\boldsymbol{r}) \ \ ;\qquad r>R\; \end{array} \right. \;, \end{equation}
with the RS index, $\alpha$, uniquely identified by the full set of quantum numbers, {\it i.e.} $\alpha(q,n,m,\ell)$.

For spherically symmetric cases, the overall normalization factor, $\mathcal{N}_{\alpha}$, can be associated with the external field coefficient as $c_{\alpha} = 1/\mathcal{N}_{\alpha}$. Given the expressions in \cref{Psiout} and \cref{Psint} for the $\Phi^{(+)}$ and $\Phi^{(1)}$ basis functions, the continuity of the transverse electric and transverse magnetic fields~\cite{Bohr98} respectfully take the form of analytic linear relationships between internal field and external coefficients, which we henceforth write as, $\gamma_{\alpha} \equiv d_{\alpha}/c_{\alpha}$:
\begin{subequations}\label{PsiRes} \begin{align}
\gamma_{\alpha(e,n,\ell)}&= \frac{\rho_{\alpha}}{\varepsilon_{\alpha}} \frac{h_{n}(z_{\alpha})}{j_{n}(\rho_{\alpha}z_{\alpha})} =\rho_{\alpha}\frac{\xi_{n}^{\prime}( z_{\alpha})} {\psi_{n}^{\prime}(\rho_{\alpha} z_{\alpha})}  \label{egamma}\\ \gamma_{\alpha(h,n,\ell)}&=\frac{h_{n}(z_{\alpha})}{j_{n}(\rho_{\alpha}z_{\alpha})}=\mu_{\alpha}\frac{\xi_{n}^{\prime} (z_{\alpha})}{\psi_{n}^{\prime}(\rho_{\alpha} z_{\alpha})}\;, \label{hgamma} \end{align} \label{gammadefs}\end{subequations}
where $z_{\alpha}\equiv k_{\alpha} R$, is the complex size parameter, $j_{n}(z)$ the spherical Bessel functions, and $h_{n}(z)$ the outgoing spherical Hankel functions while $\psi_{n}(z)\equiv z j_{n}(z)$ and $\xi_{n}\equiv z h_{n}(z)$ are their respective Ricatti-Bessel function counterparts (cf. \cref{VPWs}).
Since the respective pairs of continuity conditions for $\gamma_{\alpha}$  in \cref{gammadefs} are different from one another at arbitrary frequencies, the RS frequencies are defined as being the set of discrete frequencies for which \emph{both} conditions on the respective $\gamma_{\alpha} $ are satisfied.

The notations $\varepsilon_{\alpha}$ and $\mu_{\alpha}$, and  $\rho_{\alpha}$, introduced in \cref{gammadefs} are shorthands for designating the  constitutive parameters evaluated at the RS frequency, {\it i.e.},
\begin{align} \varepsilon_{\alpha} \equiv \frac{\varepsilon_{s}(\omega_{\alpha})}{\varepsilon_{b}} \quad, \quad \mu_{\alpha} \equiv \frac{\mu_{s}(\omega_{\alpha})}{\mu_{b}} \quad,\quad \rho_{\alpha}\equiv\sqrt{\frac{\varepsilon_{s}(\omega_{\alpha}) \mu_{s}(\omega_{\alpha})}{\varepsilon_{b}\mu_{b}}}\;, \label{const} \end{align}
which are normalized with respect to the exterior medium in accordance with the notation introduced in \cref{Maxeq}. To summarize, the exact expressions for the RSs of spherical particles can henceforth be written,
\begin{equation} \Psi_{\alpha}(\boldsymbol{r})=\frac{1}{\mathcal{N}_{\alpha}}\left( \begin{array}[c]{c} \gamma_{\alpha}\Phi_{q,n,m}^{(1)}(  \omega_{\alpha},\boldsymbol{r}) \;;\qquad r\le R\\ \ \ \qquad \Phi_{q,n,m}^{(+)}( \omega_{\alpha},\boldsymbol{r})  \;;\quad r \ge R\; \end{array} \right)  \;, \label{Psispher} \end{equation}
The value of the normalization factor, $\mathcal{N}_{\alpha}$, will be determined analytically in \cref{NormSect} for the spherical scatterer case. Alternative compact expressions for the spherical particle RS expression of \cref{Psispher} and comparisons with expressions in the literature are given in \cref{app:comp}.

One can readily verify that satisfying the equalities in \cref{gammadefs} is identical to the condition for the occurrence of poles in the electric and magnetic Mie coefficients respectively (as can be seen by examining \cref{Miecoefs} of \cref{app:Mietheory}).
We will show in \cref{sect:ResMieTh} that the determination of the values of $z_{\alpha}$ is almost all that we need in order to perfectly reconstruct Mie response theory, once the normalization factors, $\mathcal{N}_{\alpha}$, have been determined as functions of $\omega_{\alpha}$ in \cref{Nnorm}. 
\subsection{Inner products and RS orthogonalization}
\label{ssect:RSorth}

In conventional quantum mechanics, inner products are defined using complex conjugated `bra' states, but in a lossy system this would transform loss to gain which is something we need to avoid\cite{LDOS,Ch90}. We thus define `bra' states as, $\left\langle \Psi_{\alpha} \right\vert \equiv \left\langle 
\boldsymbol{E}_{\alpha},\boldsymbol{H}_{\alpha} \right\vert$, 
without complex conjugation of the fields. The inner product of any two electromagnetic states, $\Psi_{\alpha}$ and 
$\Psi_{\beta}$ is thus defined as the integral over a volume, $\mathcal{V}$, inside a surface $\mathcal{S}_{\rm o}$ that is sent to infinity:
\begin{align} \left\langle \Psi_{\beta}|\Psi_{\alpha}\right\rangle \equiv \underset{\mathcal{S_{\rm o}} \rightarrow \infty}{\lim} \int_{\mathcal{V}}d\boldsymbol{r} \left[ \Psi_{\beta}(\boldsymbol{r})\right]^{t}.\Psi_{\alpha}(\boldsymbol{r})\equiv\int_{V_{\infty}}d\boldsymbol{r}\left\{  \boldsymbol{E}_{\beta}(\boldsymbol{r})\cdot\boldsymbol{E}_{\alpha}(\boldsymbol{r}) +\boldsymbol{H}_{\beta}(\boldsymbol{r})\cdot\boldsymbol{H}_{\alpha}(\boldsymbol{r})\right\} \;.\label{prod} \end{align}

Given the fundamental exponential divergence of the RSs in the far-field discussed in the introduction, it could appear that the RS inner product integral of \cref{prod} is `ill-defined', but the Gaussian regularization method assigns unambiguous, finite, values to RS products like those of \cref{prod} in a manner which allows the differential operator, $\mathbb{L}$ of \cref{Maxeq}, to remain symmetric when acting on regularized RS products, {\it i.e.},
\begin{equation} \left\langle \Psi_{\beta} \vert\mathbb{L} \Psi_{\alpha}\right\rangle = \left\langle \mathbb{L} \Psi_{\beta} \vert\Psi_{\alpha}\right\rangle \;.\label{symrel} \end{equation}
Applying the RS equations of motion in \cref{Maxeq} to this relation provides an orthogonality relation for the product states, valid even in the case of temporally dispersive media,
\begin{align} \frac{1}{c_{b}}\left[\left\langle \Psi_{\beta} \vert\mathbb{L} \Psi_{\alpha}\right\rangle - \left\langle \mathbb{L} \Psi_{\beta} \vert\Psi_{\alpha}\right\rangle \right]=0= \left\langle \Psi_{\beta}\right\vert \left[  k_{\alpha}\Gamma(\omega_{\alpha})-k_{\beta}\Gamma(\omega_{\beta})\right]  \left\vert \Psi_{\alpha}\right\rangle \;, \label{ortharb} \end{align}
where $\Gamma(\omega)$ is the frequency dependent medium metric.

\begin{figure}[htb]\begin{center} \includegraphics[width=7cm]{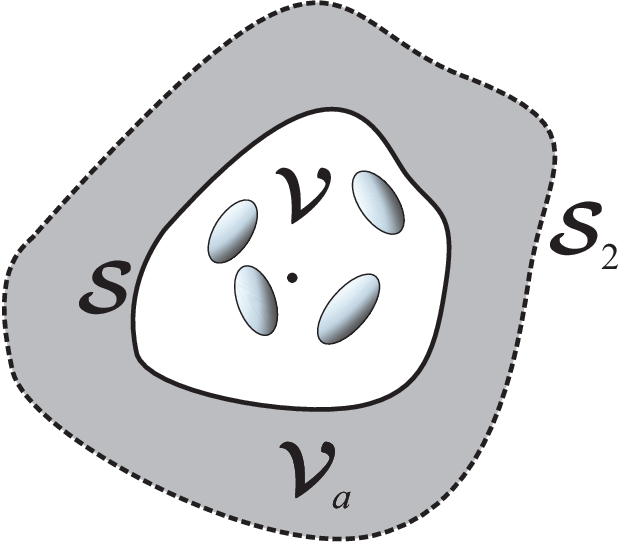} \caption{\label{fig:volume_arb} Annular volume, $\mathcal{V}_{a}$, around a scattering system, with an inner surface, $\mathcal{S}$,  surrounding the system and an exterior surface, $\mathcal{S}_{2}$, that will be sent to infinity.}
\end{center}
\end{figure}

Although the inner product orthogonality relation of \cref{ortharb} is defined as an integration over all space, it can alternatively be reexpressed as an integration over a finite volume supplemented by an appropriate surface integral. 
This is done by remarking that vector identities allow the inner product integrands of the left hand side of \cref{ortharb} to be rewritten:
\begin{align}\begin{split} \frac{1}{ic_{b}}\left[\Psi_{\beta}^{t} . \mathbb{L} \Psi_{\alpha} - \left(\mathbb{L}\Psi_{\beta}\right)^{t} .\Psi_{\alpha} \right]  &  =\left[  \boldsymbol{E}_{\beta},\boldsymbol{H}_{\beta}\right] \begin{bmatrix} 0 & \nabla\times\\ \nabla\times & 0 \end{bmatrix} \begin{bmatrix} \boldsymbol{E}_{\alpha}\\ \boldsymbol{H}_{\alpha} \end{bmatrix} -\left[  \boldsymbol{E}_{\alpha} ,\boldsymbol{H}_{\alpha}\right] \begin{bmatrix} 0 & \nabla\times\\ \nabla\times & 0 \end{bmatrix} \begin{bmatrix} \boldsymbol{E}_{\beta}\\ \boldsymbol{H}_{\beta} \end{bmatrix} \\ &  =\boldsymbol{H}_{\beta}\cdot\nabla\times\boldsymbol{E}_{\alpha} -\boldsymbol{E}_{\alpha}\cdot\nabla\times\boldsymbol{H}_{\beta}-\left[\boldsymbol{H}_{\alpha}\cdot\nabla\times\boldsymbol{E}_{\beta} -\boldsymbol{E}_{\beta}\cdot\nabla\times\boldsymbol{H}_{\alpha}\right]  \\ &  =\nabla\cdot\left(  \boldsymbol{E}_{\alpha}\times\boldsymbol{H}_{\beta}-\boldsymbol{E}_{\beta}\times\boldsymbol{H}_{\alpha}\right) \;. \label{divform}
\end{split}\end{align}
Next, let us consider an arbitrary `annular' closed volume, ${\mathcal{V}_{a}}$, with an inner surface, $\mathcal{S}$, exterior to the scattering system and an outer surface, $\mathcal{S}_{2}$, exterior to $\mathcal{S}$ as illustrated in \cref{fig:volume_arb}.

Integrating over the annular volume integral and applying the divergence theorem to \cref{divform} yields:
\begin{align}\begin{split}
\frac{1}{c_{b}} \int_{{\cal V}_{a}}d\boldsymbol{r} \left[\Psi_{\beta}^{t} . \mathbb{L} \Psi_{\alpha} - \left(\mathbb{L}\Psi_{\beta}\right)^{t} .\Psi_{\alpha}\right]  & =i\bigcirc\hspace{-1.4em}\int\hspace{-0.8em}\int_{\mathcal{S}_{2}}\left[ \boldsymbol{E}_{\alpha}(k_{\alpha}{\boldsymbol{r}})\times \boldsymbol{H}_{\beta}(k_{\beta}{\boldsymbol{r}})-\boldsymbol{E}_{\beta}(k_{\beta}{\boldsymbol{r}})\times \boldsymbol{H}_{\alpha}(k_{\alpha}{\boldsymbol{r}})\right]  \cdot d\boldsymbol{\mathcal S}_{2}\\ &  -i\bigcirc\hspace{-1.4em}\int\hspace{-0.8em}\int_{\mathcal{S}}\left[ \boldsymbol{E}_{\alpha}(k_{\alpha}{\boldsymbol{r}})\times \boldsymbol{H}_{\beta}(k_{\beta}{\boldsymbol{r}})-\boldsymbol{E}_{\beta}(k_{\beta}{\boldsymbol{r}})\times \boldsymbol{H}_{\alpha}(k_{\alpha}{\boldsymbol{r}})\right] \cdot d\boldsymbol{\mathcal S} \;.
\end{split}\label{divrel}\end{align}
Letting the outer surface tend to infinity, the result can only be consistent with \cref{ortharb} provided that,
\begin{align} \underset{S_2 \rightarrow \infty}{\text{lim}} \bigcirc\hspace{-1.4em}\int\hspace{-0.9em}\int_{\mathcal{S}_{2}}\left[ \boldsymbol{E}_{\alpha}(k_{\alpha}{\boldsymbol{r}})\times \boldsymbol{H}_{\beta}(k_{\beta}{\boldsymbol{r}})-\boldsymbol{E}_{\beta}(k_{\beta}{\boldsymbol{r}})\times\boldsymbol{H}_{\alpha}(k_{\alpha}{\boldsymbol{r}})\right]  \cdot d\boldsymbol{\mathcal S}_{2} = 0 \;. \label{outnul} \end{align}
The vanishing of the surface integral, $\mathcal{S}_{2}$, at infinity arises from distribution theory and is discussed at length in ref.\cite{McPhStout20} and reviewed in \cref{KMS}, but it can be intuitively understood as a consequence of the fact that the Gaussian regularization program suppresses the fields at distances far from the origin thanks to an $e^{-\eta r^2}$ factor. 

\Cref{outnul} allows the orthogonality relation of \cref{ortharb} to alternatively be expressed as a volume integral over any finite closed region, $\mathcal{V}$, containing the system, supplemented by an  integral on the surface, $\mathcal{S}$, of $\mathcal{V}$:
\begin{align}\begin{split} \int_{\mathcal{V}}& d\boldsymbol{r}\Psi_{\beta}^{t}(\boldsymbol{r}).\left[ k_{\alpha}\Gamma(\omega_{\alpha})-k_{\beta}\Gamma(\omega_{\beta})\right] .\Psi_{\alpha}(\boldsymbol{r})\\ & + \;i\bigcirc\hspace{-1.4em}\int \hspace{-0.8em}\int_{\mathcal{S}}\left[  \boldsymbol{E}_{\alpha}(k_{\alpha}{\boldsymbol{r}})\times\boldsymbol{H}_{\beta}(k_{\beta}{\boldsymbol{r}})-\boldsymbol{E}_{\beta}(k_{\beta}{\boldsymbol{r}})\times\boldsymbol{H}_{\alpha}(k_{\alpha}{\boldsymbol{r}})\right] \cdot d\boldsymbol{\mathcal S}=0\label{volsurf} \;.\end{split}\end{align}

For lossless particles, the medium metric, $\Gamma$, is frequency independent and real valued, so that \cref{ortharb} reads $(k_{\alpha} - k_{\beta}) \left\langle \Psi_{\beta}\right\vert \Gamma  \left\vert \Psi_{\alpha}\right\rangle = 0$,  which for $k_{\alpha}\neq k_{\beta}$ leads to a frequency independent eigenstate orthogonality condition for lossless systems:
\begin{equation} \label{nodisporth}\begin{split} 0&=\left\langle \Psi_{\beta }\right\vert \Gamma\left\vert \Psi_{\alpha}\right\rangle  =\left\langle \Psi_{\beta}|\Gamma\Psi_{\alpha}\right\rangle =\left\langle \Gamma\Psi_{\beta}|\Psi_{\alpha}\right\rangle \\ &  =\int_{V_{\infty}}d\boldsymbol{r}\left[  \boldsymbol{E}_{\beta}(\boldsymbol{r})\, ,\boldsymbol{H}_{\beta}(\boldsymbol{r})\right]. \begin{bmatrix} \varepsilon(\boldsymbol{r}) & 0\\ 0 & -\mu(\boldsymbol{r}) \end{bmatrix}. \begin{bmatrix} \boldsymbol{E}_{\alpha}(\boldsymbol{r})\\ \boldsymbol{H}_{\alpha}(\boldsymbol{r}) \end{bmatrix} \\ &  =\int_{V_{\infty}}d\boldsymbol{r}\left\{  \varepsilon(\boldsymbol{r})\boldsymbol{E}_{\beta}(\boldsymbol{r}) \cdot\boldsymbol{E}_{\alpha}(\boldsymbol{r})-\mu(\boldsymbol{r})\boldsymbol{H}_{\beta}(\boldsymbol{r})\cdot\boldsymbol{H}_{\alpha}(\boldsymbol{r})\right\}  \;, \end{split} \end{equation}
which can of course also be expressed in the finite volume-surface form of \cref{volsurf}.

It is worth remarking at this stage that the presence of the medium metric, $\Gamma$, gives a Lagrangian aspect to RS field products (as opposed to the more common Hamiltonian type products), which  will become even more apparent when we consider RS normalization in \cref{NormSect}.

\subsubsection{Regularizing multipole expansion products}
\label{sssect:RSmult}

Once the RS fields are developed according to \cref{RSdev}, then the analytic expressions for RS product integrals, given in \cref{KMS}, allow one to rapidly evaluate all the product integrals required for orthogonalization and normalization in the problematic far-field region. Concretely, for an arbitrary pair of resonant states, $\Psi_{\alpha}$ and $\Psi_{\beta}$, developed according to \cref{RSdev}, the volume integral of RS product in the region $r>R_{\rm out}$ (cf. \cref{fig:RinRout}) is: 
\begin{align} \int_{R_{\rm out}}^{\infty} d\boldsymbol{r} \left[ \Psi_{\beta}^{t}(\boldsymbol{r}) \right]^{t} .\Gamma. \Psi_{\alpha}(\boldsymbol{r}) = \sum_{q,n,m} c_{\beta(q,n,m)} c_{\alpha(q,n,m)} \int_{R_{\rm out}}^{\infty}dr \Phi_{q,n,m}^{(+)}(k_{\beta},\boldsymbol{r}).\Gamma. \Phi_{q,n,m}^{(+)}(k_{\alpha},\boldsymbol{r}) \;, \label{outint} \end{align}
where the orthogonality of the VPWs over angular integration results in there being only a single sum over basis indices rather than the double sum that would generally occur using other basis functions. Analytic expressions for all the integrals on the right hand side of \cref{outint} are given in \cref{outanalab} and \cref{outanalaa} of \cref{KMS} for $\alpha\neq \beta$ and $\alpha= \beta$ respectively.

Inside a homogeneous region, $r<R_{\rm in}$, lying completely inside a scattering object (cf. \cref{fig:RinRout}),  one can again develope the RS field on a multipole basis according to \cref{RSin}, and the RS volume integrals are:
\begin{align} \int_{0}^{R_{\rm in}} d\boldsymbol{r} \Psi_{\beta}^{t}(\boldsymbol{r}) . \Gamma .\Psi_{\alpha}(\boldsymbol{r}) = \sum_{q,n,m} d_{\beta(q,n,m)}d_{\alpha(q,n,m)}  \int_{0}^{R_{\rm in}} d\boldsymbol{r} \left[\Phi_{q,n,m}^{(1)}(\rho_{\beta}k_{\beta}, \boldsymbol{r})\right]^{t}. \Gamma . \Phi_{q,n,m}^{(1)}(\rho_{\alpha}k_{\alpha},\boldsymbol{r}) \;, \label{inint}\end{align} 
with ordinary analytic expressions for the integrals on the right hand side of \cref{inint} being given in \cref{inanalab} and \cref{inanalaa} of \cref{KMS}. Numerical integration is still required for volume product integrals of RS fields in heterogeneous regions, like the region $R_{\rm in} < r < R_{\rm out}$ in \cref{fig:RinRout}, but for spherical scatterers, one can take $R_{\rm in}=R_{\rm out}=R$, and the inner product integrals for RS orthogonalization and normalization of become entirely analytic as we demonstrate below.

\subsubsection{Orthogonalization : Example with a spherical scatterer}
Let us consider the case of a lossless sphere with relative permittivity, $\varepsilon = 16$ and null permeability contrast, $\mu=1$. A few low frequency electric mode RSs are graphically represented in \cref{fig:RSMie} and given to high numerical precision in \Cref{tablenodisp}. Since the sphere is lossless, the frequency independent RS orthogonality relation of \cref{nodisporth} applies. This orthogonality relation requires that \emph{either} the electric field and magnetic product integrals exactly cancel one another, \emph{or} that they are both null. However, we know that for a system with material losses, knowledge of the electric field completely determines the magnetic field and \emph{vice versa}, which leads us to the conclusion that RS orthogonality for dispersionless scatterers corresponds to the magnetic and electric field product integrals being independently zero. Rather than mathematically proving this statement, we next invoke the formulas of \cref{KMS} to show that this is indeed true for a spherical scatterer. 

Let us consider a \emph{magnetic} field product of two distinct \emph{electric} type RSs ($q=0$) of a \emph{dispersionless} spherical scatterer. Rescaling the radial variable by the sphere radius, $\tilde{r}\equiv r/R$, a volume integral of the (complex valued) magnetic fields inside the sphere can then be evaluated as,
\begin{subequations} \begin{align}
&  \int_{0}^{R} d\boldsymbol{r} \mu \boldsymbol{H}_{e,n,m} (k_{\beta}\boldsymbol{r})  \cdot\boldsymbol{H}_{e,n,m} (k_{\alpha}\boldsymbol{r}) =-\frac{z_{\alpha}^{3/2}z_{\beta}^{3/2}\gamma_{\alpha(e,n)} \gamma_{\beta(e,n)}} {\mathcal{N}_{\alpha}\mathcal{N}_{\beta}}\varepsilon \left\{ \int_{0}^{1} \tilde{r}^{2}j_{n}( \rho z_{\alpha}\tilde{r} )  j_{n} ( \rho z_{\beta}\tilde{r})  d\tilde{r}  \right\} \label{inteh}\\ & =\frac{z_{\alpha}^{3/2}z_{\beta}^{3/2}\gamma_{\alpha(e,n)} \gamma_{\beta(e,n)}}{\mathcal{N}_{\alpha} \mathcal{N}_{\beta}}\varepsilon \left\{ \frac{ z_{\beta} j_{n}(\rho z_{\alpha})j_{n}^{\prime} (\rho z_{\beta}) -z_{\alpha}j_{n}^{\prime}(\rho z_{\alpha}) j_{n}(  \rho z_{\beta})  }{\rho \left( z_{\alpha}^{2}-z_{\beta}^{2}\right) } \right\} \label{Wats}\\ &   =\frac{z_{\alpha}^{3/2}z_{\beta}^{3/2}}{\mathcal{N}_{\alpha} \mathcal{N}_{\beta}}\frac{z_{\alpha}h_{n}^{\prime} (z_{\alpha})h_{n}(z_{\beta})-z_{\beta}h_{n}^{\prime} (z_{\beta})h_{n}(z_{\alpha})}{z_{\alpha}^{2}-z_{\beta}^{2}} \;,\label{rltRmag}
\end{align}\end{subequations}
where \cref{inteh} is obtained after analytical integration of the angular variables, while the term in brackets of (\ref{Wats}) is a direct application of the bracketed integral in \cref{inteh} as given by Eq.(85a) of~\cite{McPhStout20} (cf. also Watson ref.~\cite{Watson80}, and \cref{watsoninteg,jjint}). The final result of \cref{rltRmag} exploits  the expressions of \cref{egamma} for $\gamma_{\alpha(e,n)}$, and employs algebraic manipulations that are \emph{only true} when $z_{\alpha}$ and $z_{\beta}$ are \emph{distinct} RS size parameter eigenvalues ({\it i.e.} $z_{\alpha}\neq z_{\beta}$).

Thanks to the Gaussian regularization program, we find an analytic expression for integral in the region outside the sphere (where $\mu(r)=1$ due to medium normalization),
\begin{subequations}\begin{align}
&  \int_{R}^{\infty} d\boldsymbol{r} \boldsymbol{H}_{e,n,m}(k_{\beta} \boldsymbol{r})   \cdot\boldsymbol{H}_{e,n,m}(k_{\alpha}\boldsymbol{r}) \longrightarrow -\frac{z_{\alpha}^{3/2}z_{\beta}^{3/2}}{\mathcal{N}_{\alpha}\mathcal{N}_{\beta}} \underset{\eta \rightarrow 0}{\text{lim}} \int_{1}^{\infty}\tilde{r}^{2}h_{n} (z_{\alpha}\tilde{r})  h_{n} (z_{\beta}\tilde{r}) e^{-\eta \tilde{r}^{2}} dr \label{exteh}\\ &   = - \frac{z_{\alpha}^{3/2}z_{\beta}^{3/2}} {\mathcal{N}_{\alpha} \mathcal{N}_{\beta}}\frac{z_{\alpha}h_{n}^{\prime}(z_{\alpha})h_{n} (z_{\beta})-z_{\beta}h_{n}(z_{\alpha})h_{n}^{\prime}(z_{\beta})} {z_{\alpha}^{2}-z_{\beta}^{2}} \;,\label{rgtRmag}
\end{align}
\label{Rmag}\end{subequations}
where we used Eq.(103a) in ref.~\cite{McPhStout20} to evaluate the integral in the first line of this equation (reproduced in \cref{hhint}). The `Killing' regularization, described in detail in ref.~\cite{McPhStout20} and \cref{KMS}, tames the integrand's exponentially divergent amplitude with a $e^{-\eta r^{2}}$ factor in order to yield the finite analytic expression of \cref{rgtRmag} in the $\eta \rightarrow 0$ limit. We ignored irrelevant overall sign factors in the above demonstration because it is now clear that the integrals in \cref{rltRmag,rgtRmag} exactly cancel, so that indeed there is a magnetic field orthogonality of RS states in spheres for dispersion free media to yield,
\begin{align} \int_{\mathcal{V}_{\infty}} d\boldsymbol{r} \mu (\boldsymbol{r}) \boldsymbol{H}_{q,n,m}(k_{\beta} \boldsymbol{r}) \cdot \boldsymbol{H}_{q,n,m}(k_{\alpha}\boldsymbol{r}) = 0 \label{magorth}\;, \end{align}
where we dropped both the multiple indices, $n$ and the field type index `$e$' $(q=0)$ in \cref{magorth} since as shown in \cref{KMS} that analogous orthogonality relations holds for magnetic, (`$h$' $(q=1)$), type RS modes.

\begin{figure}[htb]\begin{center} \includegraphics[width=12cm]{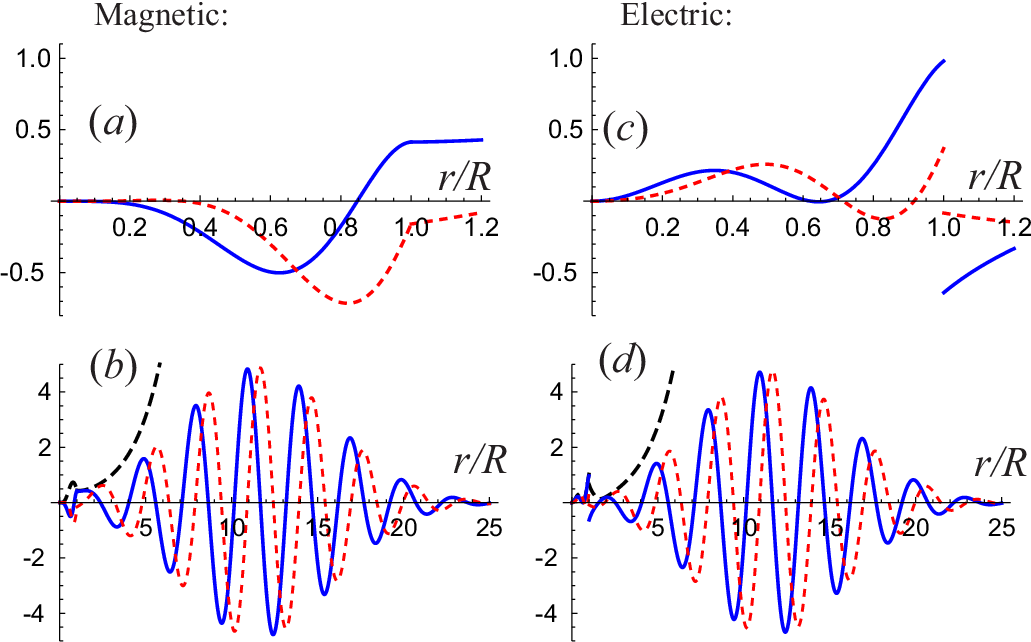} \caption{\label{fig:Orth}Numerical plots of the RS  magnetic inner product integrands of \cref{Rmag} (i.e. $r^{2}h_{n} (z_{\alpha}r)  h_{n} (z_{\beta}r) e^{-\eta r^{2}}$)  for $z_{\alpha(e,n=1,\ell=1)}\simeq 1.0395 - i 0.5009$ and $z_{\beta(e,1,\ell=2)} \simeq 1.0527 - i 0.072355$.
The real (blue-solid) and imaginary (red-dotted) parts of \cref{Rmag}, are plotted in (a) in the range ($r\in \{ 0,1.2 \}$) and for ($r\in \{ 0, 25 \}$) with a $\eta=0.025$ killing factor. Analogous plots for the electric field integrand of \cref{elecorthap} are given in (c) and (d). The envelope of the curves when $\eta=0$ is given by black dashed lines in (b) and (d).}
\end{center}
\end{figure}

In order to appreciate the complexity of what is  accomplished in \cref{magorth}, the real and imaginary parts of a RS magnetic product integrand of the first two electric dipole RSs at $z_{\alpha(e,n=1,\ell=1)}\simeq 1.0395 - i 0.5009$ and $z_{\beta(e,n=1,\ell=2)}\simeq 1.0527 - i 0.072355$ of a high index dielectric sphere ($\varepsilon=16$) are plotted in \cref{fig:Orth} (cf. \Cref{tablenodisp} and \cref{fig:RSMie}).  Specifically, the real and imaginary parts of the magnetic field product integrands given in \cref{inteh} ($r<R$) and \cref{exteh} ($r>R$) are plotted first for $r/R$ in \cref{fig:Orth}(a), up to $1.2$ and then extended in \cref{fig:Orth}(b) up to values of $r/R=25$, where a Gaussian killing factor of $\eta=0.025$ is adopted for values of $r>R$. High accuracy numerical calculations confirm that the integrals of both the real and imaginary parts of the total magnetic field integral do indeed tend towards zero when a sufficiently small value of $\eta$ in the region $r/R>1$ is adopted.

Analogous result holds for RS orthogonality in terms of an electric field despite presence of discontinuities in the electric field component normal to the scatterer surface. Analytical formulas for electric field orthogonality are given explicitly in \cref{Orthint} and one finds,
\begin{align} &\int_{\mathcal{V}_{\infty}} d\boldsymbol{r} \varepsilon(\boldsymbol{r})\boldsymbol{E}_{e,n,m}(k_{\beta}\boldsymbol{r}) \cdot\boldsymbol{E}_{e,n,m} (k_{\alpha}\boldsymbol{r}) =0 \label{elecorth} \;, \end{align}
when $\alpha\neq\beta$, a result which is illustrated graphically by plotting the electric field integrands of \cref{intee} and \cref{exteeint} for $r/R$ up to $1.2$ and \cref{fig:Orth}(d) for $r/R$ up to $25$. As shown in \cref{Orthint}, the mathematics for the electric field integration was somewhat more complicated than it was for the magnetic field integrals of \cref{rgtRmag} since the required integrals were not given in standard references like that of Watson~\cite{Watson80}, but they are provided in reference~\cite{McPhStout20}.

\begin{figure}[htb]
\begin{center}
\includegraphics[width=8cm]{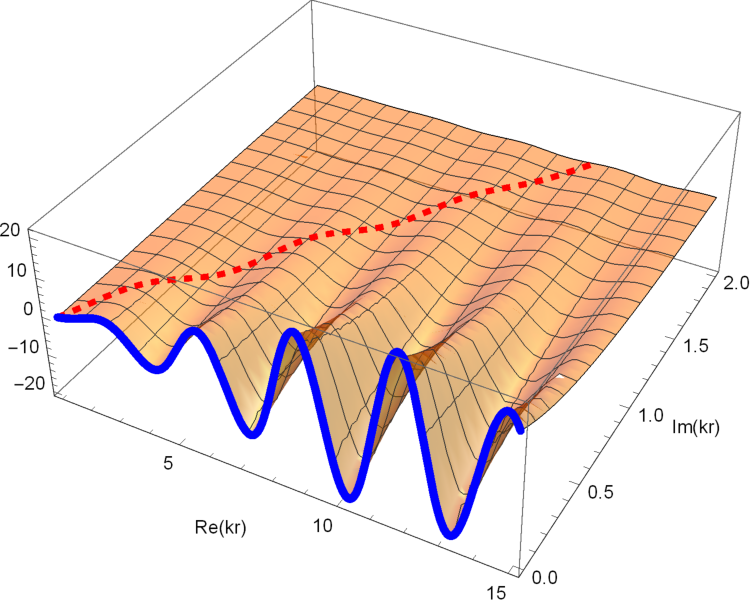}
\caption{\label{fig:PML_path}Plot of the real part of \cref{Rmag}  in the complex $kr$ plane for $\eta=0.025$. A Gaussian integration path along the real axis is given by the thick blue curve, while the red dashed curve indicates the path of a contour shifted by an angle of $\pi/4$ into the complex plane.}
\end{center}
\end{figure}

An advantage of Gaussian regularization is that it can replace numerical integrations by analytical expressions (once the multipole development of the RS fields outside the system are determined). Although Gaussian regularization could also be used as numerical method, this may not be the most efficient choice for the following reason:  a numerical evaluation of a Gaussian regularized RS integral should adopt a small value of $\eta$ , so as to be close to the $\eta\to 0$ limit, but small values of $\eta$ lead to large amplitude far-field oscillations (cf. \cref{fig:PML_path}), which renders numerical integrations time consuming. For numerical evaluations, it is often preferable to generalize the radial coordinate to complex values and deform radial integration path so that it goes to infinity at a finite angle from the real axis in the complex plane. This procedure is commonly employed in temporal analysis of applied mathematics and lies at the heart of the PML method.\cite{Hugonin}

We can readily understand the numerical advantage of the complex contour method  from \cref{fig:PML_path}, which plots the real part of \cref{Rmag}, {\it i.e.} ${\rm Re}\left[r^{2}h_{n} (z_{\alpha}r) h_{n} (z_{\beta}r) e^{-\eta r^{2}}\right]$ (with $\eta=0.025$) in the complex plane. The Gaussian regularized contour on the real axis is plotted as a thick blue line, while the red dashed curve indicates the path of a contour rotated by an angle of $\pi/4$ into the complex plane. The integrals along the two contours must be the same, but the red dashed PML type contour is clearly more manageable numerically and indeed a Gaussian regularization is superfluous for the rotated path which could be evaluated directly without regularization ({\it i.e.} for $\eta = 0$). Although it is visually clear from \cref{fig:Orth} and \cref{fig:PML_path} that RS product integrands oscillate around zero outside the scatterer, they do not quite average to zero since the RS orthogonality relation of \cref{magorth} requires the exterior product integral to be equal and opposite to the field product integral inside the sphere, (this interior integral is generally not null, as seen in this example by inspection of \cref{fig:Orth}(a) and \cref{fig:Orth}(b)).   

The utility of analytic Gaussian regularization formulas is further highlighted by remarking that if  one also evaluates the electric field product integrals (cf.  \cref{elecorthap} in \cref{Orthint}), then after some algebra one finds, when $k_{\alpha} \neq k_{\beta}$, that:
\begin{align}\begin{split} &\left( k_{\alpha}-k_{\beta} \right)\int_{{\mathcal{V}_{R}}}d\boldsymbol{r}\left[  \varepsilon(\boldsymbol{r}) \boldsymbol{E}_{q,n,m}(k_{\beta}\boldsymbol{r}) \cdot \boldsymbol {E}_{q,n,m} (k_{\alpha}\boldsymbol{r})-\mu(\boldsymbol{r})\boldsymbol{H}_{q,n,m}(k_{\beta}\boldsymbol{r}) \cdot\boldsymbol{H}_{q,n,m}(k_{\alpha} \boldsymbol{r})\right] \\ &\qquad\qquad\qquad\qquad + r^{2} (-1)^{(n-q)}\frac{ k_{\alpha}^{3/2}k_{\beta}^{3/2}}{\mathcal{N}_{\alpha} \mathcal{N}_{\beta}} \left[h_{n}(R k_{\alpha})\frac{\xi_{n}^{\prime}(R k_{\beta})}{R k_{\beta}}-h_{n}(R k_{\beta})\frac{\xi_{n}^{\prime}(R k_{\alpha})}{R k_{\alpha}} \right]  = 0 \;, \label{elecmagorth}\end{split}\end{align}
where the first term is a volume integral of fields inside a sphere of radius, $R$, denoted $\mathcal{V}_{R}$, and the second term is a quantity evaluated at the sphere's surface. It is worth remarking that although $R$ can be `naturally' taken to be the particle radius, \cref{elecmagorth} remains valid for any sphere surrounding the scatterer.

An examination of \cref{elecmagorth} shows that the volume integral and the second, `surface', contribution are none other than special cases (spherical volume and surface) of the more general volume and surface integrals derived in \cref{volsurf}. Although the explicit agreement between the analytic Gaussian regularization method with \cref{volsurf} was only shown here for the case of a dispersionless sphere, analogous manipulations confirm this agreement for the case of dispersive spheres, at the cost of some additional algebra.

\subsection{Resonant state normalization}
\label{NormSect} 

A variety of methodologies have been adopted to determine RS normalization \cite{Krist2015,Muljar2016a,Muljar2016b,yan2018rigorous,Doost14}, which has given rise to `different' normalization conditions being proposed for photonic problems, but Kristenssen {\it et al.}\cite{Krist2015} and others have argued that varying normalization formulas should be at least inter-consistent. We agree that final response functions of an RS theory should be independent of regularization schemes, but as we saw that if we adopted \cref{Gspect} as Green function response, the `correctness' of the RS normalization factors can be considered a matter of convention.


We found it advantageous to adopt a field theory based route to normalization and impose the RS norms in \cref{Gspect} be proportional to their associated RS frequencies, $\omega_{\alpha}$, which given the adimensional requirements scalar product leads to:
\begin{align}  \left\langle \Psi_{\alpha}\right\vert \left[\frac{d}{d\omega} (\omega\Gamma) \right]_{\omega_{\alpha}}\left\vert \Psi_{\alpha}\right\rangle   &=\int_{{\cal V}_\infty} d\mathbf{r}\left\{ \left. \frac{d\left[ \omega\varepsilon(\boldsymbol{r},\omega) \right]} {d\omega}\right|_{\omega_{\alpha}} \boldsymbol{E}_{\alpha}^{2}(\boldsymbol{r}) -\left. \frac{d\left[\omega\mu(\boldsymbol{r},\omega)\right]}{d\omega}\right|_{\omega_{\alpha}} \boldsymbol{H}_{\alpha}^{2} (\boldsymbol{r}) \right\} = z_{\alpha} \;, \label{psinorm} \end{align}where we recall that $z_{\alpha} = \omega_{\alpha} R/c$. Since the frequency derivatives of the constitutive factors in $d\left[\omega\varepsilon(\boldsymbol{r},\omega)\right] /d\omega$ and $d\left[\omega\mu(\boldsymbol{r},\omega)\right] /d\omega$ can be interpreted as accounting for the energy associated with material degrees of freedom \cite{JDJ99}, one can interpret the Lagrangian energy of the RS as $\mathcal{E}_{{\alpha}}^{\rm L} = \hbar c\left\langle \Psi_{\alpha}\right\vert \frac{d}{d\omega} (\omega\Gamma) \left\vert \Psi_{\alpha}\right\rangle/R =\hbar\omega_{\alpha}$. The Lagrangian interpretation of \cref{psinorm}, is more transparent in dispersionless media where it simplifies to:
\begin{align}\label{lagint} \mathcal{E}_{{\alpha}}^{\rm L} \propto \left\langle \Psi_{\alpha}\right\vert \Gamma  \left\vert \Psi_{\alpha}\right\rangle = \int_{{\cal V}_\infty} d\mathbf{r} \left\{\varepsilon(\boldsymbol{r}) \boldsymbol{E}_{\alpha}^{2}(\boldsymbol{r})-\mu(\boldsymbol{r}) \boldsymbol{H}_{\alpha}^{2} (\boldsymbol{r})\right\} \overset{\text{no disp.}}{=} z_{\alpha} \;. \end{align}


Despite the divergence of RS field amplitudes, the infinite volume integrals in \cref{psinorm} can be regularized by either Gaussian regularization or by a PML type rotation of the integration path into the complex plane as illustrated in \cref{fig:PML_path}. Yet another way to determine the normalization factors is said to avoid the RS divergences by integrating only over a finite region of space, analogous to the formula found for RS orthogonalization in \cref{ssect:RSorth}. Indeed, it suffices to divide both sides of \cref{volsurf} by $k_{\alpha}- k_{\beta}$, and then take the limit $k_{\alpha}\to k_{\beta}$ to obtain a finite volume expression of
$\left\langle \Psi_{\alpha}\right\vert \left. \frac{d\left( k\Gamma \right)}{d k}\right|_{k_{\alpha}} \left\vert \Psi_{\alpha}\right\rangle$ being equal to a constant (which we assign as $z_{\alpha}$), so that a finite volume normalization expression reads:
\begin{align}\begin{split} \left\langle \Psi_{\alpha}\right\vert &\left. \frac{d\left( k\Gamma \right)}{d k}\right|_{k_{\alpha}} \left\vert \Psi_{\alpha}\right\rangle  \rightarrow \int_{\mathcal{V}}  d\boldsymbol{r}\Psi_{\alpha}^{t}(\boldsymbol{r}). \frac{d}{d k} \left[ k \Gamma(\boldsymbol{r},k) \right]_{k_{\alpha}} .\Psi_{\alpha}(\boldsymbol{r})\\ & + i\bigcirc\hspace{-1.4em}\int \hspace{-0.8em}\int_{\mathcal{S}}\left\{  \left[\frac{d}{d k}\boldsymbol{E}_{\alpha}(k {\boldsymbol{r}})\right]_{k_{\alpha}} \times\boldsymbol{H}_{\alpha}-\boldsymbol{E}_{\alpha} \times \left[ \frac{d}{d k} \boldsymbol{H}_{\alpha}(k{\boldsymbol{r}})\right]_{k_\alpha}\right\} \cdot d\boldsymbol{\mathcal{S}} = z_{\alpha} \label{volsurfnorm} \;, \end{split}\end{align}
where $\mathcal{V}$ is a finite volume surrounding the system and $\mathcal{S}$ its surface. Although not necessary, we used $k=\omega/c_{b}$, to formulate \cref{volsurfnorm} in terms of exterior medium wavenumber, $k$, (as opposed to angular frequency, $\omega$). Several works in recent years have obtained formulas quite similar to \cref{volsurfnorm}\cite{Muljar2016a,Muljar2016b,Doost14}, but the formula with the most explicit agreement is Eq.(20) of Ref.\cite{Muljarov:18}, derived in the context a six-by-six component Green's function (except for a differing RS convention - obtained by replacing $z_{\alpha}$ on the right hand side of \cref{volsurfnorm} by $1$).

The frequency derivative (or $k$ derivative) of fields in the surface integral of \cref{volsurfnorm} could be deemed undesirable for numerical analysis, but we can take advantage of the scale invariance in the region outside the system to replace $d/dk$ by $\frac{\boldsymbol{r}\cdot\boldsymbol{\nabla}}{k}$, as done in Ref.\cite{Muljarov:18,Doost14}. Those favoring a surface integral approach to RS normalization often choose to formulate normalization in terms of volume integrals involving only the electric field (in contrast to the `Lagrangian' field formulation of this work). One can transform from one formulation to the other by invoking a Poynting vector analysis which yields:
\begin{align} \int_{\mathcal{V}} d\boldsymbol{r} \boldsymbol{E}_{\alpha} \cdot \varepsilon_{\alpha}(\boldsymbol{r} ) \cdot \boldsymbol{E}_{\alpha} + \int_{\mathcal{V}} d\boldsymbol{r} \boldsymbol{H}_{\alpha} \cdot \mu_{\alpha}(\boldsymbol{r}) \cdot \boldsymbol{H}_{\alpha}  = \frac{c_{b}}{i\omega_{\alpha}} \bigcirc\hspace{-1.4em}\int \hspace{-0.8em}\int_{\mathcal{S}} \left( \boldsymbol{E}_{\alpha} \times \boldsymbol{H}_{\alpha} \right) \cdot d\boldsymbol{\mathcal{S}} \;, \end{align}
which in the case of null magnetic permeability contrast, {\it i.e.} $\mu\rightarrow 1$ allows the normalization constraint to be written in terms of an \emph{electric} field volume integral:
\begin{align}\begin{split} & 2\int_{\mathcal{V}} d\boldsymbol{r} \boldsymbol{E}_{\alpha} \cdot \frac{d}{d k^{2}} \left[ k^{2} \varepsilon(k^{2}) \right]_{k^{2}_{\alpha}} \cdot \boldsymbol{E}_{\alpha} + \frac{i}{k_{\alpha}}\bigcirc\hspace{-1.4em}\int \hspace{-0.8em}\int_{\mathcal{S}} \left( \boldsymbol{E}_{\alpha} \times \boldsymbol{H}_{\alpha} \right) \cdot d\boldsymbol{\mathcal{S}} \\ & + \frac{i}{k_{\alpha}}\bigcirc\hspace{-1.4em}\int \hspace{-0.8em}\int_{\mathcal{S}}\left\{ \left[\boldsymbol{r}\cdot\boldsymbol{\nabla} \boldsymbol{E}_{\alpha}\right] \times\boldsymbol{H}_{\alpha}-\boldsymbol{E}_{\alpha} \times \left[ \boldsymbol{r}\cdot\boldsymbol{\nabla} \boldsymbol{H}_{\alpha}\right]\right\} \cdot d\boldsymbol{\mathcal{S}} = z_{\alpha}\label{volsurfv2} \;,\end{split}\end{align}
where we used the replacement $\frac{d}{d \omega} \left[  \varepsilon(\omega) \right]_{\omega_{\alpha}} \rightarrow 2\frac{d}{d k^{2}} \left[  \varepsilon(k^{2}) \right]_{k^{2}_{\alpha}}$. \added{The surface terms in \cref{volsurfv2} are radial derivatives, and the first appearance of such a normalization in photonics seems to be have been derived in 2010 for dispersionless media \cite{Muljarov_2010}.}

Even though the normalization formulas of \cref{psinorm,volsurfnorm,volsurfv2} may appear distinct from one another, the Gaussian regularization perspective shows their equivalence, under the proper conditions, since they are all determined by the same underlying condition. This equivalence can be demonstrated directly for a spherical surface, $\mathcal{S}$, of radius $R$ by using Hankel function recursion relations to show that the \emph{surface} integral on the right hand side of \cref{volsurfnorm} of an arbitrary spherical surface of radius $R$ is \emph{analytically identical} to the formula of \cref{EHh2} for a Gaussian regularized \emph{volume} integration of the electromagnetic field carried out in the region outside this sphere. Such analytic equivalencies can also be extended to arbitrary scatterer geometries by invoking the multipole decomposition of \cref{RSdev}.

\subsubsection{Analytic formulas for fully dispersive spherical scatterers}

Analytical RS normalization expressions are derived in \cref{Norminteg} using a Gaussian regularization of the full volume integration of \cref{psinorm} for an electrically and magnetically dispersive sphere adopting the expressions for the RSs in \cref{ssect:RSdefs} and using the formulas in \cref{KMS} and Ref.\cite{McPhStout20}, we find: 
\begin{subequations}\label{Nrmform}\begin{align}\begin{split} \mathcal{N}_{\alpha(h,n,\ell)}^{2} &  =(\varepsilon_{\alpha}-1)\xi_{n}^{2}(z_{\alpha})+(\mu_{\alpha}-1)  \left\{ \left[  \xi_{n}^{\prime}(z_{\alpha})\right]^{2}  +\frac{n(n+1) h_{n}^{2}(z_{\alpha})}{\mu_{\alpha}}\right\} \\ &  \qquad+\frac{\omega_{\alpha}}{2}\left\{\Xi_{\alpha(h,n,\ell)}^{(-)}\left. \frac{d}{d\omega}\ln\varepsilon(\omega)  \right\vert_{\omega_{\alpha}} +\Xi_{\alpha(h,n,\ell)}^{(+)}\left.  \frac{d}{d\omega} \ln\mu(\omega)  \right\vert_{\omega_{\alpha}}\right\} \end{split} \label{hNnorm}\\\begin{split} \mathcal{N}_{\alpha(e,n,\ell)}^{2} &  = (\mu_{\alpha}-1)\xi_{n}^{2}(z_{\alpha})   +(\varepsilon_{\alpha}-1)\left\{ [\xi_{n}^{\prime} (z_{\alpha}) ]^{2}+\frac{n(n+1)} {\varepsilon_{\alpha}}h_{n}^{2}(z_{\alpha})  \right\} \\ &  +\frac{\omega_{\alpha}}{2}\left\{  \Xi_{\alpha(e,n,\ell)}^{(+)}\left. \frac{d}{d\omega}\ln\varepsilon_{\alpha}(\omega) \right\vert_{\omega_{\alpha}}+\Xi_{\alpha(e,n,\ell)}^{(-)}\left. \frac{d}{d\omega} \ln\mu_{\alpha}(\omega) \right\vert_{\omega_{\alpha}}\right\}\end{split} \label{eNnorm}  \end{align} \label{Nnorm}
with the definitions:
\begin{align}
 \Xi^{(\pm)}_{\alpha(h,n,\ell)}&\equiv\mu_{\alpha}\left[\xi_{n}^{\prime}(z_{\alpha})\right]^{2} +\varepsilon_{\alpha}\xi_{n}^{2} (z_{\alpha}) -\frac{n(n+1) h_{n}^{2} (z_{\alpha})} {\mu_{\alpha}}\pm h_{n}(  z_{\alpha})\xi_{n}^{\prime}(z_{\alpha}) \label{Xihdef} \\ \Xi^{(\pm)}_{\alpha(e,n,\ell)} &\equiv \varepsilon_{\alpha}\left[  \xi_{n}^{\prime}(z_{\alpha}) \right]^{2}+\mu_{\alpha}\xi_{n}^{2}(z_{\alpha}) -\frac{n(n+1) h_{n}^{2}(z_{\alpha})} {\varepsilon_{\alpha}} \pm h_{n}(z_{\alpha}) \xi_{n}^{\prime}(  z_{\alpha})  \;.  \label{Xiedef} \end{align} \end{subequations}

For dispersionless media, light scattering becomes scale invariant and \cref{Nnorm} simplifies to expressions which only depend on the size parameter of the mode, $z_{\alpha}$, and the relative constitutive parameters, $\varepsilon$ and $\mu$ (now real-valued) as was previously derived in Ref.\cite{StMcPh17}:
\begin{subequations} \label{nodispnorm} \begin{align}\begin{split} \mathcal{N}_{\alpha(e,n,\ell)}^{2} &\overset{\text{no disp.}}{=} \; \xi_{n}^{2}(z_{\alpha}) (\mu-1) \\ &  \qquad+\left\{[\xi_{n}^{\prime}(z_{\alpha})]^{2}  +\frac{n(n+1)}{\varepsilon}h_{n}^{2}(z_{\alpha})\right\} (\varepsilon-1) \end{split} \\ \begin{split} \mathcal{N}_{\alpha(h,n,\ell)}^{2}  &  \overset{\text{no disp.}}{=} \; \xi_{n}^{2}(z_{\alpha}) (\varepsilon-1) \\ &\qquad +\left\{\left[\xi_{n}^{\prime}(z_{\alpha})\right]^{2}+\frac{n(n+1)}{\mu} h_{n}^{2}(z_{\alpha})\right\}(\mu-1) \;.
\end{split}\end{align} \end{subequations}
Comparisons of the above normalization formulas for spherical scatterers with other analytic formulations are discussed in \cref{app:comp}. 

\begin{figure}[htb]
\begin{center}
\includegraphics[width=12cm]{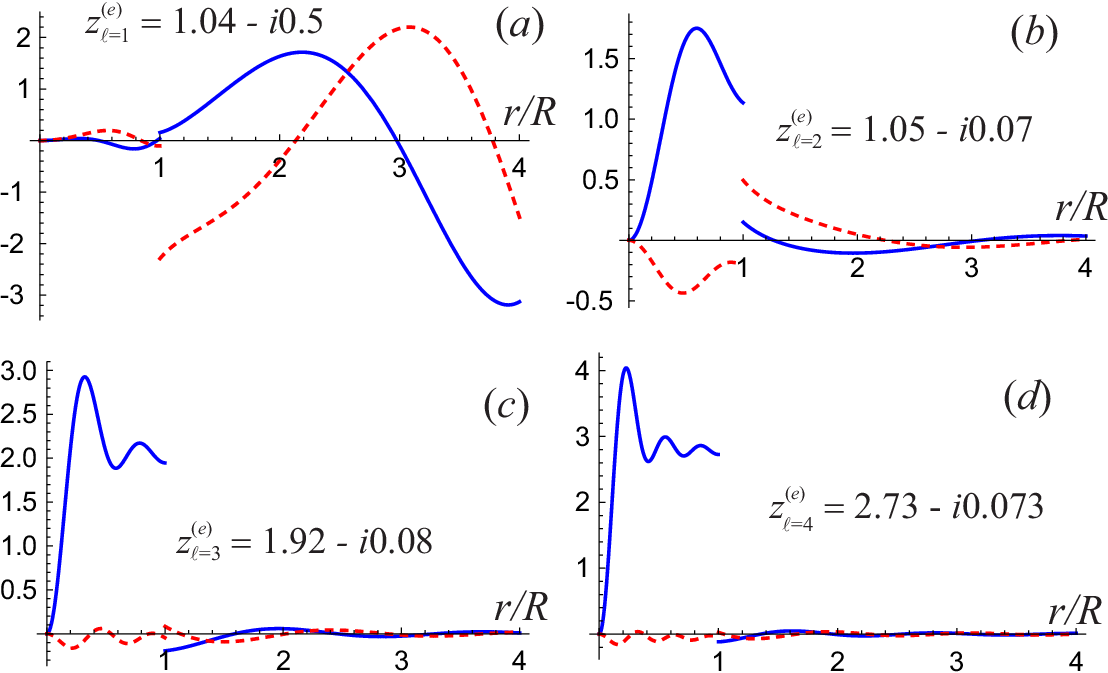} \caption{\label{fig:Norm} RS normalization integrand real parts (blue) and imaginary parts (red)  for the first four electric dipole RS size parameters, $z_{\alpha(e,n=1,\ell=1,2,3,4})$ plotted as functions of $r/R$ with a null killing factor ( $\eta=0$) in the case of a dielectric sphere with high permittivity ($\varepsilon=16$). 
} \end{center}
\end{figure}

It is insightful to plot the radial dependence of the normalization integrals of \cref{lagint} (after angular integration) for a lossless dielectric of $\varepsilon =16$  as shown in  \cref{fig:Norm}. The convergence factor was simply set to $\eta=0$ in these graphs since the exponential divergence only becomes relevant for $r/R$ significantly larger than the $r/R<4$ range in the RSs plots for the $z_{2}$, $z_{3}$, and $z_{4}$  RSs. For such low loss RSs, a numerical `verification' of \cref{lagint} or \cref{psinorm} is simple because the exponential divergence only becomes non-negligible after passing through a long region where the integrand is nearly zero. Therefore, truncations of the low loss RS integrals will approximately verify \cref{lagint} as one can guess by visual inspection of Figs.\ref{fig:Norm}(b)-(d).

The situation is quite different however for poorly confined RSs with large imaginary parts, like $z_{1}\simeq 1.04 -i0.5$. As one can see in \ref{fig:Norm}(a), the exponential divergence sets in immediately outside the sphere and a regularization of the RS normalization is essential if we look to numerically reproduce the analytic normalization result of \cref{lagint}. Similar behavior is observed for the RSs of lossy scatterers studied in \cref{RSDrude} below.

Unlike the physics of closed systems, where the phase of the modes can often be treated as arbitrary, the RS `normalization' factors, $\mathcal{N}_{\alpha}$, must correctly adjust the phase of the RSs which is a crucial information for reconstructing response functions. This point will be reinforced in \cref{sect:ResMieTh}, where we will see that the RS phase determines the \emph{complex} residues in spectral developments of response functions.

The ultimate justification for setting the right hand sides of \cref{psinorm,volsurfnorm,volsurfv2} to $z_{\alpha}$, is that this choice generates particularly simple spectral response functions presented in \cref{sect:ResMieTh} below. A more widespread choice is obtained by defining a rescaled RS wavefunction $\widetilde{\Psi}_{\alpha}(\boldsymbol{r}) \equiv \Psi_{\alpha}( \boldsymbol{r})/z_{\alpha}^{1/2}$ for which the RS normalization reads:\begin{align} \left\langle \widetilde{\Psi}_{\alpha} \left\vert \left[\frac{d}{d\omega}(\omega\Gamma) \right]_{\omega_{\alpha}} \right\vert \widetilde{\Psi}_{\alpha}\right\rangle =1 \;, \end{align}
which is reminiscent of normalization in closed systems like conventional quantum mechanics.

\subsubsection{Mode volumes}

 Recent derivations of the Purcell factor for RSs define a mode volume, $V_{\alpha}$.\cite{Sauvan2013,Krist2015,Muljar2016b,Lass18}  These derivations generally invoke a Green's function formalism and the local density of states (LDOS), which generally leads to a `complex' valued mode volume\cite{Cognee:19}. A typical definition for $V_{\alpha}$ derived along the LDOS Green function approach is:
 \begin{align} V_{\alpha} \overset{{\rm Ref.}\cite{Krist2015}}{=} \frac{\langle\langle \widetilde{\boldsymbol{\rm f}}_{\alpha} \vert  \widetilde{\boldsymbol{\rm f}}_{\alpha} \rangle\rangle}{ \varepsilon(\boldsymbol{r}_c) \widetilde{\boldsymbol{\rm f}}_{\alpha}^{2}(\boldsymbol{r}_c)}\;, \label{typvol} \end{align} where $\boldsymbol{r}_c$ is a `chosen' position (often the field maxima) while $\widetilde{\boldsymbol{\rm f}}_{\alpha}(\boldsymbol{r})$ is one of the  commonly employed photonic notations\cite{Krist2015} for the RS \emph{electric} field, and the double `$\langle\langle ... \rangle\rangle$' is an inner product notation ref.\cite{Krist2015}.
 Following the usual LDOS route to mode volume in the framework of this work leads to,
 \begin{align} V_{\alpha} \propto \frac{\left\langle \Psi_{\alpha}\right\vert \left[\frac{d}{d\omega} (\omega\Gamma) \right]_{\omega_{\alpha}}\left\vert \Psi_{\alpha}\right\rangle}{ \varepsilon(\boldsymbol{r}_c) {\boldsymbol{\rm E}}_{\alpha}^{2}(\boldsymbol{r}_c)}\;, \label{lagvol} \end{align}
 which up to possible proportionality factors agrees with more recent mode volume definitions that continue to be debated even now\cite{lalanne2020mode}.

 Mode volume assignments along the lines of \cref{typvol,{lagvol}} tell us that $|V_{\alpha}|$ is proportional $1/\left| \mathcal{N}_{\alpha}\right|^2$, so we generally expect small values of $\left| \mathcal{N}_{\alpha} \right|$ to correspond to large mode volumes. One indeed sees from \Cref{table2} that poorly confined modes like that $z_1$ (cf. \cref{fig:Norm}(a)) have smaller $\left| \mathcal{N}_{\alpha} \right|$, than the more tightly confined modes illustrated in the other graphics of \cref{fig:Norm}. 
 
 Inspection of \cref{nodispnorm} leads to the physically intuitive conclusion that weak constitutive contrasts lead to large mode volumes ({\it i.e.} small values of $\left| \mathcal{N}_{\alpha} \right|$). Nevertheless, the $\mathcal{N}_{\alpha}$ factors in \cref{nodispnorm} do not vanish in the limit of zero contrast since the vanishing $(\varepsilon-1)$ and $(\mu-1$) factors, are multiplied by factors expressed in terms of spherical Hankel functions evaluated at the RS frequencies. The fact that the $\mathcal{N}_{\alpha}$ normalization factors remain finite in the limit of null contrast is in fact a necessary ingredient for retrieving a vanishing  $T$-matrix response, (cf. \cref{sect:ResMieTh}), in the zero contrast limit. A detailed analysis of this intriguing result will be addressed elsewhere. 

\section{Resonant state expansions of Mie theory response functions}
\label{sect:ResMieTh}
One can obtain physical response functions directly from the Green's function, but we have argued in recent years the advantages of employing elegant and powerful scattering formalisms involving $S$ and $T$-matrices and reaction matrices (cf.~\cite{StMcPh17,summie,Colom18,Grig2013,St15,Lass18,Gr13} for more details and applications). An essential idea is seen by remarking that the $T$-matrix operator, $\mathbb{T}$, is related to the Green's function as:\cite{PellStout2011,Ko00,PingSheng1995,Ts80,Ne66}
\begin{align}
\mathbb{G} = \mathbb{G}_{0} +\mathbb{G}_{0} \mathbb{T} \mathbb{G}_{0} \;, \label{Tmat}
\end{align}  
where $\mathbb{G}_{0}$ is the Green function of the homogeneous background. An advantage of this formalism is that the $\mathbb{T}$-operator describes electromagnetic response in terms of fields (as opposed to source current for Green's functions). This occurs because the $\mathbb{G}_{0}$ operators on the left and right side of $\mathbb{T}$  in \cref{Tmat} respectively generate outgoing and incident multipole fields, (the normalization in \cref{psinorm} accounts for differences in the fields generated by $\mathbb{G}_{0}$ operators and the conventional multipole fields).\cite{PellStout2011,Ch90,Ts80} Once projected onto the multipole basis, the $\mathbb{T}$-`operator' takes the form of a `matrix' on the multipole basis functions (which is precisely the justification of the terminology $T$-\emph{matrix}).  

Given the observations made around  \cref{Gspect} and in the previous paragraph, we conclude that the RS residues of the scattering $T$-matrix are $\mathcal{R}_{\alpha} \equiv i/\mathcal{N}^{2}_{\alpha}$, as will be confirmed below for spherical scatterers, since the $T$-matrices in this case are diagonal on the multipole basis (with matrix elements given analytically by Mie theory for complex frequencies). This agreement is  remarkable in that $\mathcal{N}_{\alpha}$ was obtained through RS regularization, quite outside the context of traditional Mie theory. Although it is clear from \cref{Tmat} that $T$-matrices are just a reformulation of the Green function of \cref{Gspect},  we have found in our prior works that the $T$-matrix formulation has the advantage of facilitating the derivation of the non-resonant contributions to response functions~\cite{StMcPh17,summie,Colom18,Grig2013,St15,Lass18,Gr13} (notably through considerations of causality and energy conservation~\cite{Nuss72}). Relevant results of this approach, and some new derivations in the context of RS Mie theory, are reviewed below.

\subsection{RS spectral expansions}
In the interest of completeness, essential elements of Mie theory and linear response theory are recalled in \cref{app:Mietheory}. Since Mie theory predates the theory of matrix response functions, it remains common in the literature to keep the traditional notations~\cite{Bohr98} for the diagonal `$T$-matrix' elements; $T_{h,n}\equiv -b_{n}$ and $T_{e,n}\equiv -a_{n}$, and  $\Omega_{h,n}=c_{n}$ and $\Omega_{e,n}= d_{n}$ for the internal field coefficients. 

The meromorphic expansion of the $T_{n}$ functions in terms of frequency, $\omega$, have been determined to be:\cite{StMcPh17,summie,Colom18,Gr13}
\begin{subequations}\label{TMieresp}
\begin{align}
T_{q,n}(\omega,R)=\frac{A_{q,n}e^{-2iz}-1}{2}+e^{-2iz}A_{q,n}\sum\limits_{\ell=-\infty}^{\infty} \frac{r_{\alpha}}{z-z_{\alpha}}\;, \label{TmatRS} 
\end{align}
which again uses our RS state notation of the discrete index, $\alpha(q,n,\ell)$, are uniquely designated by multipole indices and the additional discrete RS index, $\ell$.  
The first, non-resonant, term on the right hand side of \cref{TmatRS} is a function of only the particle size parameter, $z=kR=\omega R/c$, while its second term sums over the spectral RS contributions with  $z_{\alpha}=k_{\alpha}R$. The $A_{q,n}$, are multipole sign factors, which for  $q=0$ (magnetic) or $q=1$ (electric), modes are:
\begin{equation} A_{q,n}\equiv(-1)^{n-q} \;. \end{equation}\label{Tlor}\end{subequations}
The residues, $r_{\alpha}$, of the RSs in the RS expansion of the $T$-matrix in \cref{TmatRS} are,
\begin{align} r_{\alpha}=e^{2iz_{\alpha}} \frac{i}{\mathcal{N}^{2}_{\alpha}}\equiv e^{2iz_{\alpha}} \mathcal{R}_{\alpha}\;, \label{resnorm} \end{align} 
where the $e^{2iz_{\alpha}}$ phase factor doesn't affect the $T$-matrix residue \cref{TmatRS} since it is canceled by the $e^{-2iz}$ multiplicative factor so the RS residues are
indeed $\mathcal{R}_{\alpha}=i/\mathcal{N}^{2}$ as expected. 

The internal field coefficients of Mie thery, $\Omega_{q,n}$, are quite similar $T$-matrix expressions with the exception of their response being expressed entirely in terms of RS states, ({\it i.e.} there are no non-resonant contributions):
\begin{equation} \Omega_{q,n}(\omega,R)  =e^{-2iz} A_{q,n}\sum\limits_{\ell=-\infty}^{\infty} \frac{r_{\alpha}^{(\Omega)}}{z-z_{\alpha}}\;, \label{OmegLor} \end{equation}
As one could deduce from \cref{Gspect} and the resonant state expression in \cref{Psispher}, the residues of internal RS coefficients, are related to the $T$-matrix residues in \cref{OmegLor}, as $r_{\alpha}^{(\Omega)} = \gamma_{\alpha} r_{\alpha}$. We will see below that this relation can also be deduced directly from Mie theory.

\subsection{Spectral residues in Mie theory}

Given the expressions for $\Omega_{n}$, and $T_{n}$ in \cref{Tlor,OmegLor} in terms of the $N_{q,n}$ and $D_{q,n}$ functions defined in \cref{numT_func,den_funcs} of \cref{ssect:Miecoefcross}, the residues in \cref{OmegLor,Tlor} can alternatively be calculated from Mie theory, as:
\begin{subequations} \begin{equation} r_{\alpha}^{(\Omega)}(R)=\frac{e^{2iz_{\alpha}}}{\left. \frac{c}{R} \frac{d}{d\omega}D_{q,n}(\omega,R)  \right\vert _{\omega=\omega_{\alpha}}}\quad,\quad r_{\alpha}(R)=-\frac{N_{q,n}\left(\omega_{\alpha},R\right)  e^{2iz_{\alpha}}}{\left. \frac{c}{R} \frac{d}{d\omega}D_{q,n}(\omega,R)  \right\vert _{\omega=\omega_{\alpha}}}\;, \label{residues}
\end{equation}
which can be rewritten as,
\begin{equation} r_{\alpha}^{(\Omega)}(R)=e^{2iz_{\alpha}}\lim_{\omega\rightarrow \omega_{\alpha}} \frac{  z-z_{\alpha}}{D_{q,n}(\omega,R)}\quad,\quad r_{\alpha}(R)=-N_{q,n}(\omega_{\alpha},R) r_{\alpha}^{(\Omega)}\;. \label{reslim} \end{equation} \label{resforms}\end{subequations}
One should note that for non-dispersive media, the scale invariance of the underlying electromagnetic equations imposes that the residues, $r_{\alpha}$ and $r_{\alpha}^{(\Omega)}$, only depend on particle radius, $R$, as a function of, $z_{\alpha}=\omega_{\alpha} R$, which results in the spectral response functions of \cref{OmegLor} and \cref{TMieresp} becoming  solutions for all particle sizes and frequencies in this case.

We remark from the second equality in \cref{reslim} that the relation between the $T$-matrix residues, $r_{n,\alpha}$, and the internal field residues, 
$r_{n,\alpha}^{(\Omega)}$, is purely analytic. This relation can be rewritten in a more 
transparent form by invoking the respective RS conditions of \cref{gammadefs} such that the expressions of $N_{e,n}$ and $N_{h,n}$ can be rewritten:
\begin{subequations} \begin{align}\begin{split} -N_{e,n}(\omega_{\alpha},R)  = &z_{\alpha} \frac{\psi_{n}^{\prime}(\rho_{\alpha}z_{\alpha})j_{n}(z_{\alpha}) -\varepsilon_{\alpha}j_{n}(\rho_{\alpha}z_{\alpha})\psi_{n}^{\prime}(z_{\alpha})}{i\rho_{\alpha}} \rightarrow\frac{\psi_{n}^{\prime}(\rho_{\alpha}z_{\alpha})}{\rho_{\alpha} \xi_{n}^{\prime}(z_{\alpha})}=\frac{1}{\gamma_{e,n,\alpha}}\;, \label{Ne} \end{split}\end{align}
for the electric modes and,
\begin{align}\begin{split} -N_{h,n}(\omega_{\alpha},R)=&z_{\alpha}\frac{\psi_{n}^{\prime}(\rho_{\alpha}z_{\alpha})j_{n}(z_{\alpha}) -\mu_{\alpha}j_{n}(\rho_{\alpha}z_{\alpha})\psi_{n}^{\prime}(z_{\alpha})}{i\mu_{\alpha}} \rightarrow \frac{\psi_{n}^{\prime}(\rho_{\alpha}z_{\alpha})}{\mu_{\alpha}\xi_{n}^{\prime}(z_{\alpha})}= \frac{1}{\gamma_{h,n,\alpha}}\;, \label{Nh} \end{split}\end{align}
\label{numerat}\end{subequations}
for the magnetic modes. In both \cref{Ne,Nh} we invoked the Wronskian relation,
\begin{equation}
\xi_{n}^{\prime}(z)\psi_{n}(z)-\xi_{n}(z)\psi_{n}^{\prime}(z)=i\;,
\end{equation}
and recalled the definitions of the $\gamma_{\alpha}$ functions defined in \cref{gammadefs}.

Mie theory therefore analytically validates the relation
\begin{align}
r_{\alpha}=\frac{r_{\alpha}^{(\Omega)}}{\gamma_{\alpha}} \label{romeg}
\end{align}
that we deduced after \cref{OmegLor} using arguments based on RS spectral expansions of the response function. Although the derivation is a bit lengthy here, one can also 
directly derive the $T$-matrix residue, $\mathcal{R}_{\alpha}=i/\mathcal{N}^{2}$, directly from Mie theory. We found this fact quite striking since Mie theory makes
no use whatsoever of the RS regularization methodology. We interpret this as striking evidence for the mathematical validity of the RS regularization program.

\section{Numerical implementations}\label{Examples}

This section provides precise calculations of RS eigenvalues, normalization, and response function reconstructions in the case of spherical scatterers.  As discussed in \cref{sect:Resstates}, generalizations to non-spherical shapes are quite analogous, but they will not be considered here since they are not amenable to such high precision analytics.

Even for spherical geometries, solving for the RS frequencies must be carried out numerically, but thanks to the field continuity conditions of \cref{gammadefs}, the RS frequencies are determined as solutions to analytic expressions that can be expressed in terms of the reduced logarithmic derivatives of the Ricatti-Bessel functions:
\begin{subequations}
\begin{align} \left[ z\ln^{\prime} \psi_{n}(z) \right]_{z=\rho_{\alpha} z_{\alpha}} -\varepsilon_{\alpha} \left[z \ln^{\prime} \xi_{n}(z) \right]_{z= z_{\alpha}} &=0  \label{elecmodecond}\\ \left[ z\ln^{\prime} \psi_{n}(z) \right]_{z=\rho_{\alpha} z_{\alpha}} -\mu_{\alpha} \left[z \ln^{\prime} \xi_{n}(z) \right]_{z=z_{\alpha}} &=0 \;, \label{magmodecond} \end{align} \label{RSdispeqs}\end{subequations}
where we recall from \cref{const} that $\varepsilon_{\alpha}$, $\mu_{\alpha}$, and $\rho_{\alpha}$ are in general frequency 
dependent constitutive parameters evaluated at the RS frequency, $\omega_{\alpha}$, such that the first terms in \cref{RSdispeqs} have a non-trivial frequency dependence. 

\subsection{Resonant states and normalizations of high refractive index dielectric spheres}\label{dielsph}
There is currently considerable interest in non-dispersive scatterers since high index dielectrics are being considered in nano-optics as an alternative to plasmonics~\cite{straude17}.
Even for such energy conserving systems, one still needs to solve \cref{RSdispeqs} numerically, but the difficulty is  greatly reduced and the problem becomes scale invariant, which means that one solves the problem for all particle sizes and/or frequencies simultaneously. Furthermore, the RS normalization coefficients are calculated with the simple expressions found in \cref{nodispnorm}.

Numerical results for a few low-order modes of a dielectric with no magnetic permeability contrast, $\mu=1$ and high, $\varepsilon=16$, permittivity contrast are shown in \Cref{tablenodisp} 
(further analysis and discussion of such modes may be also found in~\cite{summie}). There are an infinite number of modes for each value of $n$, and they follow an asymptotic pattern resembling that of zeros of Bessel functions. Beginning with electric modes, \cref{tablenodisp} 
gives the values of $z_{\alpha}$ and the normalization factors for the first two dipole and quadrupole modes. An interesting feature of the RSs in this case is that for $n$ even (odd), there is one electric (magnetic) mode with purely imaginary $z$, labeled with $\ell=0$. One can also remark from \cref{tablenodisp} that for imaginary RS eigenvalues, the interaction `strengths' (residues), $\mathcal{R}_{\alpha}$, have purely imaginary values.

\begin{table}[htb]
\begin{center}
\begin{tabular}{|c|c|c|}
\hline
$\alpha (q,n,[\ell])$ & $z_{\alpha}$ &$\mathcal{R}_{\alpha}$ \\ \hline
$(e,1,[1])$ & $1.0395 - i 0.500935$ & $-0.236682 + i 0.231492$ \\
$(e,1,[2])$ &  $1.05273 - i 0.0723549$ & $0.0659905 - i 0.0579972$ \\
$(e,1,[3])$ &  $ 1.92043 - i 0.082005 $ & $0.0748408 - i 0.0282738$ \\ 
$(e,1,[4])$ &  $ 2.7227 - i 0.073007 $ & $0.00279437 - i 0.0683107$ \\ \hline
$(e,2,[0])$ &  $- i 1.6797303$ & $i 0.146892 $ \\
$(e,2,[1])$ & $1.377484 - i 0.0118433$ & $0.00184613 -i 0.0118059$ \\ 
$(e,2,[2])$ & $2.071446 - i 0.667649$ & $-0.305381 +i 0.277002$ \\\hline
$(h,1,[0])$ & $- i 1.250038$ & $i 0.136765$ \\
$(h,1,[1])$ &  $0.7537823 - i0.0240302$ & $-0.00601759 - i 0.0229898$ \\
$(h,1,[2])$ & $1.5414631 - i 0.0459254$ & $-0.0394075 -i0.0195948$ \\ \hline
$(h,2,[1])$ & $0.870513 -i 1.75259$ &$-0.0521306 + i 0.140046$ \\
$(h,2,[2])$ & $1.0957165- i 0.00684025$ & $-0.000482964 - i0.00681678$ \\ \hline
\end{tabular}
\caption{\label{tablenodisp}RS eigenvalues, $z_{\alpha}$, and associated residues, $\mathcal{R}_{\alpha}= i/\mathcal{N}^{2}_{\alpha}$, for a non-dispersive dielectric medium with dielectric contrast of $\varepsilon=16$ and no magnetic contrast, $\mu=1$.}
\end{center}
\end{table}

The high number of significant figures given in \Cref{tablenodisp} for the RS eigenvalue size parameters, $z_{\alpha}$ and normalizations, $\mathcal{R}_{\alpha}= i/\mathcal{N}^{2}_{\alpha}$ were adopted because these values were indeed calculated up to such high order accuracy which  underscores the results of this work to serve as benchmarks for more numerically based approaches.

\subsection{Numerical verification of RS expansions of Mie theory.}
\label{guide}

\begin{figure}[htb]
\begin{center}
\includegraphics[width=10cm]{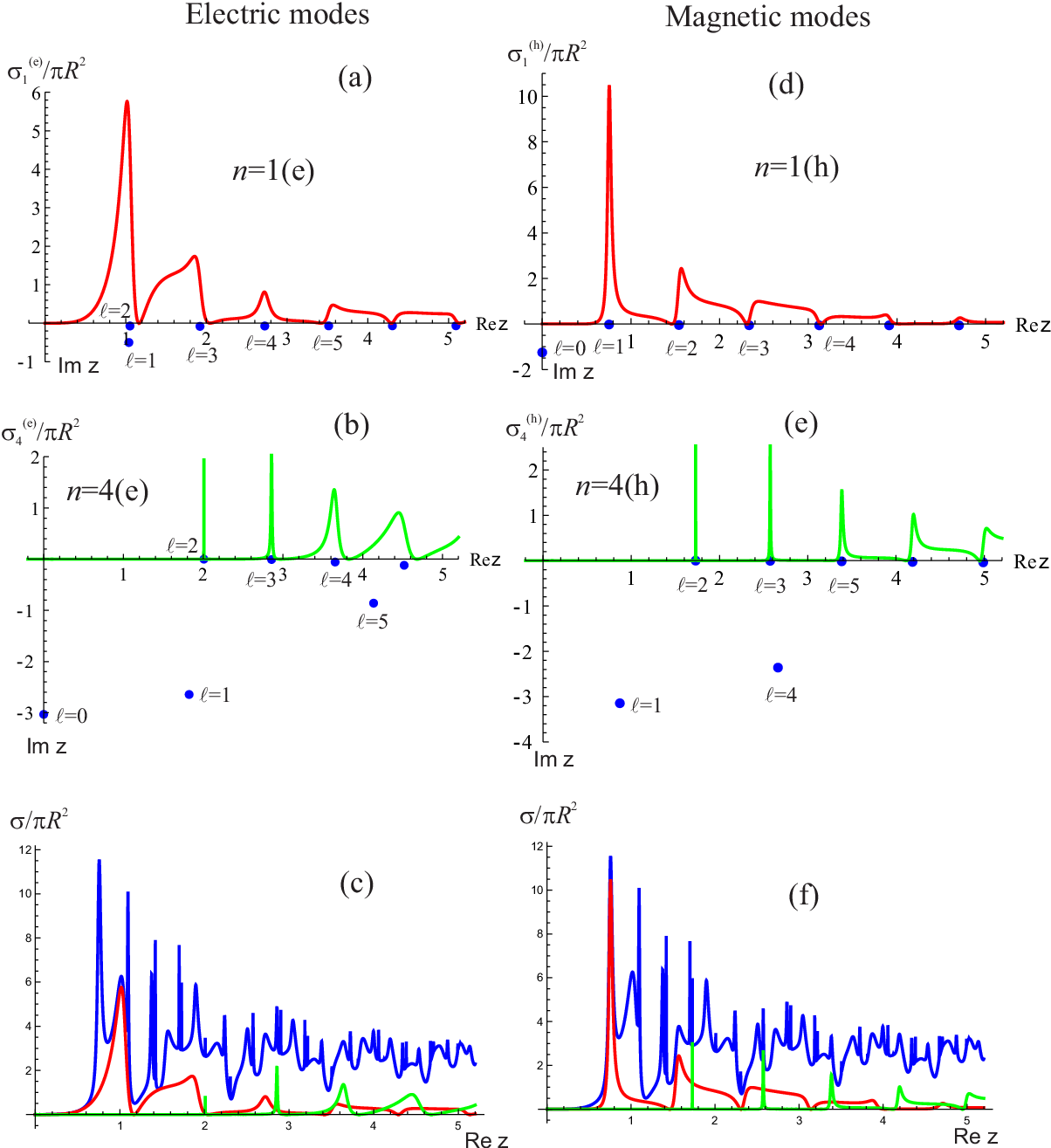}
\caption{\label{fig:RSMie}The electric RSs and their contributions to cross-sections for multipoles, $n=1$ and $n=4$ for a dispersionless sphere  with $\varepsilon=16$:  electric modes (a)-(b) and magnetic modes (c)-(f). The geometrically normalized cross section contributions, $Q\equiv\sigma/\left(\pi R^{2}\right)$, are plotted according to the positive vertical axis. Complex RSs are indicated by blue dots with 
${\rm Im}(z_{\alpha})$ on the negative vertical axis. In parts (c) and (f), the contributions of the $n=1$ (red) and $n=4$ (green) multipoles are compared with the total cross section (blue line) determined by the sum of all multipole contributions for the electric and magnetic cases respectively.}
\end{center}
\end{figure}

The steps to implement the results of this paper up to the construction of Mie theory response functions can be summarized as follows :
\begin{enumerate}
\item For each required multipole order $n$, and mode type $q=(0,1)$, ({\it i.e.} $(h)$ and $(e)$ respectively), the \emph{only} non-trivial step is to find a sufficient number of the resonant state size parameters, $z_{\alpha}$, by numerically solving the transcendental  \cref{gammadefs} (or equivalently \cref{RSdispeqs}). The \emph{sufficient} number of RSs depends on the particle radius $R$ (or frequency of interest) and the dispersion relations of the constitutive parameters, but can be as low as 1 for some applications in dispersive materials. When only a few RSs are required, graphical solutions with numerical refinements may suffice (see \cref{RSDrude} for examples, but one should exercise caution with such techniques since important RSs can occur far from the real $z$ axis as can be seen in \cref{fig:RSMie}(b) and \cref{fig:RSMie}(e)).
\item Calculate the square of the complex-valued RS norms, $\mathcal{N}_{\alpha(q,n,\ell)}^{2}$, using \cref{Nnorm}.
\item Insert the values of $\mathcal{N}_{\alpha(q,n,\ell)}^{2}$ found in step 2 into \cref{resnorm} to determine the residues, $r_{\alpha(q,n,\ell)}$, for the meromorphic expression of \cref{Tlor} for the scattered field Mie coefficients, $a_{n}=-T_{e,n}$ and 
$b_{n}=-T_{h,n}$. Cross section contributions, $\sigma_{n}$, can then be determined by the standard formulas recalled in \cref{crosssect} of \cref{ssect:Miecoefcross}.
\item If one needs to determine the fields inside the sphere, use the values of $r_{\alpha}$ found in step 3 and the $\gamma_{\alpha(q,n,\ell)}$ coefficients of \cref{PsiRes} to calculate the residues, $r_{\alpha(q,n,\ell)}^{\left(\Omega\right)}$, of \cref{romeg} that appear in the meromorphic expression of \cref{OmegLor}, for the internal field Mie coefficients $c_{n}=\Omega_{h,n}$ and $d_{n}=\Omega_{e,n}$ (defined in \cref{intMie}).
\end{enumerate}

The results displayed in \Cref{tablenodisp} of \cref{dielsph} were obtained by applying steps 1 and 2 of the above steps. An illustration of carrying out step 3 above, with the same material parameters ($\varepsilon=16$, $\mu=1$) used in \cref{fig:Norm},
is illustrated graphically for both electric and magnetic mode contributions and orbital quantum numbers, $n=1$ and $n=4$. The first few RSs for each mode are given as blue dots in the lower half plane with ${\rm Im}(z)$ on the vertical axis, while the associated contributions to the cross section efficiencies, $Q\equiv \sigma/(\pi R^{2})$, are plotted on the positive vertical scale. In \cref{fig:RSMie}(c) and \cref{fig:RSMie}(f), we verified that this procedure reproduces Mie theory up to arbitrary computational accuracy for all multipole orders \added{(the curves in \cref{fig:Norm} were generated both from RS calculations and Mie theory, but are identical since they agreed here up to six figure accuracy). }

Similar calculations can reproduce Mie theory for dispersive media, but in that case, the problem is no longer scale invariant and the position of the RS eigenfrequencies will depend on the particle size and will be seen in \cref{RSDrude} below.

\subsection{Resonant states and normalizations of `Drude' model conductors}
\label{RSDrude}
The calculation of the complex wavenumbers of RS's depends on an extrapolation of the experimental data on complex permittivities (or permeabilities in cases involving magnetic materials) from the real axis of frequencies or wavelengths. Such an extrapolation depends critically on both the accuracy of the experimental data and on the accuracy of the analytic model used for the extrapolation. 

Data on the complex refractive index of gold as a function of frequency are available in Volume 1 of he comprehensive collection due to Palik~\cite{palik}. The simplest widely used analytic expression for  this data is that of the Drude model, which may be written in terms of the angular frequency, $\omega$, for gold~\cite{Sauvan2013} and silver~\cite{yang-Ag} as:
\begin{subequations}\begin{align}
\varepsilon (\omega)&=\varepsilon_\infty-\frac{\omega_p^2}{\omega^2+i \omega \Gamma_D} \;. \nonumber \\ \varepsilon_\infty^{\rm Ag}&=5 \quad , \quad \omega_p^{\rm Ag}=1.35 \times  10^{16} [{\rm s}^{-1}]\;(8.89 {\rm\, eV})\quad , \quad \Gamma_D^{\rm Ag}= 5.88\times 10^{13} [{\rm s}^{-1}]]\;(0.0387\, {\rm eV}) \; \label{drudeAgeq}
\\ \varepsilon_\infty^{\rm Au}&=1 \quad , \quad
\omega_p^{\rm Au}=1.26 \times  10^{16} [{\rm s}^{-1}] \; (8.29 {\rm\, eV})
\quad , \quad \Gamma_D^{\rm Au}=1.41\times 10^{14} [{\rm s}^{-1}]\;(0.0928\, {\rm eV})\;. \label{drudeeq}
\end{align}\end{subequations}
The comparison with experimental data given in \cref{fig:cfAg} of \cref{DrudeMod} shows that the Drude model works relatively well for silver at wavelengths longer than around $0.30 \mu {\rm m}$, while the comparison of the Drude model for gold in \cref{fig:cfgold} shows that the Drude model for gold only works well for wavelengths longer than $0.60 \mu {\rm m}$.

To achieve a better fit for gold valid down to shorter wavelengths, extra Lorentz resonant terms need to be added to the Drude model of \cref{drudeeq}. For example, Sikdar and Kornyshev~\cite{goldnpssrep} give the parameters for a model for gold taking into account inter-band transitions \textit{via} two additional resonances:
\begin{equation}
\varepsilon (\omega)=\varepsilon_\infty-\frac{\omega_{pD}^2}{\omega^2+i \omega \Gamma_{D}} -s_1 \frac{\omega_{p1,L}^2}{\omega^2-\omega_{p1,L}^2+i \Gamma_{1,L}\omega} -s_2 \frac{\omega_{p2,L}^2}{\omega^2-\omega_{p2,L}^2+i \Gamma_{2,L}\omega} \;. \label{drudep2}
\end{equation}
Here the parameters are given in eV:  $\epsilon_\infty=5.9752$, $\hslash \omega_{pD}=8.8667 {\rm eV}$,  $\hslash\Gamma_D=0.03799  {\rm eV}$, $s_1=1.76$, 
$\hslash\omega_{p1,L}=3.6 {\rm eV}$, $\hslash\Gamma_{1,L}=1.3 {\rm eV}$, $s_2=0.952$, $\hslash\omega_{p2,L}=2.8 {\rm eV}$ and 
$\hslash\Gamma_{2,L}=0.737 {\rm eV}$. The conversion of $\omega$ to electron volts is achieved by dividing the frequency results by $e/\hbar = 1.51927\times 10^{15}$.
As can be seen from \cref{fig:cfgold}, this model works well down to around $0.40 \mu {\rm m}$.
An interesting feature of the dispersion model of \cref{drudep2} is that it supports bulk plasmons.~\cite{Maier07} These occur for complex wavelengths or frequencies  at which $\varepsilon (\omega)=0$.  For the dispersion model of \cref{drudep2} they are at: $\lambda_{\rm Lg} = 0.257778 + i 0.0206709\mu {\rm m}$, $\lambda_{\rm Lg}=0.395618 + i 0.0536046\mu {\rm m}$ , $\lambda_{\rm Lg} = 0.502518 + i 0.048786 \mu {\rm m}$.
Bulk plasmons are longitudinal waves, which do not couple to transverse electromagnetic waves and cannot be excited by or scattered into them. This translates as the fact that one obtains a normalization residue factor, $\mathcal{R}_{n,\alpha}$, of zero for these longitudinal `resonant states'.

\begin{figure}[htb]
\includegraphics[width=\linewidth]{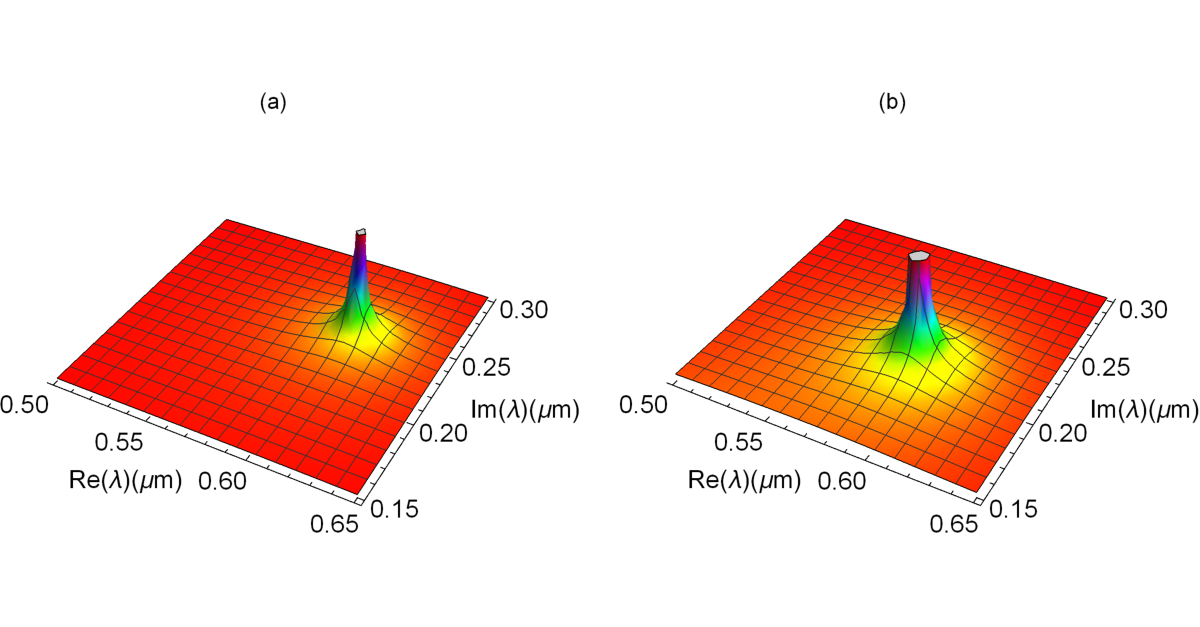}
\caption{\label{fig:cf100}The inverse of the modulus of the dispersion equation (\ref{elecmodecond}) for the electric dipole mode of a gold sphere of radius $100 {\rm nm}$. Left: the Drude model of \cref{drudeeq}; right- the extended Drude-Lorentz model of \cref{drudep2}.}
\end{figure}

\Cref{fig:cf100} shows the modulus of the dispersion equation of \cref{elecmodecond} for $n=1$ and for a gold sphere of radius $100 {\rm nm}$, both for the
Drude model of \cref{drudeeq} and for the more elaborate model of \cref{drudep2}. The former has its minimum at $\lambda_{\alpha}=0.606976+ i 0.239112\mu{\rm m}$,
in good agreement with the value reported in Sauvan et al.~\cite{Sauvan2013}. The latter has a slightly different value:
$\lambda_{\alpha}=0.592227+ i 0.210097\mu{\rm m}$. From \Cref{table2} 
the values of $\varepsilon_{\alpha}$ at the resonance and consequently the normalization factor disagree more significantly for the more elaborate model.

\begin{figure}[tbh]
\includegraphics[width=\linewidth]{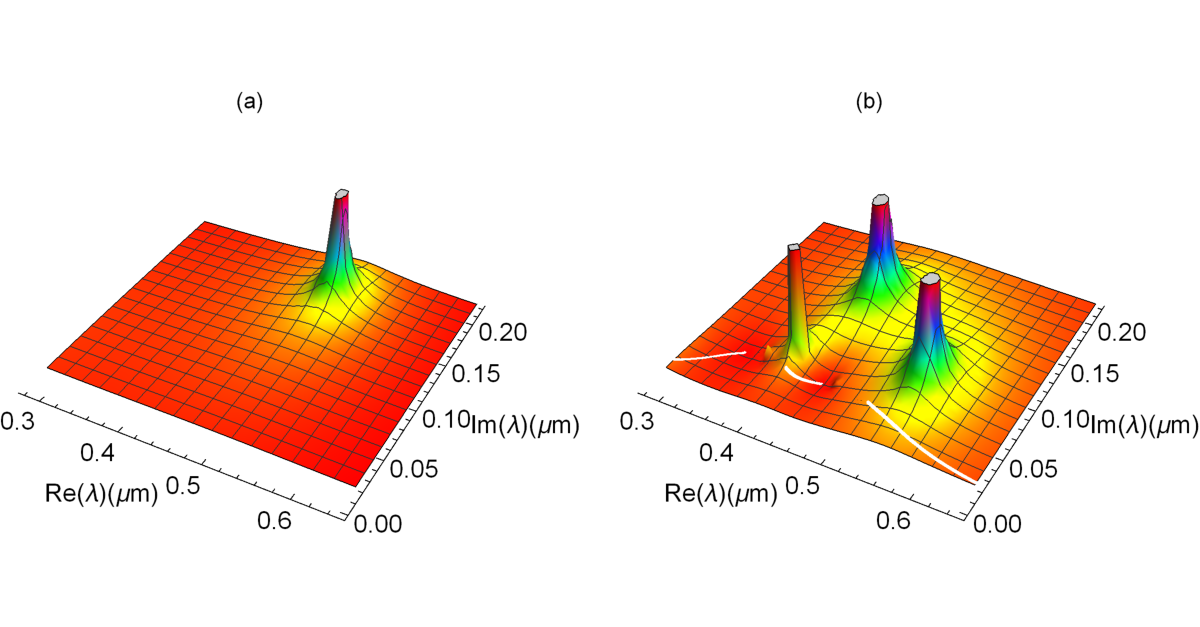}
\caption{The inverse of the modulus of the $n=1$ dispersion equation, \cref{elecmodecond}, for   a gold sphere of radius $80 {\rm nm}$. Left: the Drude model of \cref{drudeeq}; Right: the model of \cref{drudep2}.}
\label{figcf80}
\end{figure}

\Cref{figcf80} shows the inverse modulus of the $n=1$ dispersion relation of \cref{RSdispeqs} for a gold sphere of radius $80 {\rm nm}$. The plot is more complicated for the more elaborate dispersion model of \cref{drudep2} since the decrease in radius moves the wavelength range of interest into that for which the zeros, poles and branch cuts of this dispersion model become evident (see the data listed above for the wavelengths corresponding to bulk plasmons). The complex wavelength of the RS for the Drude model is $\lambda_{\alpha}=0.505163+ i 0.174433\mu{\rm m}$. With the model of \cref{drudep2}, this becomes  
$\lambda_{\alpha} = 0.547815 + i 0.080062 \mu{\rm m}$. Note the significant decrease in the imaginary part around this wavelength, as seen in \cref{fig:cfgold} shows, since the Drude model is only a crude approximation to the experimental data. 

The aforementioned electric dipole RSs for the different dispersion models are given in \Cref{table2} for RSs in silver and gold spheres of radius 100 and 80nm. In order to calculate the mode normalization factors, numerical differentiation, as described in \cref{resforms}, can be used to find the necessary dispersion derivatives.
The data given in \Cref{table2} for the resonances of silver spheres of radius 100 and 80nm  shows complex wavelengths and normalization factors which are broadly similar to the corresponding values for gold.

\begin{table}[htb]
\begin{center}
\begin{tabular}{|c|c|c|c|c|c|}
\hline
 $R$ (nm) & $\ell$ &Disp. & $\lambda_{\alpha}$ (nm) & $\varepsilon_{\alpha}$  & $\mathcal{R}_{\alpha}$ \\ \hline
 100 & 1 & (\ref{drudeeq}) & $606.976+ i 239.112$  &$-13.7606 - i 12.5419$  &$-0.217942+i 0.034889$  \\
 100 & 1 & (\ref{drudep2}) & $592.227+ i 210.097$  & $-5.90132- i 12.9326$  & $-0.309295-i 0.00587748$  \\ 
 80 & 1 & (\ref{drudeeq}) &$505.163+ i 174.433$  &$-9.48069 - i 7.58828$ &  $- 0.2111 -i 0.00385708$  \\
 80 & 1 & (\ref{drudep2}) &$547.815 + i 80.062$ & $-3.64747 - i 3.52209$ & $0.0167796-i 0.10005$ \\
\hline
 100 & 1 & (\ref{drudeAgeq}) & $600.211 + i 231.333$ & $ -11.1356  - i 14.0631$ & $ -0.236629 + i 0.0266621 $ \\ $\cdot$  & 2 & $\cdot$ & $290.678 + i33.6325$ & $0.704812 - i0.966352$ & $ 0.178962 + i0.124692$ \\ $\cdot$ & 3 & $\cdot$ & $217.845 + i20.8224$ & 
 $2.58005 - i 0.449824$ & $ 0.0139492 - i0.235957$
 \\
 80 & 1 & $\cdot$ & $500.306 + i156.443$ & $-6.78066 -i7.89519 $ & $ -0.235268 - i0.0509908$   \\
 $\cdot$ & 2 & $\cdot$ & $281.965 + i50.9047 $ & $1.03036 - i1.44187$ & $0.12746 + i0.24896 $  \\ $\cdot$ & 3 & $\cdot$ & $ 192.72 + i20.4675 $ & $ 3.11019 - i 0.394101 $ & $ 0.10372 - i0.210644$ \\
\hline
\end{tabular}
\end{center}
\caption{\label{table2}Illustrative electric dipole RSs, $\alpha(e,n=1,\ell)$, for plasmonic particles: for the dispersion models of \cref{drudeeq} and \cref{drudep2} for gold and \cref{drudeAgeq} for silver. The complex wavelengths and dielectric constants
are given in addition to the Mie coefficient residue, $\mathcal{R}_{n,\alpha}$, associated with RS normalization is given in the last column.}
\end{table}



\section{Conclusions and perspectives}
\label{Conclusions}
We used Gaussian regularization to demonstrate that resonant states can be assigned \emph{finite} normalization and orthogonality properties that are closely analogous to definitions adopted for the eigenstates of closed systems. A notable difference however is that the RS normalization factors, $\mathcal{N}_{\alpha}$, are complex valued, (in contrast to their real-valued counterparts in closed systems). Although, ``irrelevant'' phase factors are a familiar feature of the time evolution of eigenstates in dispersionless closed systems, phase plays a crucial physical role in an open system's response to an external excitation (and consequently can no longer be treated as arbitrary). The RS normalization factor must therefore ensure the physically correct RS phase, in addition to its amplitude.  Recognizing the importance of phase factors from the outset helps one to understand the fact that when RS physics is imposed on closed system concepts, like the Purcell factor, the complex normalization factor inevitably leads to the introduction of apparently incongruous notions like \emph{complex valued} mode volumes.\cite{lalanne2020mode,Cognee:19,Krist2015}

\replaced{For systems of arbitrary geometry,}{Given the applicability of RSs to systems of arbitrary geometries} the determination of the RS eigenvalues and the regularization of their inner products are often carried out numerically. This work combined a multipolar expansion of arbitrary RSs with Gaussian regularization in order to analytically evaluate the inner product integrals in the region lying outside the system of interest. Our analytical analysis can both serve as benchmarks for purely numerical treatments, as shown in \cref{Examples}, and also increase the speed of numerical approaches by replacing certain numerical integrals with analytical formulas. The results of this work for RS normalization have been shown to be consistent with other formulas in the literature for non-spherical scatterers and we have explicitly generalized previous analytical results of the literature for dispersive spherical scatterers with \emph{both} permittivity and permeability contrasts.

The analytic formulas obtained herein enabled us to completely reformulate the `Mie theory' response functions of dispersive spherical objects in terms of resonant state expansions, as summarized in \cref{sect:ResMieTh}, and illustrated in \cref{fig:Norm}. A notable by-product of the response function formulation was that it allowed an analytical determination of the controversial `non-resonant' contributions that have repeatedly appeared as snags in RS reconstructions of response functions. Given the vast popularity and practical utility of Mie theory in applications, there can be little doubt that the ability to reformulate Mie theory in terms of the conceptually and mathematically simple RS expansions will find future, and likely unforeseen, implementations.

\section{Acknowledgements}
Research conducted within the context of the International Associated Laboratory for Photonics
between France and Australia (LIA ALPhA). This work has been carried out thanks to the support of the A*MIDEX
project (no. ANR-11-IDEX-0001-02) funded by the Investissements d'Avenir French Government program, managed by the French National Research Agency (ANR). We would like to thank Jean-Paul Hugonin for calling our attention to the numerical equivalence between the Gaussian normalization and the PML deformation into the complex plane as demonstrated in Figure 2. We would also thank the referees whose remarks helped improve this manuscript, most notably concerning notation and more thorough the comparisons with results in literature.

\appendix

\section{Alternative normalizations and literature comparisons} \label{app:comp}

Given the wide variety of constantly evolving notations and conventions in the QNM/RS literature, it is not realistic to carry out exhaustive comparisons, so in this appendix we show that our normalization formulas agree with other analytic formulations in the special case of purely dielectric scatterers that have previously been given in the literature, notably those by Muljarov {\it et al.} and in particular in Refs.\cite{Muljar2016b,Doost14}. 

Although we adopted a six component electromagnetic field formulation in contrast to the more common formulations entirely in terms of the 3-component electric field, such differences should \emph{not} affect the values of RS normalization. One remarks that Ref.\cite{Doost14} labeled RS indices `$n$' as opposed to our generalized index $\alpha$, and they used an index, $l$, for multipole number as opposed to our `Mie theory' choice of index $n$. 
Comparisons between our formulas and those given in Refs.\cite{Doost14,Muljar2016b} are facilitated by introducing the definitions of Ref.\cite{Doost14}. These authors used radial dependent constitutive parameters of a spherical scatterer are defined as:
\begin{equation} \varepsilon(\omega,r)  \equiv\left\{ \begin{array}[c]{ccc} \varepsilon(\omega) & \text{for} & r\leq R\\ 1 & \text{for} & r>R \end{array} \right.  \qquad\mu(\omega,r) \equiv\left\{ \begin{array}[c]{ccc} \mu(\omega)  & \text{for} & r\leq R\\ 1 & \text{for} & r>R \end{array} \right. \;, \label{constr} \end{equation}
and defined a function $R_n$:
\begin{equation} R_{n}(\omega,r)  \equiv\left\{ \begin{array}[c]{ccc} j_{n}(\rho_{\omega}kr)/j_{n}(\rho_{\omega}kR) & \text{for} & r\leq R\\ h_{n}(kr)/h_{n}(kR) & \text{for} & r>R \end{array} \right. \;,\label{Rexp} \end{equation}
together with
\begin{equation} \frac{\partial}{\partial r}rR_{n}( \omega,r)  =\left\{ \begin{array}[c]{ccc} \psi_{n}^{\prime}(\rho_{\omega}kr)/j_{n}(\rho_{\omega}kR) & \text{for} & r\leq R\\ \xi_{n}^{\prime}(kr)/h_{n}(kR) & \text{for} & r>R \end{array} \right.  \;,\label{DRexp} \end{equation}
where the various spherical Bessel functions, $j_{n},\psi_{n},h_{n}$, and $\xi_{n}$  are defined here in \cref{VPWs}.  

Invoking our definitions of \cref{PsiRes} together with the RS relations for spherical scatterers in \cref{RSdispeqs}, our six component RSs  \cref{Psispher} can finally be compactly reexpressed:
\begin{align} \Psi_{\alpha(h,n,m,\ell)}(\boldsymbol{r})  &  =\frac{i^{n}k_{\alpha}^{3/2}}{\widetilde{\mathcal{N}}_{\alpha}}\left( \begin{array}[c]{c} R_{n}(\omega_{a},r)  \boldsymbol{X}_{h,n,m}\left( \theta,\phi\right)  \\ -i\frac{\left[ \frac{\partial}{\partial r}rR_{n}(\omega_{a},r) \right]  \boldsymbol{Z}_{h,n,m}\left(  \theta,\phi\right)  +R_{n}\left( \omega_{a},r\right)  l(l+1)  \boldsymbol{Y}_{h,n,m}(\theta,\phi)  }{\mu_{\alpha}(r)  k_{\alpha}r} \end{array} \right)  \equiv z_{\alpha}i^{n}\overline{A}_{\alpha}^{\mathrm{TE}}\left( \begin{array}[c]{c}\cdot\\ \cdot \end{array} \right)  \nonumber\\ \Psi_{\alpha(e,n,m,\ell)}(\boldsymbol{r}) &  =\frac{i^{n-1}k_{\alpha}^{3/2}}{\widetilde{\mathcal{N}}_{\alpha}}\left( \begin{array}[c]{c}\frac{\left[  \frac{\partial}{\partial r}rR_{n}(\omega_{a},r) \right]  \boldsymbol{Z}_{e,n,m}\left(  \theta,\phi\right)  +R_{n}\left( \omega_{a},r\right)  n(n+1)  \boldsymbol{Y}_{e,n,m}(\theta,\phi)  }{\varepsilon_{\alpha}(r)  k_{\alpha}r}\\ -iR_{n}(\omega_{a},r)  \boldsymbol{X}_{e,n,m}(\theta ,\phi) \end{array} \right)  \equiv z_{\alpha}i^{n-1}\overline{A}_{\alpha}^{\mathrm{TM}}\left(\begin{array}[c]{c} \cdot\\ \cdot \end{array} \right)  \;,\label{EHtilde} \end{align}
where the final set of parentheses are identical to the parenthesis of the central expressions; thus establishing the connection between the normalization coefficients $A^{\rm TE}_{\alpha}$ and $A^{\rm TM}_{\alpha}$ of Ref.\cite{Doost14}, with our normalization factors, $\mathcal{N}_{\alpha}$. Note that we normalized the coefficients of Ref.\cite{Doost14}  $\overline{A}_{\alpha}\equiv A_{\alpha} \sqrt{n(n+1)}$, in order to compensate for the $1/\sqrt{n(n+1)}$ normalization factor of spherical harmonics that was included in our vector spherical harmonics $\boldsymbol{X}_{q,n,m}$, $\boldsymbol{Y}_{q,n,m}$, and $\boldsymbol{Z}_{q,n,m}$, given in \cref{VPWs}  (also note that $\sqrt{n(n+1)}$ reads $\sqrt{l(l+1)}$ in Ref.\cite{Doost14}). 

The electric fields in the parenthesis of \cref{EHtilde} can be directly identified with the RS electric fields in Eq.(26) and Eq.(27), of Ref.\cite{Doost14} (again up to the $1/\sqrt{n(n+1)}$ factor).  In order to make our RS fields of \cref{Psispher} match the RS field expressions of Ref.\cite{Doost14}, we had to divide our multipolar fields $\Phi_{\alpha}$ by $h_{n}(z_{\alpha})$, which we then had to compensate for {\it via} a modified normalization factor, $\widetilde{\mathcal{N}}_{\alpha}$:
\begin{equation} \widetilde{\mathcal{N}}_{\alpha}\equiv \frac{\mathcal{N}_{\alpha}}{h_{n}(z_{\alpha})}\;. \end{equation}
which results in particularly compact and symmetric expressions:
\begin{align}\begin{split} \widetilde{\mathcal{N}}_{\alpha(e,n,\ell)}^{2} &  =z_{\alpha}^{2}(\mu_{\alpha}-1)+(\varepsilon_{\alpha}-1)\left\{  \left[ \varphi_{n}^{(+)}(z_{\alpha})\right]^{2}+\frac{n(n+1)}{\varepsilon_{\alpha}}\right\}\\ &  +\frac{\omega_{\alpha}}{2}\left\{  \widetilde{\Xi}_{\alpha(e,n,\ell)}^{(+)}\left. \frac{d}{d\omega}\ln\varepsilon_{\alpha}(\omega)\right\vert_{\omega_{\alpha}}+\widetilde{\Xi}_{\alpha(e,n,\ell)}^{(-)}\left. \frac{d}{d\omega}\ln\mu_{\alpha}(\omega)\right\vert_{\omega_{\alpha}}\right\} \\ \widetilde{\mathcal{N}}_{\alpha(h,n,\ell)}^{2}  &  =z_{\alpha}^{2}(\varepsilon_{\alpha}-1)+(\mu_{\alpha}-1)\left\{  \left[ \varphi_{n}^{(+) }(z_{\alpha})\right]^{2}-\frac{n(n+1)}{\mu_{\alpha}}\right\} \\ & \qquad+\frac{\omega_{\alpha}}{2}\left\{  \widetilde{\Xi}_{\alpha(h,n,\ell)}^{(-)}\left.  \frac{d}{d\omega}\ln\varepsilon(\omega)\right\vert_{\omega_{\alpha}}+\widetilde{\Xi}_{\alpha(h,n,\ell)}^{(+)}\left. \frac{d}{d\omega}\ln\mu(\omega)\right\vert _{\omega_{\alpha}}\right\}  \;, \\ \widetilde{\Xi}_{\alpha(e,n,\ell)}^{(\pm)} &  \equiv\varepsilon_{\alpha}\left[  \varphi_{n}^{(+)}(z_{\alpha})\right]^{2}+\mu_{\alpha}z_{\alpha}^{2}-\frac{n(n+1)}{\varepsilon_{\alpha}}\pm\varphi_{n}^{(+)}(z_{\alpha})\\ \widetilde{\Xi}_{\alpha(h,n,\ell)}^{(\pm)} &  \equiv\mu_{\alpha}\left[ \varphi_{n}^{(+)}(z_{\alpha})\right]^{2}+\varepsilon_{\alpha}z_{\alpha}^{2}-\frac{n(n+1)}{\mu_{\alpha}}\pm\varphi_{n}^{(+)}(z_{\alpha})\;, \end{split}\end{align}
where $\varphi_{n}^{(+)}(z)$ is the reduced
logarithmic derivative of the outgoing Riccati Hankel function, $\xi_{n}(z)$:
\begin{equation} \varphi_{n}^{(+)}(z)\equiv\frac{d}{dz}\left[ z\xi_{n}(z)\right] =\frac{\xi_{n}^{\prime}(z)}{h_{n}(z)}\;. \end{equation}

Finally, we remark that we had to multiply the $A_{\alpha}$ coefficients of Ref.\cite{Doost14} by an additional $z_{\alpha}$ factor in order to account for the fact that we normalized the RS to a $z_{\alpha}$ in \cref{psinorm} as opposed to a normalization of the RSs to unity. Finally, reading off the relation between the $\overline{A}_{\alpha}$ and $\widetilde{\mathcal{N}}_{\alpha},$ in \cref{EHtilde} we find our expressions of $\widetilde{\mathcal{N}}_{\alpha}$ factors provided we first set $\mu_{\alpha}(\omega,r)=1$ and $\varepsilon_{\alpha} \rightarrow \varepsilon=\rho^{2}$ with $\varepsilon$ and $\rho$ real valued:
\begin{subequations}\begin{align} \left( \overline{A}_{\alpha}^{\mathrm{TE}}\right)^{2} &  =\frac{k_{\alpha }^{3}}{z_{\alpha}\widetilde{\mathcal{N}}_{\alpha(h,n,\ell)}^2} =\frac{1}{R^{3}(\varepsilon-1)}\overset{{\rm Ref.}\cite{Doost14}}{=}\frac{1}{R^{3}(n_R^{2}-1)}\\ \left(  \overline{A}_{\alpha}^{\mathrm{TM}}\right)^{2} &  =\frac{k_{\alpha}^{3}}{z_{\alpha}\widetilde{\mathcal{N}}_{\alpha(e,n,\ell)}^{2}} =\frac{k_{\alpha}^{3}}{z_{\alpha}(\varepsilon-1)\left\{  \left[ \varphi_{n}^{(+)}(z_{\alpha})\right]^{2}+\frac{n(n+1)}{\varepsilon}\right\}  }=\frac{z_{\alpha}^{2}}{R^{3}\left\{ \left[\frac{\xi_{n}^{\prime}(z_{\alpha})}{h_{n}^{2}(z_{\alpha})}\right]^{2} +\frac{n(n+1)}{\varepsilon}\right\}  (\varepsilon - 1)}\;. \label{ATM} \end{align}\end{subequations}
The agreement with the $A_{\alpha}^{\mathrm{TE}}$ factor with our formulas of Ref.\cite{Doost14} is immediate by remaking that $\varepsilon\rightarrow n_{R}^{2}$ where $n_{R}$ is the notation of Ref.\cite{Doost14}  for relative index contrast. Our expression in \cref{ATM} for $\overline{A}_{\alpha}^{\mathrm{TM}}$ also agrees with that of Ref.\cite{Doost14} even though this is not  immediately clear by inspection. In order reveal this agreement, we first recall that the magnetic mode RS condition of \cref{magmodecond} can be expressed:
\begin{equation}\frac{\xi_{n}^{\prime}(z_{\alpha})}{h_{n}(z_{\alpha})}=\frac{1}{\varepsilon_{\alpha}}\frac{\psi_{n}^{\prime}(\rho_{\alpha}z_{\alpha} )}{j_{n}(\rho_{\alpha}z_{\alpha})}\;, \end{equation}
which allows us to write,
\begin{align} \begin{split}\label{ATMinv} \frac{1}{\left(  \overline{A}_{\alpha}^{\mathrm{TM}}\right)^{2}} & =R^{3}\frac{\varepsilon-1}{\varepsilon}\left\{  \frac{1}{\varepsilon}\left[  \frac{\psi_{n}^{\prime}(\rho z_{\alpha})}{z_{\alpha}j_{n}(\rho z_{\alpha})}\right]^{2}+\frac{n(n+1)}{z_{\alpha}^{2}}\right\}  \\ &  =R^{3}\frac{\varepsilon-1}{\varepsilon} \left\{  \frac{1}{\varepsilon}\left[ \frac{\rho j_{n-1}\left(  \rho z_{\alpha}\right)  }{j_{n}(\rho z_{\alpha })}-\frac{n}{z}\right]^{2}+\frac{n(n+1)}{z_{\alpha}^{2}}\right\}  \;,\end{split}\end{align} and if we express this in a manner of Ref.\cite{Doost14}, \cref{ATMinv} becomes,
\begin{align} \varepsilon \left(  \frac{A_{\alpha}^{\mathrm{TE}}}{A_{\alpha}^{\mathrm{TM}}}\right)^{2}=\frac{1}{\varepsilon}\left[ \frac{\rho j_{n-1}\left( \rho_{\alpha}z_{\alpha}\right)}{j_{n}(\rho z_{\alpha})}-\frac{n}{z_{\alpha}}\right]^{2}+\frac{n(n+1)}{z_{\alpha}^{2}} \;, \end{align}
which in their notation reads,
\begin{equation} n_{R}\frac{A_{n}^{\mathrm{TE}}}{A_{n}^{\mathrm{TM}}}\overset{{\rm Ref.}\cite{Doost14}}{=}\sqrt{\left[ \frac{j_{l-1}( n_{R}kR)}{j_{l}(n_{R}kR)}-\frac{l}{n_{R} kR}\right]^{2}+\frac{l(l+1)}{k^{2}R^{2}}} \;, \end{equation}
so our expressions finally agree exactly provided we take $\mu(\omega,r)=1$ and set $n_{R}\rightarrow\sqrt{\varepsilon}=\rho$.

In the above, we only compared the formulas  for dispersionless media normalization of Ref.\cite{Doost14}, but proceeding along the same lines permits the retrieval of the comparison with later formulas derived for dispersive media \cite{Muljar2016b}.


\section{Vector partial waves and vector spherical harmonics}\label{VPWs} 

The regular Vector Partial Waves (VPWs) are defined as:
\begin{subequations}\label{RgMetN} \begin{align} \boldsymbol{M}_{q,n,m}^{(1)}(k\boldsymbol{r})  &  \equiv j_{n}(kr)  \boldsymbol{X}_{q,n,m}(\theta,\phi)\label{Hwave}\\ \boldsymbol{N}_{q,n,m}^{(1)}(k\boldsymbol{r})  &  \equiv\frac{1}{kr} \left[ \sqrt{n(n+1)}j_{n}(kr) \boldsymbol{Y}_{q,n,m}(\theta,\phi)+\psi_{n}^{\prime}(kr) \boldsymbol{Z}_{q,n,m}(\theta,\phi)\right] \;, \label{Ewave}\end{align}\end{subequations}
while for the outgoing $(+)$ and incoming ($-$) waves,

\begin{subequations}\label{OutMetN}\begin{align} \boldsymbol{M}_{q,n,m}^{(\pm)}(k\boldsymbol{r})  &  \equiv h_{\pm,n}(kr) \boldsymbol{X}_{q,n,m}(\theta,\phi)\\ \boldsymbol{N}_{q,n,m}^{(\pm)}(k\boldsymbol{r})  &  \equiv \frac{1}{kr}\left[\sqrt{n(n+1)}h_{\pm,n}(kr) \boldsymbol{Y}_{q,n,m}(\theta,\phi) +\xi_{\pm,n}^{\prime}(kr) \boldsymbol{Z}_{q,n,m}(\theta,\phi)\right]  \;. \end{align}

In \cref{RgMetN,OutMetN}, we used the Riccati-Bessel functions $\psi_{n}$ and
$\xi_{\pm,n}$ that are functions of a complex variable $z$, defined by,
\end{subequations}
\begin{equation} \psi_{n}(z)  \equiv zj_{n}(z)  \text{ \ \ and \ \ \ \ \ }\xi_{\pm,n}(z)  \equiv z h_{\pm,n}(z)  \;, \end{equation}
and the prime is the derivative with respect to the argument, \textit{i.e.},
\begin{align}\begin{split} \psi_{n}^{\prime}(z) &  =j_{n}(z) +xj_{n}^{\prime}(z) \\\xi_{\pm,n}^{\prime}(z)   &  =h_{\pm,n}(z)  +z h_{\pm,n}^{\prime}(z)\; . \end{split}\end{align}

The angle dependent vector spherical harmonics, $\boldsymbol{X}$, $\boldsymbol{Y}$, and $\boldsymbol{Z}$
of the `electric' mode type ($q=1$) are defined for $m=0,1,...,n:$
\begin{align}\begin{split} \boldsymbol{Y}_{e,n,m}(\theta,\phi)&\equiv \boldsymbol{Y}_{1,n,m}(\theta,\phi)\equiv\widehat{\boldsymbol{r}}\overline{P}_{n}^{m}(\cos\theta)\cos m\phi \\ \boldsymbol{X}_{e,n,m}(\theta,\phi) &  \equiv\boldsymbol{X}_{1,n,m}(\theta,\phi)\equiv-\overline{u}_{n}^{m}(\cos\theta)\sin(m\phi)  \widehat{\boldsymbol{\theta}}-\overline{s}_{n}^{m}(\cos \theta)\cos( m\phi) \widehat{\boldsymbol{\phi}} \\ \boldsymbol{Z}_{e,n,m}(\theta,\phi) & \equiv \boldsymbol{Z}_{1,n,m}(\theta,\phi)  \equiv\overline{s}_{n}^{m}(\cos\theta)\cos(m\phi)  \widehat{\boldsymbol{\theta}}-\overline{u}_{n}^{m}(\cos\theta)\sin( m\phi) \widehat{\boldsymbol{\phi}} \;, \label{SPHe} \end{split}\end{align}
while the magnetic VSHs are defined for $m=1,...,n$,
\begin{align}\begin{split} \boldsymbol{Y}_{h,n,m}(\theta,\phi)&\equiv \boldsymbol{Y}_{0,n,m}(\theta,\phi) \equiv \overline{P}_{n}^{m}(\cos\theta)\sin m\phi \\ \boldsymbol{X}_{h,n,m}(\theta,\phi) & \equiv \boldsymbol{X}_{0,n,m}(\theta,\phi) \equiv\overline{u}_{n}^{m}(\cos\theta)\cos\left(m\phi\right)  \widehat{\boldsymbol{\theta}}-\overline{s}_{n}^{m}(\cos\theta)\sin( m\phi)  \widehat{\boldsymbol{\phi}}\\ \boldsymbol{Z}_{h,n,m}(\theta,\phi) &  \equiv \boldsymbol{Z}_{0,n,m} (\theta,\phi)\equiv \overline{s}_{n}^{m}(\cos\theta) \sin\left(m\phi\right)  \widehat{\boldsymbol{\theta}} +\overline{u}_{n}^{m}(\cos\theta)\cos( m\phi)  \widehat{\boldsymbol{\phi}} \;. \label{SPHo}
\end{split}\end{align}
We remark that in the classic text of ref.~\cite{Bohr98}, the magnetic and electric modes were referred to as
`odd' and `even' modes respectively, but we found the terminology magnetic and electric to have more physical relevance. We also point a recent convention of associating the magnetic (`odd') modes with negative $m$ and the electric (`even') modes with non-negative $m$ values. Although such a notation has the advantage of eliminating the need of a separate index to distinguish electric and magnetic type modes, we found it useful to keep the separate index here for reasons of clarity in explanations and notation.

The normalized associated Legendre functions are defined,
\begin{align}\begin{split} \overline{P}_{n}^{m}\left(  \cos\theta\right) &\equiv  \gamma_{nm}\sqrt{n(n+1)}P_{n}^{m}\left(  \cos\theta\right) \\ \overline{u}_{n}^{m}(\cos\theta)  &  \equiv\frac{1}{\sqrt{n(n+1)}}\frac{m}{\sin\theta} \overline{P}_{n}^{m}(\cos\theta)\\ \overline{s}_{n}^{m}(\cos\theta)  &  \equiv\frac{1}{\sqrt{n(n+1)}}\frac{d}{d\theta}\overline{P}_{n}^{m}(\cos\theta)\;. \end{split}\end{align}
with a normalization factor,
\begin{equation} \gamma_{nm}\equiv\sqrt{\frac{(2n+1)(n-m)!}{4\pi n(n+1)(n+m)!}} \;. \end{equation}
and we adopted a common definition of (positive $m$) associated Legendre functions as:
\begin{equation} P_{n}^{m}(x)=(-1)^{m}\left( 1-x^{2}\right)  ^{m/2}\frac{d^{m}}{dx^{m}} P_{n}(x) \;. \end{equation}

 \section{Killing Mie softly}
\label{KMS}

This section demonstrates how an integral of special functions with diverging amplitudes can yield finite results by first taming the relevant integrals by a Gaussian factor, $\exp(-\eta x^2)$ and then taking of the limit $\eta \rightarrow 0$. The analytic arguments presented here and in~\cite{McPhStout20} may also be of help to those developing purely numerical methods for nanophotonics. As described in~\cite{McPhStout20}, if the integrand oscillates about a non-zero mean value, this needs to be treated separately, as it may generate a delta function contribution (such contributions cancel out in the cases treated here). The oscillations about a mean of zero will in general give a zero contribution for the integration variable tending to infinity, as in the spirit of generalized function theory.

 We now give an example of the results in the references~\cite{McPhStout20,MDS}. The integral chosen is that over the product of the spherical Bessel functions $j_{n}(K x) $ and $y_{n}(k x)$, where $K$ and $k$ are (possibly complex) wavenumbers: see equation \cref{snr18}.
 \begin{eqnarray} {\cal I}_{j y}(n,K,k,\eta) &\equiv& \int_0^\infty x^2 \exp(-\eta x^2) j_{n}(K x) y_{n}(k x) d x \nonumber \\ &=& \frac{\pi}{2}\left\{ \frac{\exp [-(K^2+k^2)/(4 \eta)]}{2\pi \eta} \left[ -{\cal H}(n+1/2,k,K,\eta) +(n+1/2) h_{-1,n+1/2}\left(\frac{-K k}{2\eta}\right) \right]\right\}. \nonumber\\ && \label{snr18} \end{eqnarray}
 The analytic result for this integral from~\cite{MDS} is given by the rightmost expression in (\ref{snr18}). Here $h_{-1,b}$ denotes an associated Bessel function~\cite{Luke62}, and ${\cal H}(b,k,K,\eta)$ is the following finite-range integral:
\begin{equation} {\cal H}(b,k,K,\eta)=\int_1^{K/k} u^{(b - 1)} \exp[\frac{K k}{4\eta} (u + 1/u)] du \;. \label{mds4} \end{equation}
Using expansions given by Luke~\cite{Luke62} and evaluating the finite integral in (\ref{mds4}) by direct numerical integration, the expression (\ref{snr18}) may be verified for arbitrary choices of the parameters $K$, $k$ and $\eta$.

The asymptotic treatment given in~\cite{MDS} which takes the limit as $\eta \rightarrow 0$ is  lengthy and complicated. However, it yields a simple result, which is in keeping with the well-known result from Watson~\cite{Watson80}:
 \begin{equation} \int^{z} z {\cal C}_\mu (k z)  {\cal D}_\mu (l z) d z= \frac{z\left\{k {\cal C}_{\mu+1} (k z) {\cal D}_\mu (l z) -l {\cal C}_\mu (k z) {\cal D}_{\mu+1} (l z)\right\} }{k^2-l^2}, \label{watsoninteg} \end{equation}
where ${\cal C}_\mu$ and ${\cal D}_\mu$ are cylinder functions of integer or real order $\mu$. For the integral (\ref{snr18}) we need to take ${\cal C}_\mu=J_{n+1/2}$
and ${\cal D}_\mu=Y_{n+1/2}$. The expression (\ref{watsoninteg}) is used (with a minus sign) to give the contribution from the lower limit to the integral (\ref{snr18}). In the case chosen, the asymptotics show that the contribution from the upper limit (infinity) is precisely zero. (For other cases, like the integrals $ {\cal I}_{j j}$ and $ {\cal I}_{y y}$, there is a delta function contribution which comes from the upper limit; this arises if the asymptotic expansion of the integrand in expressions like (\ref{snr18}) with $\eta=0$ as $x \rightarrow \infty$ contain a constant term in addition to terms oscillating around zero.) The result is in this case:
\begin{equation} \lim_{\eta\rightarrow 0} {\cal I}_{j y}(n,K,k,\eta)=\frac{ \quad \left(\frac{K}{k}\right)^{n+1/2}}{k^2-K^2}.\label{snr21}
\end{equation}

\begin{figure}[tbh]
\includegraphics[width=\linewidth]{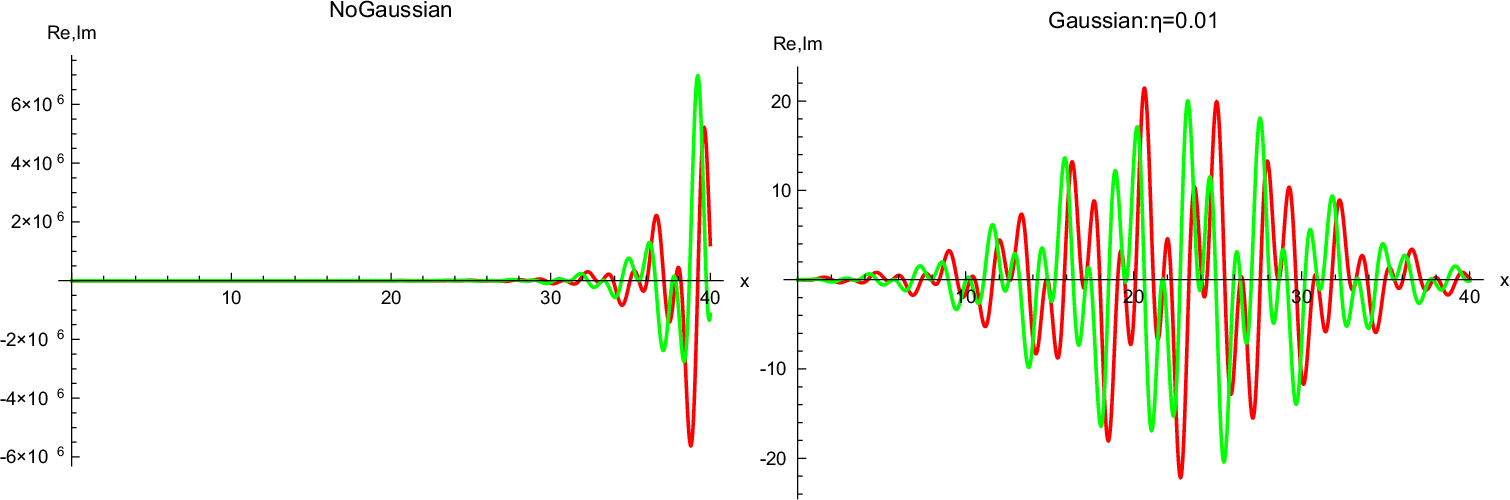}
\caption{\label{fig:kill}The effect of a Gaussian ``killing" function on ${\cal I}_{jy}$: $n=1$, $k=1.37$, $K=2.96 + 0.457 i$. At left: without the Gaussian factor, the integrand (red: real part; green: imaginary part)  diverges strongly; at right, with the Gaussian factor with $\eta=0.01$ the  integrand converges to zero for large $x$.}
\end{figure}

We now give an example of the effectiveness of the Gaussian ``killing" function technique for the integral (\ref{snr18}): see \cref{fig:kill}. Even with a Gaussian with $\eta$ only equal to 0.01, the divergent integrand is replaced by one which can be integrated accurately.
The numerical integration of the Gaussian form with $\eta =0.01$ gives $0.0164787 - 0.0138487 i$, for integration with upper limit 80 or beyond. The analytic value for the integral (\ref{snr21}) is $0.0163332 - 0.0135188 i$. For $\eta=0.005$, the numerical integral gives $0.0164062 - 0.0136812 i$,  slightly closer to the exact answer, while for $\eta=0.001$,  the numerical integration in Mathematica fails.

\begin{figure}[tbh]
\includegraphics[width=\linewidth]{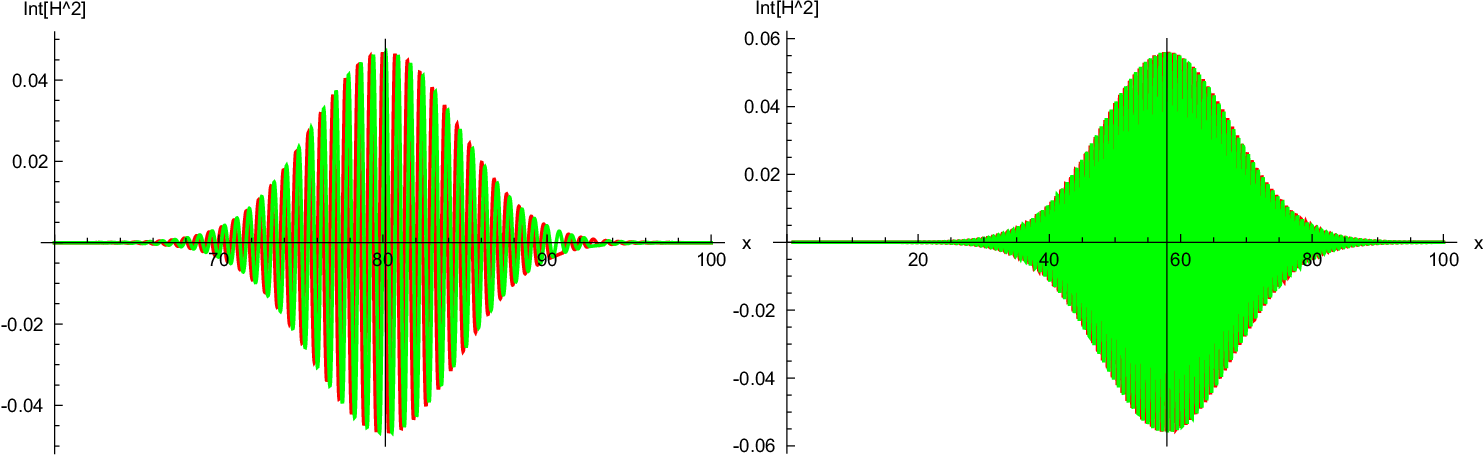}
\caption{\label{fig:kill2}The effect of  Gaussian ``killing" functions on ${\cal I}_{hh}$: $n=1$, $k=4 (1.0395-0.5009 i$ (left), $k=4 (1.0527-0.072355 i)$
(right).  The integrand (red: real part; green: imaginary part)  has the  Gaussian factors with $\eta=0.025 $ (left) and $0.005$ (right).
The black vertical lines mark the estimates $-{\rm Im}(k)/\eta$ for the abscissa corresponding to the peak of the Gaussian envelopes.}
\end{figure}

As a second example, we give in \cref{fig:kill2} plots of the  integrand in the following normalization integral:
\begin{equation} {\cal I}_{hh} \equiv \lim_{\eta \rightarrow 0} \left[ \int_R^\infty x^2 h_n^{(1)} (k x)^2 e^{-\eta x^2} \right] dx= -\frac{R^3}{2} \left[ h_n^{(1)} (k R)^2  -h_{n-1}^{(1)} (k R) h_{n+1} ^{(1)} (k R)\right]\;. \label{hhintex} \end{equation}
As $\eta$ gets smaller, the oscillating real and imaginary parts of the integrand become for larger $x$ concentrated between Gaussian envelopes. The envelopes peak round $x=-\Im (k)/\eta$, and have a $1/e$ full width of $2/\sqrt{\eta}$. The peak modulus of the envelope is $\exp \{[\Im (k)]^2/\eta\}$, and the oscillations within the envelope go as $\exp [2i \Re (k) x]$.  The oscillatory behavior means that the value of the integral over this peaked region alternates between positive and negative values, and has on average the value zero. To carry out the integral directly for small $\eta$ becomes increasingly difficult, particularly if $\Re (k)$ is small in magnitude.

The peak value of the envelope increases rapidly as $\eta$ decreases, while there are more and more cancelling positive and negative contributions to the integral. In order to obtain an accurate numerical estimate for the integral, more and more decimal places have to be used in evaluations of the elements of the integrand, if the oscillations are to cancel the diverging peak value.

This means, for calculations with a fixed accuracy (say 10-12 decimal places) there will be a smallest value of $\eta$ for which the numerical estimate is good. The calculations will get more and more difficult as the ratio $|\Im(k)/\Re(k)|$  increases. These remarks emphasize the value of the analytic studies~\cite{McPhStout20} which have shown there is a simple explicit and general answer for normalization integrals such as (\ref{hhintex}).

In  region outside the sphere, the specific integrals of the more general exposition of Ref.~\cite{McPhStout20} that are needed for this work are:
\begin{subequations}
\label{outanalab}
\begin{align}\begin{split} & \int_{R}^{\infty}d\boldsymbol{r} \left[ \Phi_{q,n,m}^{(+)}(k_{\alpha}\boldsymbol{r})\right]^{t}.  \Gamma.\Phi_{q^{\prime},\nu,\mu}^{(+)}(k_{\beta}\boldsymbol{r}) =  \delta_{q,q^{\prime}} \delta_{n,\nu} \delta_{m\mu}(-1)^{n-q} k_{\alpha}^{3/2}k_{\beta}^{3/2} \times \\ & \qquad \left\{ \int_{R}^{\infty}d\boldsymbol{r} \boldsymbol{M}_{q,n,m}^{(+)}(k_{\alpha}\boldsymbol{r})\cdot \boldsymbol{M}_{q,n,m}^{(+)}(k_{\beta}\boldsymbol{r}) +  \int_{R}^{\infty}d\boldsymbol{r} \boldsymbol{N}_{q,n,m}^{(+)}(k_{\alpha}\boldsymbol{r})\cdot \boldsymbol{N}_{q,n,m}^{(+)}(k_{\beta}\boldsymbol{r}) \right\}
\end{split}\end{align}
which for $\alpha\neq \beta$ are analytically: \begin{align}\begin{split} \int_{R}^{\infty}d\boldsymbol{r} \boldsymbol{M}_{q,n,m}^{(+)}(k_{\alpha}\boldsymbol{r})\cdot  \boldsymbol{M}_{q,n,m}^{(+)}(k_{\beta}\boldsymbol{r}) &=  \underset{\eta \rightarrow 0}{\text{lim}}\int_{R}^{\infty}r^{2}h_{n}(  k_{\alpha}r)  h_{n}( k_{\beta}r)  e^{-\eta r^{2}}dr \\ &= - R \frac{ z_{\beta} h_{n}(z_{\alpha})  h_{n}^{\prime} (z_{\beta}) -z_{\alpha}h_{n}^{\prime}(  z_{\alpha}) h_{n}(  z_{\beta})  }{k_{\alpha}^{2}-k_{\beta}^{2}} \end{split}\\ \begin{split} \int_{R}^{\infty}d\boldsymbol{r} \boldsymbol{N}_{q,n,m}^{(+)}(k_{\alpha}\boldsymbol{r})\cdot \boldsymbol{N}_{q,n,m}^{(+)}(k_{\beta}\boldsymbol{r})&=\underset{\eta \rightarrow 0}{\text{lim}}\int_{R}^{\infty} \frac{ n(n+1) h_{n}(k_{\alpha}r) h_{n}( k_{\beta}r)  + \xi_{n}^{\prime}( k_{\alpha}r) \xi_{n}^{\prime}(  k_{\beta}r)}{k_{\alpha}k_{\beta}}  e^{-\eta r^{2}}dr \\ & =-R \frac{z_{\alpha} h_{n}(z_{\alpha})   h_{n}^{\prime}(z_{\beta})-z_{\beta} h_{n}(z_{\beta})   h_{n}^{\prime}(z_{\alpha})}{k_{\alpha}^{2}-k_{\beta}^{2}} - R \frac{h_{n}(z_{\alpha})h_{n}(z_{\beta})}{k_{\alpha}k_{\beta}} \;,\end{split}\end{align}\end{subequations}
while for $\beta=\alpha$ one finds:
\begin{subequations}\label{outanalaa} \begin{align} \begin{split} \int_{R}^{\infty}d\boldsymbol{r} \left[ \boldsymbol{M}_{q,n,m}^{(+)}(k_{\alpha}\boldsymbol{r})\right]^{2} &=\underset{\eta \rightarrow 0}{\text{lim}}\int_{R}^{\infty}r^{2}h_{n}^{2}(k_{\alpha}r)  e^{-\eta r^{2}}dr \\&=-R \frac{\left[\xi_{n}^{\prime}(z_{\alpha})\right]^{2} +\xi_{n}^{2}(z_{\alpha})-n(n+1) h_{n}^{2}(z_{\alpha}) -h_{n}(z_{\alpha})  \xi_{n}^{\prime}(z_{\alpha})}{2k_{\alpha}^{2}} \end{split}\\ \begin{split} \int_{R}^{\infty}d\boldsymbol{r} \left[\boldsymbol{N}_{q,n,m}^{(+)}(k_{\alpha}r)\right]^{2} &= \underset{\eta \rightarrow 0}{\text{lim}}\int_{R}^{\infty}\frac{ n(n+1)h_{n}^{2}( k_{\alpha}r) + \left[\xi_{n}^{\prime}\left(  k_{\alpha}r\right)\right]^{2} }{k^{2}_{\alpha}} e^{-\eta r^{2}} dr \\ & =- R \frac{\left[  \xi_{n}^{\prime}( z_{\alpha})\right]^{2} +\xi_{n}^{2}(  z_{\alpha})  -n(n+1)  h_{n}^{2}(z_{\alpha}) + h_{n}(z_{\alpha})  \xi_{n}^{\prime}(z_{\alpha})}{2k^{2}_{\alpha}} \;, \label{hhint} \end{split}\end{align} \end{subequations}
where $z_{\alpha}\equiv k_{\alpha}R$, with $h_{n}(z)$ the outgoing spherical Hankel functions, and $\xi_{n}(z)\equiv z h_{n}(z)$ the outgoing Ricatti-Hankel functions. We recall that the above multipole integrals are the only one for a finite sized system at these analytically regularized results are the only 

The integrals of the regular partial waves inside a dispersionless homogeneous sphere of arbitrary radius $R$ inside the particle can also be carried out analytically without the need for regularization: 
\begin{subequations}\label{inanalab}\begin{align}\begin{split} \int_{0}^{R}d\boldsymbol{r} & \left[\Phi_{q,n,m}(k_{\alpha}\boldsymbol{r})\right]^{t}.\Gamma. \Phi_{q,n,m}(k_{\beta}\boldsymbol{r})  = \varepsilon k_{\alpha}^{3/2}k_{\beta}^{3/2}(-1)^{n-q} \times \\ &\left\{ \int_{0}^{R}d\boldsymbol{r} \boldsymbol{M}_{q,n,m}^{(1)}(k_{\alpha}\boldsymbol{r})\cdot \boldsymbol{M}_{q,n,m}^{(1)}(k_{\beta}\boldsymbol{r}) + \int_{0}^{R}d\boldsymbol{r} \boldsymbol{N}_{q,n,m}^{(1)}(k_{\alpha}\boldsymbol{r})\cdot \boldsymbol{N}_{q,n,m}^{(1)}(k_{\beta}\boldsymbol{r})\right\} \;, \end{split}
\end{align}
which when $\alpha\neq \beta$, the analytic multipole integrals are:
\begin{align}\begin{split}\int_{0}^{R}d\boldsymbol{r} \boldsymbol{M}_{q,n,m}^{(1)}(k_{\alpha}\boldsymbol{r})\cdot  \boldsymbol{M}_{q,n,m}^{(1)}(k_{\beta}\boldsymbol{r}) &=\int_{0}^{R}r^{2}j_{n}( k_{\alpha} r) j_{n}( k_{\beta}r) dr\\ &= R \frac{ z_{\beta} j_{n}(z_{\alpha})  j_{n}^{\prime}(z_{\beta}) -z_{\alpha}j_{n}^{\prime}(  z_{\alpha}) j_{n}(z_{\beta})}{k_{\alpha}^{2}-k_{\beta}^{2}} \end{split} \end{align}
\begin{align}\begin{split} & \int_{0}^{R} d\boldsymbol{r} \boldsymbol{N}_{q,n,m}^{(1)}(k_{\alpha}\boldsymbol{r})\cdot \boldsymbol{N}_{q,n,m}^{(1)}(k_{\beta}\boldsymbol{r}) =\int_{0}^{R} \frac{  n(n+1) j_{n}(k_{\alpha}r)  j_{n}(k_{\beta}r)  +\psi_{n}^{\prime}( k_{\alpha}r)
\psi_{n}^{\prime}(  k_{\beta}r) }{k_{\alpha}k_{\beta}} dr \\ & \qquad\qquad=R \frac{z_{\alpha} j_{n}(z_{\alpha})  j_{n}^{\prime}(z_{\beta})-z_{\beta} j_{n}(z_{\beta}) j_{n}^{\prime}(z_{\alpha})}{k_{\alpha}^{2}-k_{\beta}^{2}} + R \frac{j_{n}(z_{\alpha})j_{n}(z_{\beta})}{k_{\alpha}k_{\beta}} \label{jjint} \\ \end{split}\end{align}\end{subequations}
while when $\beta=\alpha$, one finds:
\begin{subequations} \label{inanalaa} \begin{align}\begin{split} \int_{0}^{R}d\boldsymbol{r} \left[ \boldsymbol{M}_{q,n,m}^{(1)}(k_{\alpha}\boldsymbol{r})\right]^{2} & =\int_{0}^{R}r^{2}j_{n}^{2}(  k_{\alpha}r)  dr  \\ & = R \frac{\left[\psi_{n}^{\prime}(z_{\alpha})  \right]^{2} +\psi_{n}^{2}(z_{\alpha})-n(n+1)  j_{n}^{2}(z_{\alpha}) -j_{n}(z_{\alpha})  \psi_{n}^{\prime}(z_{\alpha})}{2k_{\alpha}^{2}} \end{split}\\ \begin{split}  \int_{0}^{R}d\boldsymbol{r} \left[ \boldsymbol{N}_{q,n,m}^{(1)}(k_{\alpha}\boldsymbol{r})\right]^{2}   & = \int_{0}^{R} \frac{  n(  n+1)  j_{n}^{2}(  k_{\alpha}r) +\left[\psi_{n}^{\prime}\left(  k_{\alpha}r\right)\right]^{2} }{k_{\alpha}^{2}}  dr\\ & = R\frac{\left[  \psi_{n}^{\prime}( z_{\alpha})\right]^{2} +\psi_{n}^{2}(  z_{\alpha})-n(n+1) j_{n}^{2}(z_{\alpha}) + j_{n}(z_{\alpha}) \psi_{n}^{\prime}(z_{\alpha})}{2k^{2}_{\alpha}} \;, \end{split}\end{align}
where $j_{n}$ are the spherical Bessel functions with $\psi_{n}(z)\equiv z j_{n}(z)$.
\end{subequations}
These formulas are particularly useful for spherical scatterers but could also prove useful in non-spherical particles and multiple scattering situations.

\section{Mie theory}\label{app:Mietheory}

\subsection{Multipolar decomposition of fields}

Mie theory relies on developing fields in different regions in terms of vector partial waves (VPWs), which are homogeneous media electromagnetic wave solutions in an angular momentum eigenstate basis. 
The excitation field, $\boldsymbol{E}_{\mathrm{e}}$, is the field created by source currents in the absence of the scatterer and it can be developed on the complete basis of `regular' multipolar fields denoted by a superscripted $(1)$,
\begin{subequations} \begin{equation} \boldsymbol{E}_{\mathrm{e}}(k\boldsymbol{r})   =E\sum_{n=1}^{\infty}\sum_{m=0}^{n}\left[ e_{h,n,m} \boldsymbol{M}_{h,n,m}^{(1)}(k\boldsymbol{r})+e_{e,n,m} \boldsymbol{N}_{e,n,m}^{(1)}(k\boldsymbol{r})\right] \;, \end{equation}
where $n$ is the electromagnetic angular momentum quantum number, $m$ an absolute value of the angular momentum projection. The real parameter, $E$, determines the strength of the incident field. The real-valued multipole wave functions, $\boldsymbol{N}_{q,n,m}(k\boldsymbol{r})$ and $\boldsymbol{M}_{q,n,m}(k\boldsymbol{r})$, expressed in \cref{RgMetN} are either of the magnetic, $h(q=0)$, type or electric, $e(q=1)$, type (denoted `even' and `odd' multipolar types in Bohren and Huffman~\cite{Bohr98}).

The total field inside a spherical particle can also be developed in terms of regular VPWs provided that position vector in the VPWs is weighted by the field frequency dependent wavenumber inside the sphere, $k\rho_{\omega}$ using the notation of  \cref{param},
\begin{equation}
\boldsymbol{E}_{\mathrm{int}}(k\boldsymbol{r})   =E\sum_{n=1}^{\infty}\sum_{m=0}^{n}\left[  s_{h,n,m}\boldsymbol{M}_{h,n,m}^{(1)}(k \rho_{\omega} \boldsymbol{r})+s_{e,n,m} \boldsymbol{N}_{e,n,m}^{(1)}(k\rho_{\omega} \boldsymbol{r})\right]  \;.\end{equation}
Mie theory also involves the `scattered field' which has its traditional definition as the incident field subtracted from the total field in presence of the scatterer, which can be developed in terms of outgoing VPWs,
\begin{equation} \boldsymbol{E}_{\mathrm{s}}(k\boldsymbol{r})  =E\sum_{n=1}^{\infty}\sum_{m=0}^{n}\left[  f_{h,n,m} \boldsymbol{M}_{h,n,m}^{(+)}(k\boldsymbol{r})+f_{e,n,m}\boldsymbol{N}_{e,n,m}^{(+)}(k\boldsymbol{r}) \right] \;, \end{equation} \end{subequations}
which can be viewed as replacing the spherical Bessel functions in the definitions of the regular VPWs by outgoing Hankel functions as shown in \cref{OutMetN}.

\subsection{Mie coefficients and cross sections}
\label{ssect:Miecoefcross}

A consequence of linear response of any scatterer is that if one expresses the scattered and internal fields coefficients, $f$ and $s$, as infinite dimensional column matrices, then the scattering and internal field coefficients can be expressed in terms of the column matrix of excitation field coefficients {\it via} infinite \emph{matrices}, $T$ and $\Omega$, such that, $f=T.e$ and $s=\Omega.e$. The fact that scattering by spheres does not change the angular momentum, both $T$ and $\Omega$ become diagonal matrices for  spherically symmetric scatterers and traditional `Mie' theory provides algebraic expressions for these diagonal elements, henceforth denoted $T_{q,n}$ and $\Omega_{q,n}$. For the different multipole orders, one has,
\begin{subequations}
\begin{align}\begin{split} f_{q,n,m} & = T_{q,n}e_{q,n,m}  = - \frac{N_{q,n}(\omega,R)}{D_{q,n}(\omega,R)} e_{q,n,m} \\ & b_{n} = -T_{h,n}  \quad ,\quad a_{n} = -T_{e,n} \label{TMie} \end{split}\end{align}
where we specified the relation between $T_{q,n}$ and the time honored Mie coefficients, traditionally denoted $a_{n}$ and $b_{n}$. The `numerator' and `denominator' functions in \cref{TMie} are functions of the angular frequency, $\omega$ and particle radius, $R$ are written out explicitly in \cref{numT_func} and \cref{den_funcs}.

Mie theory also provides expressions for the lesser known $c_{n}$ and $d_{n}$ coefficients between the internal field and the incident field decomposition which can be directly expressed in terms of denominator functions,
\begin{align}\begin{split} s_{q,n,m} &\equiv \Omega_{q,n} e_{q,n,m} = \frac{1}{D_{q,n}(\omega,R)}e_{q,n,m} \\ &  \quad c_{n} = \Omega_{h,n} \quad ,\quad d_{n} = \Omega_{e,n} \;. \label{intMie} \end{split}\end{align}\label{Miecoefs} \end{subequations}

The `numerator' and `denominator' functions in \cref {Miecoefs} are functions of the angular frequency, $\omega$ and particle radius, $R$:
\begin{subequations} \begin{align}
N_{h,n}(\omega, R)  &  \equiv z\frac{\mu_{\omega}j_{n}(\rho_{\omega}z) \psi_{n}^{\prime}(z)-\psi_{n}^{\prime}(\rho_{\omega}z)j_{n}(z)}{i\mu_{\omega}}\\ N_{e,n}(\omega,R)  &  \equiv z\frac{\varepsilon_{\omega}j_{n}(\rho_{\omega}z) \psi_{n}^{\prime}(z)-\psi_{n}^{\prime}(\rho_{\omega}z)j_{n}(z)}{i\rho_{\omega}} \;,\end{align} \label{numT_func}\end{subequations}
and,
\begin{subequations}\label{den_funcs}\begin{align}
D_{h,n}(\omega, R)  &  \equiv z\frac{\mu_{\omega}j_{n}(\rho_{\omega}z) \xi_{n}^{\prime}(z)-\psi_{n}^{\prime}(\rho_{\omega}z)h_{n}(z)}{i\mu_{\omega}} \label{dene_func} \\ D_{e,n}(\omega,R)  &  \equiv z\frac{\varepsilon_{\omega}j_{n}(\rho_{\omega}z) \xi_{n}^{\prime}(z) - \psi_{n}^{\prime}(\rho_{\omega}z)h_{n}(z)}{i\rho_{\omega}} \label{denh_func}\;, \end{align}\end{subequations}
where we have again used $z=kR=\omega R/c$ and used the condensed notation of \cref{param} for
$\varepsilon_{\omega}$, $\mu_{\omega}$, and $\rho_{\omega}$, for the (possibly frequency dependent) constitutive parameters:
An inspection of \cref{numT_func,den_funcs} readily shows for non-dispersive media, the, $N_{q,n}$ and $D_{q,n}$ reduce to functions of size parameter, $z=kR$, only.

Cross sections like those plotted in \cref{fig:RSMie} are obtained from the Mie coefficients of \cref{TMie} via well-known relations:
\begin{subequations}\begin{align} Q_{\mathrm{ext}}  & \equiv \frac{\sigma_{\mathrm{ext}}}{\pi R^{2}} = \frac{1}{\pi R^{2}} \sum_{n=1}^{\infty} \left(\sigma_{\mathrm{ext},n}^{(h)} +\sigma_{\mathrm{ext},n}^{(e)} \right) =-\frac{2}{(kR)^{2}}\sum_{n=1}^{\infty} (2n+1)  \operatorname{Re}\left[ T_{h,n}+T_{e,n}\right] \\ Q_{\mathrm{scat}}  &  =\frac{2}{( kR)^{2}}\sum_{n=1}^{\infty}(2n+1)\left[ \left\vert T_{h,n}\right\vert^{2} +\left\vert T_{e,n}\right\vert^{2} \right] \\ Q_{\mathrm{abs}} &= Q_{\mathrm{ext}} - Q_{\mathrm{scat}}\;. \end{align} \label{crosssect} \end{subequations}


\section{Resonant state electric field  orthogonalization}\label{Orthint}

In the main text, we showed the orthogonality of electric mode RS product integrals for a spherical scatterer when integrating the magnetic field product of the RSs products over all space when $\alpha \neq \beta$. Here we show that RS orthogonality also holds for products of the \emph{electric} fields of the electric type modes (n.b. the same integrals also arise in magnetic field integration of magnetic type modes). The volume inner product integral inside the sphere for two RSs $\alpha$ and $\beta$ is zero unless they share the same $q,n,m$ numbers, and in the case where they do share the same $q,n,m$, their product integral inside the sphere is, 
\begin{subequations}
\begin{align}
& \int_{0}^{R}d\boldsymbol{r} \varepsilon \boldsymbol{E}_{\alpha(e,n,m,\ell)} (\boldsymbol{r})\cdot \boldsymbol{E}_{\beta(e,n,m,\ell^{\prime})} (\boldsymbol{r}) \notag
\\ &=
\frac{z_{\alpha}^{1/2}z_{\beta }^{1/2}\gamma_{\alpha(e,n)}
\gamma_{\beta(e,n)}}{\mathcal{N}_{\alpha}
\mathcal{N}_{\beta }}\varepsilon
\int_{0}^{1}\frac{n(n+1)j_{n}(\rho z_{\alpha }\tilde{r})j_{n}(\rho z_{\beta }\tilde{r})+ \psi_{n}^{\prime }(\rho z_{\alpha }\tilde{r})\psi_{n}^{\prime }(\rho z_{\beta }\tilde{r})}
{\rho^{2}}d\tilde{r}  \label{intee} \\
& =\frac{z_{\alpha }^{1/2}z_{\beta }^{1/2}\gamma_{\alpha(e,n)}
\gamma_{\beta(e,n)}}{\mathcal{N}_{\alpha}\mathcal{N}_{\beta}} \varepsilon \frac{z_{\alpha }^{2}j_{n}(\rho z_{\alpha })\psi_{n}^{\prime}
(\rho z_{\beta})-z_{\beta }^{2}j_{n}(\rho z_{\beta })\psi _{n}^{\prime}(\rho z_{\alpha})}{ \rho^{2} \left(z_{\alpha}^{2}-z_{\beta}^{2}\right)}  \label{inteeint} \\
& =\frac{z_{\alpha}^{3/2}z_{\beta}^{3/2}}{\mathcal{N}_{\alpha}\mathcal{N}_{\beta}}\left[  \frac{z_{\alpha}h_{n}(z_{\alpha})h_{n}^{\prime}(z_{\beta})-z_{\beta}h_{n}(z_{\beta})h_{n}^{\prime}(z_{\alpha})}{z_{\alpha}^{2}-z_{\beta}^{2}}+\frac{h_{n}(z_{\beta})h_{n}(z_{\alpha})}
{z_{\alpha}z_{\beta}}\right] \;, \label{rltRelec}
\end{align}
where the RHS of \cref{intee} is obtained after analytical integration
of the angular variables, while \cref{inteeint} follows from \cref{intee}
via the analytic result of Eq.(106) in ref.~\cite{McPhStout20}. The final
result of \cref{rltRelec} exploits the expressions of \cref{egamma}
for $\gamma_{e,n,\alpha }$ and is only true provided that both $z_{\alpha}$
and $z_{\beta}$ are RS (size parameter) eigenvalues.

One finishes the electric field inner product by carrying out the volume integral in the region outside the scatterer (\textit{i.e.} for $r>R$),
\begin{align} & \int_{R}^{\infty}d\boldsymbol{r}\boldsymbol{E}_{\alpha(e,n,m,\ell) }(\boldsymbol{r})\cdot \boldsymbol{E}_{\beta(e,n,m,\ell^{\prime})}(\boldsymbol{r})\notag \\ &=\frac{z_{\alpha}^{1/2}z_{\beta}^{1/2}}{\mathcal{N}_{\alpha} \mathcal{N}_{\beta}}\underset{\eta \rightarrow 0}{\text{lim}}\int_{1}^{\infty} \left\{n(n+1)h_{n}(z_{\alpha}\tilde{r})h_{n}(z_{\beta}\tilde{r})+\xi _{n}^{\prime}(z_{\alpha}\tilde{r}) \xi_{n}^{\prime }(z_{\beta }\tilde{r})\right\} e^{-\eta \tilde{r}^{2}}d\tilde{r} \label{exteeint} \\ & =-\frac{z_{\alpha}^{3/2}z_{\beta}^{3/2}}{\mathcal{N}_{\alpha} \mathcal{N}_{\beta}}\left[  \frac{z_{\alpha}h_{n}(z_{\alpha}) h_{n}^{\prime}(z_{\beta})-z_{\beta}h_{n}(z_{\beta})h_{n}^{\prime}(z_{\alpha})} {z_{\alpha}^{2}-z_{\beta}^{2}}+\frac{h_{n}(z_{\alpha})h_{n}(z_{\beta})}{z_{\alpha}z_{\beta}}\right]  \;, \label{rgtRelec} \end{align}\label{elecorthap}\end{subequations}
where we used Eq.(111a) in ref.~\cite{McPhStout20} to evaluate the integral in
the first line of this equation.  Simply adding \cref{rltRelec,rgtRelec} then immediately leads to the orthogonality of the electric
type RSs in lossless media when integrating their electric fields over all space, 
\begin{equation}
\int_{\mathcal{V}_{\infty}}d\boldsymbol{r} \varepsilon (\boldsymbol{r}) \boldsymbol{E}_{\alpha }(\boldsymbol{r})\cdot  \boldsymbol{E}_{\beta}(\boldsymbol{r})=0\;. \end{equation}

All the calculations for the magnetic modes when $\alpha\neq\beta$ are completely analogous to those of this section and those of \cref{sssect:RSmult} in the main text up to overall sign factors (which we did not explicit here for sake of clarity).

\section{Resonant state normalization for Mie theory: dispersive media}
\label{Norminteg}

\subsection{Magnetic mode normalization}

The normalization for magnetic (TE) mode fields is found through,
\begin{align}
z_{\alpha(h,n,m,\ell)}=\int_{\mathcal{V}_{\infty}} d\mathbf{r}\left\{\left. \frac{d\left[ \omega\varepsilon
(\boldsymbol{r},\omega)  \right]} {d\omega}\right|_{\omega_{\alpha}}\boldsymbol{E}_{\alpha}^{2} (\boldsymbol{r}) 
-\left. \frac{d\left[  \omega\mu (\boldsymbol{r},\omega)  \right]}
{d\omega}\right|_{\omega_{\alpha}} \boldsymbol{H}_{\alpha}^{2} (\boldsymbol{r}) \right\}  \;. \label{Hnrmcond}
\end{align}
For clarity, the overall magnetic multipole mode sign factor, $(-1)^{n}$, resulting from the $i^{n}$ phase factor in multipole fields (cf. \cref{Psiout,Psint}) is omitted in this section since it has no effect on normalization and is explicitly included in the response functions expansions of \cref{app:Mietheory}.

Denoting by, $\varepsilon(\omega)$, the permittivity of the sphere normalized with respect to exterior medium, the contribution to the electric field integral from the volume inside the sphere is,
\begin{subequations}
\begin{align}\begin{split}
&  \left. \frac{d\left[ \omega\varepsilon (\omega) \right] } {d\omega}\right|_{\omega_{\alpha}} \int_{0}^{R} d\mathbf{r}\boldsymbol{E}_{\alpha(h,n,m,\ell)}^{2}(\boldsymbol{r}) =\frac{z_{\alpha}^{3} \gamma_{\alpha(h,n)}^{2}}{\mathcal{N}_{\alpha}^{2}} \left. \frac{d\left[ \omega\varepsilon(\omega)  \right]} {d\omega}\right|_{\omega_{\alpha}} \int_{0}^{1}\tilde{r}^{2}j_{n}^{2} (\rho_{\alpha}z_{\alpha}\tilde{r}) d\tilde{r}\\ &  =\frac{z_{\alpha} \gamma_{\alpha(h,n)}^{2}} {2\mathcal{N}_{\alpha}^{2}} \left. \frac{d\left[ \omega\varepsilon(\omega)\right]} {d\omega}\right|_{\omega_{\alpha}} \frac{\left[ \psi_{n}^{\prime}(\rho_{\alpha}z_{\alpha}) \right]^{2} +\psi_{n}^{2}(\rho_{\alpha}z_{\alpha}) -n(n+1)  j_{n}^{2}(\rho_{\alpha}z_{\alpha}) -j_{n}( \rho_{\alpha}z_{\alpha})\psi_{n}^{\prime} ( \rho_{\alpha}z_{\alpha})}{\varepsilon_{\alpha}\mu_{\alpha}}\\ &  = \frac{z_{\alpha}}{2\mathcal{N}_{\alpha}^{2}}  \left. \frac{d\left[ \omega\varepsilon(\omega)  \right]}{d\omega}\right|_{\omega_{\alpha}} \frac{\Xi_{h,n}^{(-)}} {\varepsilon_{\alpha}}= \frac{z_{\alpha}}{2} \frac{\Xi_{h,n}^{(-)}  +\Xi_{h,n}^{(-)} \omega_{\alpha} \frac{d}{d\omega}\left. \ln\varepsilon(\omega)\right|_{\omega_{\alpha}}} {\mathcal{N}_{\alpha}^{2}} \;,\label{Eh1}
\end{split}\end{align}
where we used \cref{hgamma} for $\gamma_{\alpha(n,h)}$ and the expression of \cref{Xihdef} for $\Xi^{(h,\pm)}_{n}$.

Denoting by $\mu(\omega)$, the magnetic permeability of the sphere divided by permeability of the external medium,
the magnetic field integral of \cref{Hnrmcond} is,
\begin{align}
&  -\left. \frac{d\left[ \omega\mu(\omega)\right]} {d\omega} \right|_{\omega_{\alpha}} \int_{0}^{R}d\mathbf{r}\boldsymbol{H}_{\alpha(h,n,m,\ell)}^{2}(\boldsymbol{r}) =\left. \frac{d\left[ \omega\mu(\omega)\right]}{d\omega} \right|_{\omega_{\alpha}} \frac{z_{\alpha} \gamma_{\alpha(h,n)}^{2}}{\mathcal{N}_{\alpha}^{2}}\frac{\varepsilon_{\alpha}}{\mu_{\alpha}}\int_{0}^{1}\frac{n(n+1) j_{n}^{2}(\rho_{\alpha}z_{\alpha} \tilde{r}) +\left[ \psi_{n}^{\prime}(\rho_{\alpha}z_{\alpha} \tilde{r}) \right]^{2} } {\rho_{\alpha}^{2}}d\tilde{r}\nonumber\\ &  =\left. \frac{d\left[ \omega\mu(\omega)\right]}{d\omega} \right|_{\omega_{\alpha}} \frac{z_{\alpha} \gamma_{\alpha(h,n)}^{2}}{\mathcal{N}_{\alpha}^{2}} \frac{\varepsilon_{\alpha}}{\mu_{\alpha}} \frac{\left[\psi_{n}^{\prime}(\rho_{\alpha}z_{\alpha})\right]^{2}+\psi_{n}^{2} (\rho_{\alpha}z_{\alpha})-n(n+1) j_{n}^{2}(\rho_{\alpha}z_{\alpha}) + j_{n}( \rho_{\alpha}z_{\alpha})\psi_{n}^{\prime}(\rho_{\alpha}z_{\alpha})} {2\varepsilon_{\alpha}\mu_{\alpha}} \label{Hh1} \\ &  =\frac{z_{\alpha}}{2 \mathcal{N}_{\alpha}^{2}} \left. \frac{d\left[ \omega\mu(\omega)\right]}{d\omega} \right|_{\omega_{\alpha}} \frac{\Xi^{(+)}_{h,n}}{\mu_{\alpha}} = \frac{z_{\alpha}}{2} \frac{\Xi^{(+)}_{h,n}  +\Xi^{(+)}_{h,n}\omega_{\alpha}\left. \frac{d}{d\omega}\ln\mu (\omega)\right|_{\omega_{\alpha}}}{\mathcal{N}_{\alpha}^{2}} \;,\nonumber \end{align} \label{} \end{subequations}
where the $\Xi^{(+)}_{h,n}$ function was again defined in \cref{Xihdef}.

The exterior integrals are again finite despite their divergent kernels following our regularization approach,
\begin{subequations}\begin{align}\begin{split} \int_{R}^{\infty}d\mathbf{r} \boldsymbol{E}_{\alpha(h,n,m,\ell)}^{2} (\boldsymbol{r})  & \longrightarrow z_{\alpha}^{3} \lim_{\eta \rightarrow 0} \int_{1}^{\infty} \tilde{r}^{2} h_{n}^{2}(z_{\alpha} \tilde{r})  e^{-\eta \tilde{r}^2} d\tilde{r} \\ &=\frac{z_{\alpha}}{2} \frac{ n(n+1) h_{n}^{2}(z_{\alpha}) -\left[  \xi_{n}^{\prime}(z_{\alpha}) \right]^{2} -\xi_{n}^{2}(z_{\alpha}) + h_{n}(z_{\alpha}) \xi_{n}^{\prime}(z_{\alpha})} {\mathcal{N}_{\alpha}^{2}} \:, \label{Eh2} \end{split}\end{align}
and,
\begin{align}\begin{split}
-\int_{R}^{\infty}d\mathbf{r}\boldsymbol{H}_{\alpha(h,n,m,\ell)}^{2} (\boldsymbol{r}) & \longrightarrow z_{\alpha} \lim_{\eta \rightarrow 0} \int_{1}^{\infty} \left\{ n(n+1) h_{n}^{2}(z_{\alpha} \tilde{r}) +\left[ \xi_{n}^{\prime}(z_{\alpha} \tilde{r}) \right]^{2}\right\} e^{-\eta \tilde{r}^2} d\tilde{r}\\ &= \frac{z_{\alpha}}{2}\frac{n(n+1)h_{n}^{2}(z_{\alpha}) -\left[\xi_{n}^{\prime}(z_{\alpha})\right]^{2} -\xi_{n}^{2}(z_{\alpha})  - h_{n}(z_{\alpha}) \xi_{n}^{\prime}(z_{\alpha})} { \mathcal{N}_{\alpha}^{2}} \;,\label{Hh2} \end{split}\end{align} \label{extint} \end{subequations}
and the sum of the above two integrals yields:
\begin{align} \lim_{\eta \rightarrow 0} \int_{R}^{\infty} e^{-\eta \tilde{r}^2} d\mathbf{r} \left\{ \boldsymbol{E}_{\alpha}^{2}  (\boldsymbol{r}) -\boldsymbol{H}_{\alpha}^{2}(\boldsymbol{r}) \right\}   = z_{\alpha} \frac{ n(n+1) h_{n}^{2}(z_{\alpha}) -\left[ \xi_{n}^{\prime}(z_{\alpha}) \right]^{2} -\xi_{n}^{2}(z_{\alpha})} {\mathcal{N}_{\alpha}^{2}} \;. \label{EHh2} \end{align}

Finally, putting together the results of \cref{Eh1,Hh1,EHh2} into \cref{Hnrmcond}, one finds the normalization factor of \cref{hNnorm} for magnetic type resonant states for spherical scatterers with full temporal dispersion. 

\subsection{Electric mode normalization}

The normalization condition for electric (TM) modes is strictly analogous to those of the magnetic (TE) modes above with:
\begin{align}\begin{split} z_{\alpha(e,n,m,\ell)} = \int_{\mathcal{V}_{\infty}} d\boldsymbol{r}\left\{ \left. \frac{d\left[\omega \varepsilon(\boldsymbol{r},\omega) \right]}{d\omega} \right|_{\omega_{\alpha}} \boldsymbol{E}_{\alpha}^{2}(\boldsymbol{r})  -\left. \frac{d\left[  \omega\mu \left(  \boldsymbol{r},\omega\right)  \right]} {d\omega}\right|_{\omega_{\alpha}}  \boldsymbol{H}_{\alpha}^{2}(\boldsymbol{r}) \right\}\;,\label{Enrmcond}
\end{split}\end{align}
where the overall electric multipole mode sign factor, $(-1)^{n-1}$, will be henceforth suppressed for simplicity like we did above for the magnetic modes above.

The $E$-field integration inside the sphere is,
\begin{subequations}\begin{align}\begin{split} & \left. \frac{d\left[ \omega\varepsilon(\omega)\right]}{d\omega}\right|_{\omega_{\alpha}}  \int_{0}^{R}d\boldsymbol{r} \boldsymbol{E}_{\alpha(e,n,m,\ell)}^{2}(\boldsymbol{r}) =\frac{z_{\alpha} \gamma_{\alpha(e,n)}^{2}} {\mathcal{N}_{\alpha}^{2}} \left. \frac{d\left[ \omega\varepsilon(\omega) \right]} {d\omega}\right|_{\omega_{\alpha}}  \int_{0}^{1}\frac{ n(n+1) j_{n}^{2}(\rho_{\alpha}z_{\alpha} r)   +\left[ \psi_{n}^{\prime}(\rho_{\alpha}z_{\alpha} r) \right]^{2} } {\rho_{\alpha}^{2}}dr\\ &=\frac{z_{\alpha}\gamma_{\alpha(e,n)}^{2}}{\mathcal{N}_{\alpha}^{2}}  \left. \frac{d\left[ \omega\varepsilon(\omega)  \right]}{d\omega}\right|_{\omega_{\alpha}} \frac{\left[  \psi_{n}^{\prime}(\rho_{\alpha}z_{\alpha})\right]^{2}+\psi_{n}^{2}(\rho_{\alpha}z_{\alpha})  -n(n+1)  j_{n}^{2}(\rho_{\alpha}z_{\alpha})  + j_{n}(\rho_{\alpha}z_{\alpha}) \psi_{n}^{\prime}(  \rho_{\alpha}z_{\alpha})} {2\varepsilon_{\alpha}\mu_{\alpha}}\\ &  =\frac{z_{\alpha}}{2\mathcal{N}_{\alpha}^{2}} \frac{\Xi^{(e,+)}_{n}}{\varepsilon_{\alpha}} \left. \frac{d\left[ \omega\varepsilon(\omega)\right]}{d\omega}\right|_{\omega_{\alpha}} =\frac{z_{\alpha}}{2} \frac{\Xi_{n}^{(e,+)} +\Xi_{n}^{(e,+)}\omega_{\alpha} \left. \frac{d}{d\omega} \ln\varepsilon_{\alpha} (\omega)\right|_{\omega_{\alpha}} }{\mathcal{N}_{\alpha}^{2}} \;,\label{Ee1} \end{split}\end{align}
where we used \cref{egamma} for $\gamma_{e,n,\alpha}$
and used the definition of \cref{Xiedef} for $\Xi^{(e,\pm)}_{n}$.


The $H$-field integral inside the sphere is,
\begin{align}\begin{split} &  - \left. \frac{d\left[ \omega\mu(\omega)\right]}{d\omega} \right|_{\omega_{\alpha}} \int_{0}^{R} d\boldsymbol{r}\boldsymbol{H}_{\alpha(e,n,m,\ell)}^{2}(\boldsymbol{r}) =  \left. \frac{d\left[ \omega\mu(\omega)\right]}{d\omega} \right|_{\omega_{\alpha}} \frac{z_{\alpha}^{3} \gamma_{\alpha(e,n)}^{2}}{\mathcal{N}_{\alpha}^{2}}\frac{\varepsilon_{\alpha}}{\mu_{\alpha}} \int_{0}^{1} \tilde{r}^{2}j_{n}^{2} (\rho_{\alpha} z_{\alpha}\tilde{r}) d\tilde{r}\\ &=\frac{z_{\alpha}\gamma_{\alpha(e,n)}^{2}} {\mathcal{N}_{\alpha}^{2}} \frac{1}{\mu_{\alpha}} \left. \frac{d\left[ \omega\mu(\omega)\right]}{d\omega}  \right|_{\omega_{\alpha}} \varepsilon_{\alpha} \frac{\left[  \psi_{n}^{\prime}(\rho_{\alpha}z_{\alpha})\right]^{2} +\psi_{n}^{2}(\rho_{\alpha}z_{\alpha}) -n(n+1) j_{n}^{2}(\rho_{\alpha}z_{\alpha}) -j_{n}(\rho_{\alpha}z_{\alpha})  \psi_{n}^{\prime}(\rho_{\alpha}z_{\alpha})} {2\varepsilon_{\alpha}\mu_{\alpha}}\\ &  =\frac{z_{\alpha}}{2\mathcal{N}_{\alpha}^{2}}\frac{\Xi_{n}^{(e,-)}}{\mu_{\alpha}} \left. \frac{d\left[ \omega\mu(\omega)\right]}{d\omega} \right|_{\omega_{\alpha}}=\frac{z_{\alpha}}{2} \frac{\Xi_{n}^{(e,-)}+\omega_{\alpha} \Xi_{n}^{(e,-)}\left. \frac{d}{d\omega}\ln\mu_{s}(\omega) \right|_{\omega_{\alpha}}}{\mathcal{N}_{\alpha}^{2}} \;,\label{He1} \end{split} \end{align} \end{subequations}
where $\Xi_{e,n}^{(-)}$ is defined in \cref{Xiedef}.

The integrals outside the sphere are identical to those of \cref{extint} with the roles of electric and magnetic fields reversed, so we obtain
the same result as \cref{EHh2} for the field integrals in a region exterior to a sphere of radius $R$. Inserting the results of \cref{Ee1,He1,EHh2} into \cref{Enrmcond}, one finds the normalization factor of \cref{eNnorm} for electric type resonant states for spherical scatterers with full temporal dispersion. 


\section{Drude models for fitting dispersion relations}
\label{DrudeMod}

\begin{figure}[htb]
\includegraphics[width=\linewidth]{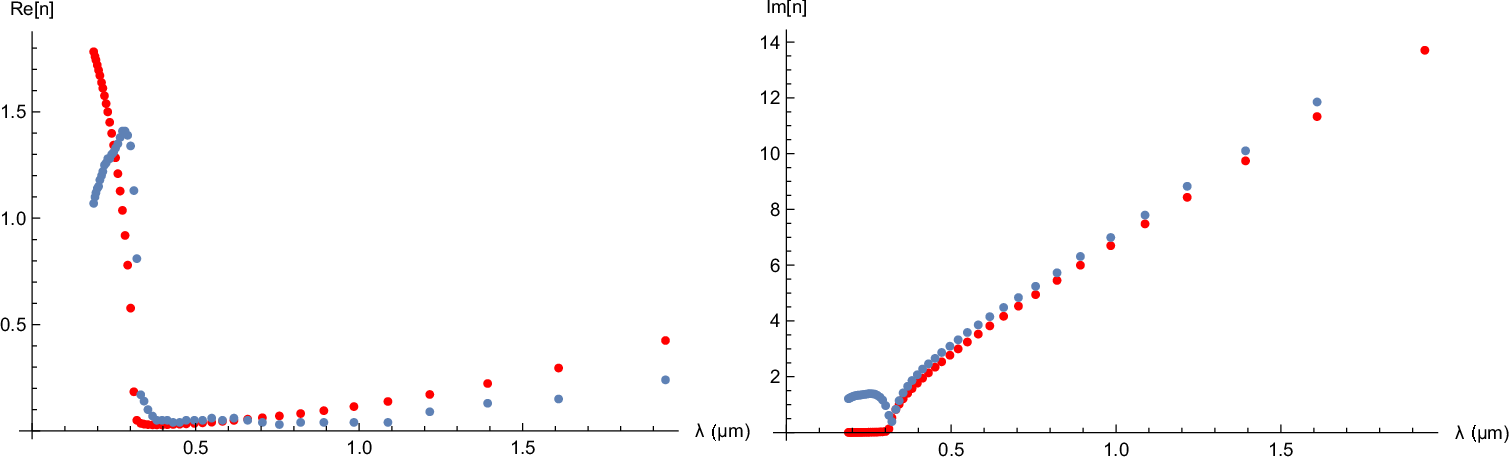}
\caption{\label{fig:cfAg}The real (left) and imaginary (right) parts of the complex refractive index of silver as a function of wavelength. Blue:
experimental data from Johnson and Christy~\cite{Johnson_Christy1972}; red-the Drude model of equation (\ref{drudeAgeq}).}
\end{figure}

\begin{figure}[htb]
\includegraphics[width=\linewidth]{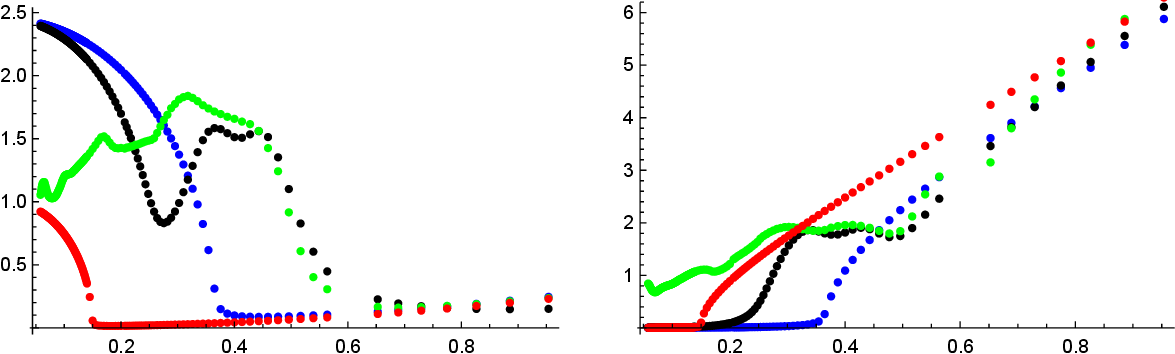}
\caption{\label{fig:cfgold}The real (left) and imaginary (right) parts of the complex refractive index of gold as a function of wavelength. Green:
experimental data from Palik; red- the Drude model of equation (\ref{drudeeq}); blue an alternative Drude model with $\epsilon_{\rm DC} =5.9752$; black - the Drude-Lorentz model of equation (\ref{drudep2}).}
\end{figure}

Drude models and its extensions can be used to fit experimental measurements like those of Johnson and Christy for silver and gold as shown respectively in \cref{fig:cfAg}, and \cref{fig:cfgold}.

\vfill\eject


\end{document}